\documentclass[usenatbib]{mnras}
\usepackage{graphicx}
\usepackage{amsmath}
\usepackage{breqn}
\usepackage{amssymb}
\usepackage{hyperref}
\usepackage{xcolor}

\newcommand{\msun}{\mathrm{M_{\odot}}}

\title[Simulating high-z BH growth from heavy seeds]{Growth of high redshift supermassive black holes from heavy seeds in the \texttt{BRAHMA} cosmological simulations: Implications of overmassive black holes}
\author[Bhowmick et al.]{Aklant K. Bhowmick$^{1}$,
Laura Blecha$^{1}$,
Paul Torrey$^{2}$, Rachel S Somerville$^{8}$ \newauthor
Luke Zoltan Kelley$^{4}$, 
Mark Vogelsberger$^{5}$,
Rainer Weinberger$^{6}$,
Lars Hernquist$^{7}$,
Aneesh Sivasankaran$^{1}$
\\
$^{1}$Dept. of Physics, University of Florida, Gainesville, FL 32611, USA\\
$^{2}$ Department of Astronomy, University of Virginia, Charlottesville, VA 22904, USA\\
$^{3}$ NSF AI Planning Institute for Physics of the Future, 
Carnegie Mellon  University, Pittsburgh, PA 15213, USA \\
$^{4}$Department of Astronomy, University of California at Berkeley, Berkeley, CA 94720, USA\\
$^{5}$Dept. of Physics, Kavli Institute for Astrophysics and Space Research, Massachusetts Institute of Technology,
Cambridge, MA 02139, USA \\
$^{6}$Leibniz Institute for Astrophysics Potsdam (AIP), An der Sternwarte 16, 14482 Potsdam, Germany\\
$^{7}$Harvard-Smithsonian Center for Astrophysics, 60 Garden Street, Cambridge, MA 02138, USA\\
$^{8}$Center for Computational Astrophysics, Flatiron institute, New York, NY 10010, USA\\
}
\begin{document}
\maketitle
\begin{abstract}
JWST has recently revealed a large population of accreting black holes (BHs) in the early Universe. Even after accounting for possible systematic biases, the high-z $M_*-M_{\rm \rm bh}$ relation derived from these objects by \citet[][P23 relation]{2023ApJ...957L...3P} is above the local scaling relation by $>3\sigma$. To understand the implications of potentially overmassive high-z BH populations, we study the BH growth at $z\sim4-7$ using the $[18~\mathrm{Mpc}]^3$ \texttt{BRAHMA} suite of cosmological simulations with systematic variations of heavy seed models that emulate direct collapse black hole~(DCBH) formation. In our least restrictive seed model, we place $\sim10^5~M_{\odot}$ seeds in halos with sufficient dense and metal-poor gas. To model conditions for direct collapse, we impose additional criteria based on a minimum Lyman Werner flux~(LW flux $=10~J_{21}$), maximum gas spin, and an environmental richness criterion. The high-z BH growth in our simulations is merger dominated, with a relatively small contribution from gas accretion. For the most restrictive simulation that includes all the above seeding criteria for DCBH formation, the high-z $M_*-M_{\rm bh}$ relation falls significantly below the P23 relation~(by factor of $\sim10$ at $z\sim4$). Only by excluding the spin and environment based criteria, and by assuming $\lesssim750~\mathrm{Myr}$ delay times between host galaxy mergers and subsequent BH mergers, are we able to reproduce the P23 relation. Overall, our results suggest that if high-z BHs are indeed systematically overmassive, assembling them would require more efficient heavy seeding channels, higher initial seed masses, additional contributions from lighter seeds to BH mergers, and / or more efficient modes for BH accretion.       
\end{abstract} 


\section{Introduction}
The James Webb Space Telescope~(JWST) is transforming the observational landscape of supermassive black holes~(SMBHs). Prior to JWST, the observed BH population at high redshifts~($z\sim4-7.5$) was confined to the most luminous quasars powered by BHs between $\sim10^9-10^{10}~M_{\odot}$~\citep{2001AJ....122.2833F,2010AJ....139..906W,2011Natur.474..616M,2015MNRAS.453.2259V,2016ApJ...833..222J,2016Banados,2017MNRAS.468.4702R,2018ApJS..237....5M,2018ApJ...869L...9W,2018Natur.553..473B,2019ApJ...872L...2M,2019AJ....157..236Y,2021ApJ...907L...1W}. JWST is pushing these frontiers by revealing a large population of fainter broad line~(BL) Active Galactic Nuclei~(AGN) candidates at $z\sim4-11$~\citep{2023ApJ...942L..17O,2023ApJ...959...39H,2023ApJ...954L...4K,2023arXiv230801230M,2023ApJ...953L..29L,2023arXiv230905714G,2024arXiv240403576K,2024A&A...685A..25A}.  About $\sim20\%$ of these BL AGN are compact and heavily obscured, characterised by a steep red continuum in the rest frame optical along with relatively blue colors in the rest frame UV~\citep{2023ApJ...959...39H,2023ApJ...954L...4K,2023arXiv230801230M,2023arXiv230905714G,2023arXiv231203065K,2024ApJ...968...38K}. These objects are now commonly referred to as ``Little red dots" or LRDs ~\citep{2024ApJ...963..129M}.  \cite{2023arXiv230905714G} used follow-up NIRSpec/PRISM spectroscopy to demonstrate that 
$>80\%$ of LRDs in the UNCOVER sample contain AGN signatures in the form of broad emission lines 
after the brown dwarf contaminants are excluded~\citep{2023ApJ...957L..27L}.

For the spectroscopically confirmed AGNs, BH masses can be estimated based on the widths of the H$\alpha$ emission line, with measurements ranging from $\sim10^6-10^{8}~M_{\odot}$~\citep{2023ApJ...954L...4K,2023arXiv230801230M,2023ApJ...959...39H,2023ApJ...953L..29L,2023ApJ...942L..17O,2023A&A...677A.145U}. Concurrently, the unprecedented resolution and sensitivity of NIRCam imaging has also made it possible to detect the starlight from the host galaxies of some high-z BHs~\citep{2023Natur.621...51D,2023arXiv230904614Y,2023arXiv231018395S}. By fitting the resulting SEDs while accounting for the contributions from AGN, it has been possible to also measure the stellar masses of the host galaxies. All these developments have resulted in the first ever measurements of the $M_*$ vs $M_{\rm bh}$ relations at $z\gtrsim4$. Remarkably, the BH mass and host stellar mass measurements~(albeit with large uncertainties at present) indicate the possible existence of a substantial population of ``overmassive" BHs that are $\sim10-100$ times heavier than expectations from local BH scaling relations~\citep{2023ApJ...957L...7K,2024NatAs...8..126B,2024ApJ...960L...1N,2024ApJ...968...38K,2024arXiv240403576K,2024arXiv240610329D}. 

Despite these exciting developments, we still need to be cautious as the BH masses may be overestimated if for example, a portion of broadening in the emission lines is due to galactic scale outflows. At the same time, it is possible that the stellar masses are underestimated as it is often difficult to separate the stellar and AGN components in the SED~\citep{2020MNRAS.499.4325R}. In addition, \cite{2024arXiv240108782P} has recently used the MIRI-SMILES data to show that the LRD SED shapes may also be explained by compact starburst galaxies along with some~(sub-dominant) contribution from an AGN. Finally, even if the BH masses and host galaxy stellar masses are not \textit{systematically} biased, we may still be observing a biased population of high-z AGN that have significantly higher luminosities~(and therefore also BH masses) compared to typical AGN populations hosted by galaxies of a given mass. This is a consequence of the detection limits of surveys, and is referred to as \textit{Lauer bias}~\citep{2007ApJ...670..249L}. \cite{2023ApJ...957L...3P} showed that despite the possible systematic biases as well as measurement uncertainties, the $M_*$ vs $M_{\rm bh}$ measurements~(as they currently stand) of the spectroscopically confirmed AGNs still imply an \textit{intrinsic} high-z $M_*$ vs $M_{\rm bh}$ relation that is above the local scaling relations at the $>3\sigma$ confidence level. However, \cite{2024arXiv240300074L} did a similar analysis of systematic biases and concluded that the intrinsic high-z relation is consistent with the local scaling relations. All these developments imply that despite these JWST measurements, it is still not firmly established whether high-z BH populations are indeed overmassive compared to their local counterparts. Overall, while we are still in the earliest stages of characterizing the high-z AGN populations, it is clear that JWST is well is on its way towards revolutionizing our understanding of early BH growth.

Along with the brightest high-z quasars discovered in the pre-JWST era, the possibility of overmassive high-z BH populations is expected to have strong implications for BH seeding. Possible candidates for the first seeds of SMBHs include ``light seeds"~($\sim10^2-10^3~\mathrm{M_{\odot}}$) as Population III stellar remnants~\citep{2001ApJ...550..372F,2001ApJ...551L..27M,2013ApJ...773...83X,2018MNRAS.480.3762S}, ``intermediate-mass seeds"~($\sim10^3-10^4~\mathrm{M_{\odot}}$) as remnants from runaway stellar and BH collisions in dense nuclear star clusters~\citep{2011ApJ...740L..42D,2014MNRAS.442.3616L,2020MNRAS.498.5652K,2021MNRAS.503.1051D,2021MNRAS.tmp.1381D}, and ``heavy seeds"~($\gtrsim10^4~M_{\odot}$) as direct collapse black holes or DCBHs~\citep{2003ApJ...596...34B,2006MNRAS.370..289B,2014ApJ...795..137R,2016MNRAS.458..233L,2018MNRAS.476.3523L,2019Natur.566...85W,2020MNRAS.492.4917L,2023MNRAS.526L..94B,2023arXiv230402066M}. The heaviest DCBH seeds have long been considered to be too rare to explain the entirety of the observed SMBH populations. However, the presence of overmassive BHs was predicted to be one of their key observational signatures~\citep{2013MNRAS.432.3438A,2017ApJ...838..117N,2024ApJ...960L...1N}.  
Therefore, these JWST detections have sparked a renewed interest in heavy DCBH seeding channels~\citep{2023ApJ...957L...3P,2024arXiv240218773J}.   

In contrast to the light and intermediate-mass seeds that form within star forming regions, the heavy DCBH seeds form when fragmentation and star formation is prevented during gravitational collapse of a gas cloud. Instead of fragmenting, the gas must undergo a nearly isothermal collapse at temperatures above $\sim10^4~\mathrm{K}$. In addition, large gas inflow rates~($\gtrsim0.1~M_{\odot}\mathrm{yr}^{-1}$ at a few tens of $\mathrm{pc}$ scales sustained for $\sim10~\mathrm{Myr}$) are required to form a massive compact object~\citep[e.g.,][]{2010MNRAS.402..673B,2012ApJ...756...93H,2013ApJ...778..178H,2013A&A...558A..59S,2020OJAp....3E...9R,2021A&A...652L...7H}. To keep the gas from cooling below $\sim10^4~\mathrm{K}$, we first need pristine environments to prevent efficient metal cooling. In addition, we also need to prevent cooling due to molecular Hydrogen~($H_2$). This could be potentially achieved by destroying the $H_2$ with sufficient ultraviolet~(UV) radiation in the Lyman-Werner (LW) band~($11.2-13.5~\rm eV$) provided by nearby star forming regions. However, the estimated values of the critical LW flux~($J_{\mathrm{crit}}$) are extremely high~($\gtrsim 1000~J_{21}$ where $\rm J_{21}=10^{-21}~erg~s^{-1}
~cm^{-2}~Hz^{-1}
sr^{-1}$) according to radiation hydrodynamic simulations~\citep{2010MNRAS.402.1249S} as well as one-zone chemistry models~\citep{2014MNRAS.445..544S,2017MNRAS.469.3329W}. Several previous works have shown that these critical fluxes are exceedingly difficult to achieve, particularly in pristine gas environments with no prior star formation history ~\citep{2008MNRAS.391.1961D,2016MNRAS.463..529H,2022MNRAS.510..177B}. In addition to molecular hydrogen cooling, having high angular momentum may also impede the gas from achieving the required inflow rates of $\gtrsim0.1~M_{\odot}\mathrm{yr}^{-1}$. This further restricts the number of feasible sites for DCBH formation~\citep{2006MNRAS.371.1813L}. However, in \cite{2022MNRAS.510..177B}, we showed that the supercritical LW flux requirement is generally much more restrictive compared to that of low gas spins.            

More recently, it has been found that dynamical heating during major mergers can compete against $H_2$ cooling and significantly lower the critical LW flux requirement~($\sim1-10~J_{21}$) compared to previous estimates~\citep{2019Natur.566...85W,2020OJAp....3E..15R,2020MNRAS.492.3021R}. \cite{2020OJAp....3E..15R} demonstrated that the combination of mild LW radiation and dynamical heating within metal free halos can lead to the formation of multiple BHs via direct gas collapse, with masses ranging from $\sim300-10^4~M_{\odot}$. These BHs can eventually sink to the halo centers and merge with one another to form $\sim10^5~M_{\odot}$ DCBHs, which is close to the seed mass used in several large cosmological simulations~\citep[for e.g.][]{2015A&C....13...12N,2015MNRAS.450.1349K,2016MNRAS.455.2778F,2017MNRAS.470.1121T,2017MNRAS.467.4739K}. While the reduced critical LW fluxes can substantially improve the prospect of DCBH formation, they will still be limited to environments where major mergers occur. Therefore, while the possibility of overmassive high-z BHs may hint at heavy seeding origins, it is yet to be determined whether the existing DCBH formation mechanisms are sufficient for explaining these BH populations.    

Cosmological simulations allow us to predict the BH populations that assemble from a given seeding origin and directly compare against observations. However, while many cosmological simulations resolve down to the postulated DCBH seed masses~($\sim10^4-10^5~M_{\odot}$, see review by \citealt{2020NatRP...2...42V}), they cannot resolve the underlying physics that leads to the direct collapse. 
As a result, many cosmological simulations simply seed $\sim10^5-10^6~M_{\odot}$ BHs based on a halo or galaxy mass threshold~\citep[e.g.][]{2012ApJ...745L..29D,2014Natur.509..177V,2015MNRAS.450.1349K,2015MNRAS.452..575S,2019ComAC...6....2N}. With these prescriptions, the primary goal is not to emulate the specific conditions of DCBH formation but rather to capture the observational expectation that massive galaxies harbor SMBHs. 
Therefore, the main focus of these simulations has typically been on understanding the influence of SMBHs on the evolution of galaxies ~\citep{2018MNRAS.478.5063H,2020ApJ...895..102L}, instead of the origins of the SMBHs themselves. Recently, several simulations have incorporated more realistic seeding prescriptions to emulate DCBH forming criteria that are based on local gas properties. For example, \cite{2017MNRAS.470.1121T} and \cite{2019MNRAS.482.2913B} seed $\sim10^5~M_{\odot}$ BHs if an individual gas element has sufficiently high density, low metallicity and high temperatures. However, seeding solely based on individual gas cells could compromise the resolution convergence of seed formation~\citep[see Figure 10 of][]{2015MNRAS.448.1835T}. The resolution convergence may be improved by the slightly modified approach taken by \cite{2024arXiv240218773J}, which avoids the over-production of seeds at higher resolutions by requiring the gas cells in the entire neighborhood~(SPH smoothing kernel) of the seed formation site to satisfy the seeding criteria~(high density, low metallicity and high temperatures).     

As mentioned above, many cosmological simulations have modeled DCBH formation in environments with high gas densities and low gas metallicities. However, it is crucial also to consider other conditions that are potentially relevant for DCBH formation, including low gas angular momentum, sufficient LW radiation, and the presence of dynamical heating due to major mergers. To the best of our knowledge, currently there are no cosmological simulations that simultaneously consider all of the above conditions in their seed models. As a result, it is currently unclear how all these different conditions come together to impact the formation and growth of DCBHs and their ensuing feasibility in producing the overmassive JWST BHs. To further complicate matters, the growth of these DCBHs would also be impacted by their dynamics and gas accretion. Similar to seeding, all these different aspects of BH physics are also extremely challenging to model in cosmological simulations due to resolution limitations. Moreover, they may have a degenerate impact on BH growth. However, in order to disentangle the complex interplay between BH seeding, dynamics and accretion, it is often useful to first study them in isolation. To that end, in this work, we largely focus on exploring BH seeding within a fixed set of assumptions about BH accretion, feedback, and dynamics.

This paper introduces a set of multiple cosmological simulation boxes wherein we systematically vary the seeding prescriptions while all other aspects of the galaxy formation~(including BH dynamics and accretion) model are adopted from the \texttt{Illustris-TNG} simulation suite~\citep{2018MNRAS.475..676S,2018MNRAS.475..648P,2018MNRAS.475..624N,2018MNRAS.477.1206N,2018MNRAS.480.5113M,2019ComAC...6....2N}. We incrementally stack the different seeding conditions relevant for DCBH formation~(i.e. high density, low metallicty, low gas angular momentum, sufficient LW radiation, rich environment), and study their impact on the resulting BH populations to compare against the JWST results. All the seeding conditions have been developed and thoroughly tested for numerical convergence in our previous series of papers using cosmological zoom simulations~\citep{2021MNRAS.507.2012B,2022MNRAS.510..177B,2023arXiv230915341B}. These new simulations are part of the \texttt{BRAHMA} simulation suite introduced in our previous paper~\citep{2024arXiv240203626B}. We envision \texttt{BRAHMA} to eventually become a large suite of simulations encompassing a wide range of possible scenarios for BH seeding. While \cite{2024arXiv240203626B} focused on low mass seed models that emulated Pop III or NSC seeding conditions, this paper focuses on heavy seed models that emulate DCBH seeding conditions. 

The structure of this paper is as follows. 
Section \ref{methods} introduces the methods, including the basic simulation setup and the detailed implementation of all the seeding criteria used. 
Section \ref{results} describes the predictions of our different simulation boxes for the seed formation rates, AGN luminosity functions, merger rates, and finally the $M_*$-$M_{\rm bh}$ relations. 
Finally, Section \ref{conclusions} describes the main conclusions of our work.

\section{Methods}
\label{methods}
To run our simulations, we used the \texttt{AREPO} gravity + magneto-hydrodynamics~(MHD) solver~\citep{2010MNRAS.401..791S,2011MNRAS.418.1392P,2016MNRAS.462.2603P,2020ApJS..248...32W}. The gravity solver uses the PM Tree~\citep{1986Natur.324..446B} and the evolution of the gas is described by the ideal MHD equations solved over a dynamic unstructured grid generated via a Voronoi tessellation of the domain. All the simulations are characterised by the \cite{2016A&A...594A..13P} cosmology i.e.~$\Omega_{\Lambda}=0.6911, \Omega_m=0.3089, \Omega_b=0.0486, H_0=67.74~\mathrm{km}~\mathrm{sec}^{-1}\mathrm{Mpc}^{-1},\sigma_8=0.8159, n_s=0.9667$. Halos are identified using the friends of friends~(FOF) algorithm~\citep{1985ApJ...292..371D} with a linking length of 0.2 times the mean particle separation. Subhalos are computed using the \texttt{SUBFIND}~\citep{2001MNRAS.328..726S} algorithm.  

\subsection{Initial conditions}

We run our simulations using two distinct types of initial conditions~(ICs). Most of the paper will focus on simulations that adopt the usual approach wherein the ICs are generated from a random Gaussian field. These ``unconstrained" initial conditions are produced at $z=127$ using \texttt{MUSIC}~\citep{2011MNRAS.415.2101H}. These simulations have a comoving volume of $[18~\mathrm{Mpc}]^3$ and $512^3$ dark matter~(DM) particles. However, because these boxes are relatively small, they do not probe the entire range of galaxy masses detected by JWST. Therefore, we also run additional simulations where the initial conditions are ``constrained" to produce more overdense regions. These constrained initial conditions were generated at $z=99$ using the \href{https://github.com/yueyingn/GaussianCR}{\texttt{Gaussian-CR}} code~\citep{2022MNRAS.509.3043N} over a $[9~\mathrm{Mpc}]^3$ volume and $360^3$ DM particles~(see \citealt{2022MNRAS.509.3043N,2022MNRAS.516..138B} for more details).  As we shall see, even though the unconstrained simulations are slightly larger than the constrained simulations, the constrained simulations produce more massive galaxies as they simulate a much more overdense region~(i.e. $4\sigma$ overdensity at a scale of $1~\mathrm{Mpc}$). We chose the number of DM particles and initial gas cells such that for all of our simulations, the resulting DM mass resolution is $1.5\times10^6~\msun$ and the gas mass resolution is $\sim10^5~\msun$. 

\subsection{\texttt{Illustris-TNG} galaxy formation model}
With the exception of BH seeding, the \texttt{BRAHMA} simulations essentially adopt all the features of its predecessor \texttt{IllustrisTNG}~(TNG) simulation suite~\citep{2018MNRAS.475..676S,2018MNRAS.475..648P,2018MNRAS.475..624N,2018MNRAS.477.1206N,2018MNRAS.480.5113M,2019ComAC...6....2N} \citep[see also][]{2018MNRAS.479.4056W,2018MNRAS.474.3976G,2019MNRAS.485.4817D,2019MNRAS.484.5587T,2019MNRAS.483.4140R,2019MNRAS.490.3234N,2019MNRAS.490.3196P,2019MNRAS.484.4413H,2021MNRAS.500.4597U,2021MNRAS.503.1940H}. Here we summarize the core features of the \texttt{TNG} model that are most consequential to the seeding of BHs. 

The radiative cooling is implemented by including contributions from primodial species~($\mathrm{H},\mathrm{H}^{+},\mathrm{He},\mathrm{He}^{+},\mathrm{He}^{++}$ based on \citealt{1996ApJS..105...19K}) as well as metals. The metal cooling rates are interpolated from pre-calculated tables as in \cite{2008MNRAS.385.1443S} in the presence of a spatially uniform, time dependent UV background. The cooling of gas leads to the formation of dense gas, wherein star formation occurs at densities exceeding $0.13~\mathrm{cm}^{-3}$ with a time scale of $2.2~\mathrm{Gyr}$. The star forming gas cells represent an unresolved multiphase interstellar medium described by an effective equation of state~\citep{2003MNRAS.339..289S,2014MNRAS.444.1518V}. Star particles represent unresolved single stellar populations that are characterised by their ages and metallicities. The underlying initial mass function is adopted from \cite{2003PASP..115..763C}. The subsequent stellar evolution is modeled based on \cite{2013MNRAS.436.3031V} with modifications for \texttt{IllustrisTNG} as described in \cite{2018MNRAS.473.4077P}. The stellar evolution leads to chemical enrichment of stars, which is modelled by following the evolution of seven species of metals~(C, N, O, Ne, Mg, Si, Fe) in addition to H and He. Stellar and Type Ia/II Supernova feedback are modelled as galactic scale winds~\citep{2018MNRAS.475..648P} that deposit mass, momentum and metals on to the gas surrounding the star particles. This leads to the enrichment of gas, which is otherwise assigned an initial metallicity of $7\times10^{-8}~Z_{\odot}$. For readers interested in further details, please refer to \cite{2018MNRAS.473.4077P}.

BH accretion in \texttt{IllustrisTNG} is modeled based on the Eddington-limited Bondi-Hoyle formalism. The Eddington limit and the bolometric luminosities of the accreting BHs are computed based on an assumed radiative efficiency of $\epsilon_r=0.2$. \texttt{IllustrisTNG} implements a two-mode AGN feedback model. For high Eddington ratios, `thermal feedback' is implemented wherein a fraction of the radiated luminosity is deposited to the neighboring gas. For low Eddington ratios, feedback is in the form of kinetic energy that is injected onto the gas at irregular time intervals along a randomly chosen direction. Readers interested in further details about the TNG feedback model may refer to \cite{2017MNRAS.465.3291W} and \cite{2020MNRAS.499..768Z}~(see also \citealt{2017ApJ...837L..18K}). 

Our simulations cannot adequately capture the small-scale BH dynamics because the limited mass resolution prevents them from fully resolving the BH dynamical friction force. This is particularly true for the seed populations as the background DM particles are $\sim10$ times more massive. To prevent the seeds from encountering spuriously large kicks by the massive DM particles, BHs are ``re-positioned" to the nearest potential minimum within its ``neighborhood" defined by 64 nearest neighboring gas cells. The BHs are promptly merged when at least one of them is within the ``neighbor search radius"~($R_{\mathrm{Hsml}}$) of the other. Note that the resulting BH merger rates are inevitably overestimated compared to the actual event rates that may be detectable by facilities such as the Laser Interferometer Space Antenna or LISA~\citep{2017arXiv170200786A}. In Section \ref{Impact of delayed BH mergers}, we account for the impact of potential delays on the BH merger rates as well as the $M_{\rm bh}$ vs. $M_*$ relations at various redshifts.  

\subsection{Black hole seed models}
\label{Black hole seed models}

The key novel feature of the \texttt{BRAHMA} simulations is the implementation of a comprehensive BH seeding model. The full \texttt{BRAHMA} suite of simulations is comprised of several runs that span a wide range of seeding prescriptions with seed masses ranging from $\sim10^3-10^6~M_{\odot}$. In this work, we primarily focus on those simulations that model $M_{\mathrm{seed}}=1.2 \times 10^5~M_{\odot}$ seeds formed via the direct collapse of gas. Our seeding criteria are motivated by conditions that are believed to be ideal for DCBH formation, namely pristine dense gas with low angular momentum wherein cooling to temperatures below $\lesssim10^4$ K is suppressed by LW radiation and dynamical heating during halo mergers. To identify and seed BHs in these environments, we designed the following set of seeding criteria.
\begin{itemize}
\item \textit{Dense and metal-poor gas mass criterion}: Seeds are placed in halos that exceed a critical threshold of gas mass that is simultaneously dense~($>0.13~\mathrm{cm}^{-3}$; i.e. the star formation threshold) and metal-poor~($Z<10^{-4}~Z_{\odot}$). The threshold is chosen to be $5~M_{\mathrm{seed}}$.

\item \textit{Lyman-Werner flux criterion}: When this criterion is applied, the dense and metal poor gas mass is also required to be illuminated by a LW flux that exceeds the critical value $J_{\mathrm{crit}}$. Additionally, star formation is suppressed in seed forming regions with supercritical fluxes. In this work, we assume a critical LW flux of $10~\mathrm{J_{21}}$. We consider relatively low $J_{\mathrm{crit}}$ compared to values~($\gtrsim1000~J_{21}$) predicted by small scale radiation hydrodynamics simulations and one-zone chemistry models~\citep{2010MNRAS.402.1249S,2014MNRAS.445..544S}, as they have been shown to be too restrictive~\citep{2022MNRAS.510..177B} to form BH seeds. As noted earlier, such low $J_{\mathrm{crit}}$ values may be viable for DCBH formation in environments where the gas is also dynamically heated during halo mergers. Note that in the absence of an explicit treatment of radiative transfer, we use a semi-empirical approach to compute the LW radiation as detailed in Section 2.1.2 of \cite{2022MNRAS.510..177B}.   

\item \textit{Gas spin criterion}: Seeds are placed in halos where the net spin of the gas is smaller than the maximum threshold for the onset of Toomre instability i.e. \begin{equation}
    \lambda =\frac{|\vec{\mathbf{J}}_{\mathrm{spin}}|}{\sqrt{2}M_{\mathrm{gas}} R_{\mathrm{vir}} V_{\mathrm{vir}}} < \lambda_{\mathrm{max}}
    \label{lambda_eqn}
\end{equation}
where $\vec{\mathbf{J}}_{\mathrm{spin}}$ is the spin of gas which is expressed in dimensionless units as $\lambda$. $M_{\mathrm{gas}}$, $R_{\rm vir}$ and $V_{\rm vir}$ are the gas mass, virial radius and circular velocity respectively. $\lambda_{\rm max}$ is the Toomre instability threshold. This criterion was based on the stability analysis of pre-galactic disks and the collapse of unstable disks to form DCBHs, as done in \cite{2006MNRAS.371.1813L}. It was subsequently implemented in \cite{2012MNRAS.422.2051N} and \cite{2020MNRAS.491.4973D}. For further details underlying the implementation of this criterion, please refer to Section 2.1.1 of \cite{2022MNRAS.510..177B}.

\item \textit{Halo environment criterion}: This criterion ensures that seeds are placed in halos that have at least one neighboring halo of comparable or higher mass within a distance of $5$ times its virial radius. The choice of this distance is somewhat arbitrary, but it is small enough to ensure that the BHs are only forming in rich environments. We could make this distance smaller and make the criterion more restrictive, but as we shall see, our current choice already leads to an underprediction of the BH masses compared to the JWST results. Further details underlying the implementation of this criterion are described in \cite{2023arXiv230915341B}. We apply it to emulate the impact of dynamical heating of gas that may occur during major mergers of halos. This dynamical heating can contribute to the suppression of $H_2$ cooling, thereby potentially allowing DCBH formation under the relatively low critical LW flux of $10~J_{21}$ assumed by us~\citep{2019Natur.566...85W,2020OJAp....3E..15R,2020MNRAS.492.3021R}.  To that end, we note that our seed model is fully self-consistent only when both the \textit{LW flux criterion} and the \textit{halo environment criterion} are applied together. Nevertheless, we do run a simulation which only applies the \textit{LW flux criterion} in order to isolate its individual impact and compare against that of the \textit{halo environment criterion}.
\end{itemize}
All of the above seeding criteria were developed and tested within the \texttt{AREPO} code and the baseline TNG galaxy formation model using zoom simulations, over a series of prior papers~\citep{2021MNRAS.507.2012B,2022MNRAS.510..177B, 2023arXiv230915341B}.

\subsection{Simulation suite and nomenclature}

Recall that we are using two different types of simulations based on whether they use unconstrained or constrained ICs. We run simulations with both ICs for each of our seed models. 

To reasonably capture the impact of all the complex unresolved physics of DCBH formation in our simulations, we require all four seeding criteria described in the previous subsection. However, to test the importance of each criterion, we apply and stack them one at a time to produce four simulation boxes. The \texttt{SM5} box solely applies the \textit{dense and metal-poor gas mass criterion}. The \texttt{SM5_LW10} box adds the \textit{Lyman-Werner Flux criterion}. The \texttt{SM5_LW10_LOWSPIN} box additionally applies the \textit{gas-spin criterion}. Finally, the \texttt{SM5_LW10_LOWSPIN_RICH} box includes all four criteria by also adding the \textit{halo environment criterion}. In addition to the above simulations, we also show results from our predecessor \texttt{IllustrisTNG} simulations, particularly the highest resolution versions of the 100 Mpc~(\texttt{TNG100} with $1820^3$ DM particles) and 300 Mpc boxes~(\texttt{TNG300} with $2500^3$ DM particles)

\section{Results}
\label{results}

\begin{figure*}
\includegraphics[width=18 cm]{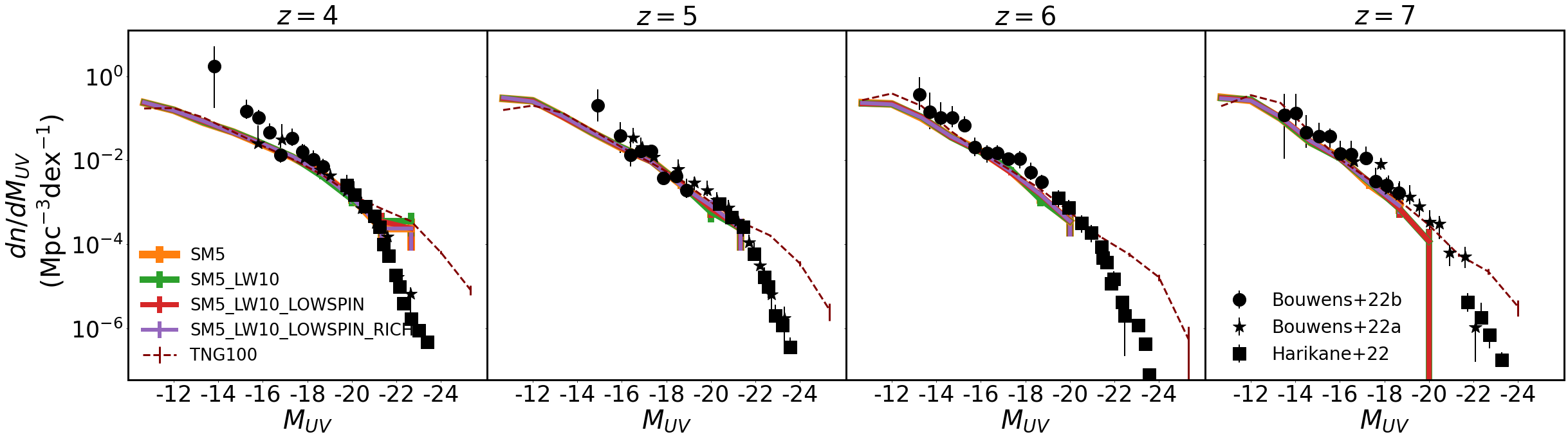}
\caption{Galaxy UV luminosity functions at $z=4,5,6~\&~7$ predicted by the unconstrained \texttt{BRAHMA} boxes, compared against measurements using Hubble Frontier Field observations~(black data points) from \protect\cite{2022ApJ...927...81B}, \protect\cite{2022ApJS..259...20H} and \protect\cite{2022ApJ...940...55B}. The maroon dashed lines show predictions from \texttt{TNG100}. The BH seed models have no significant consequence on the UV luminosity functions. The simulations and observations are broadly consistent with each other at $M_{UV}\gtrsim-20$). At the bright end~($M_{UV}\lesssim-20$), the absence of dust modeling causes \texttt{TNG100} to overestimate the UV LFs.}
\label{UVLFs_fig}
\includegraphics[width=18 cm]{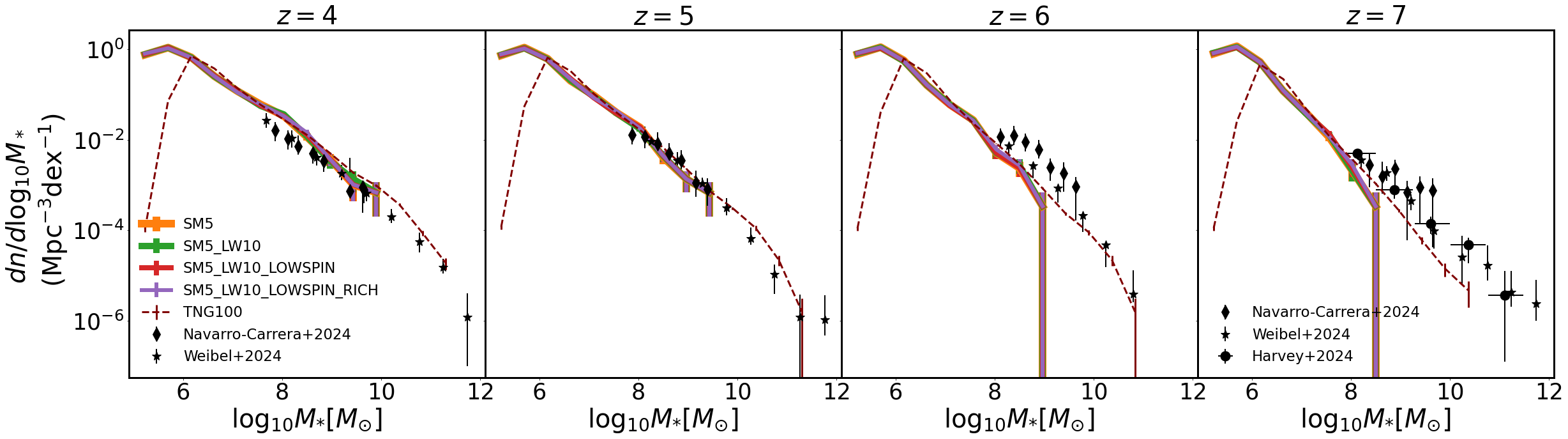}
\caption{Galaxy stellar mass functions at $z=4,5,6~\&~7$ predicted by the unconstrained \texttt{BRAHMA} boxes, compared against recent measurements from JWST observations~(black data points) from \protect\cite{2024ApJ...961..207N} and \protect\cite{2024arXiv240308872W}. The maroon dashed lines show predictions from \texttt{TNG100}. The BH seed models have no significant consequence on the stellar mass functions. Both \texttt{BRAHMA} and \texttt{TNG100} predictions are broadly consistent with the observations at $z\sim4~\&~5$, but the simulations tend to underpredict compared to the observations at $z\sim6~\&~7$.}
\label{stellar_mass_functions_fig}
\end{figure*}

\subsection{High Redshift Galaxy populations in \texttt{BRAHMA}}
\label{Galaxy populations in BRAHMA}
Since the stellar mass vs BH mass~($M_*$ vs. $M_{\rm bh}$) relations can be influenced not only by the BH masses but also by their host galaxy stellar masses, it is instructive to first look at how our simulations compare with observations in terms of the galaxy populations. In Figures \ref{UVLFs_fig} and \ref{stellar_mass_functions_fig}, we compare the UV luminosity functions~(LFs) and galaxy stellar mass functions~(GSMFs) between our unconstrained simulations and observations. The UV luminosities are computed from the global star formation rates of the galaxies as \begin{eqnarray}
M_{UV} = -2.5\log_{10}F_{UV} - 48.6 \\
L_{UV} = \log_{10}(\mathrm{SFR})+28.1427
 \end{eqnarray} where $M_{UV}$, $F_{UV}$, $L_{UV}$ and SFR are the absolute UV magnitude, UV flux, intrinsic UV luminosity and the star formation rate respectively, and the conversion from SFR to UV luminosity is taken from \citet{MD:2014} assuming a Chabrier IMF. Despite the small volumes, our unconstrained boxes are able to probe a substantial portion of the observed galaxy populations, with UV magnitudes up to $M_{UV}\sim-20$ and galaxy stellar masses up to $\sim10^{9.5}~M_{\odot}$ at $z\sim5$. This is particularly encouraging as it has a significant overlap with the range of measured stellar masses of the JWST AGN hosts~($\sim10^{7.5}-10^{11}~M_{\odot}$). Recall that we use the constrained boxes to further extend the range of galaxy masses probed. This makes our simulations an ideal arena to study the typical BH populations that are expected to reside in these galaxies under various assumptions for BH seeding. 
 
Firstly, we note that the different BH seed models have no significant impact on the predicted UV LFs and GSMFs; this may be because the BH accretion rates are not large enough to induce significant AGN feedback on the host galaxies. In fact, we also find that the  predictions between the unconstrained \texttt{BRAHMA} and \texttt{TNG100} simulations~(that seed BHs only based on halo mass) are very similar; this is not surprising as both \texttt{TNG100} and \texttt{BRAHMA} use the same underlying galaxy formation model~(except BH seeding). 
The \texttt{BRAHMA} and \texttt{TNG100} UV LFs are broadly consistent with pre-JWST observational constraints at $M_{UV}\gtrsim-20$~(the measurements shown as black points in Figure \ref{UVLFs_fig} are from the Hubble Frontier Field surveys). This is a testament of the remarkable success of the \texttt{Illustris-TNG} galaxy formation model as shown in \cite{2020MNRAS.492.5167V} in this redshift range. At the bright end ~($M_{UV}\lesssim-20$) that cannot be probed by the limited volume of \texttt{BRAHMA}, the overestimation of the \texttt{TNG100} UV LFs is due to the absence of a correction for dust attenuation in our calculation of the $M_{UV}$. \cite{2020MNRAS.492.5167V} showed that with the inclusion of dust, the simulations also reproduce the bright end of the observed UV LFs.   

The advent of JWST has now made it possible to also constrain the stellar mass functions at these redshifts~(see Figure \ref{stellar_mass_functions_fig}). At $z\sim4~\&~5$ respectively, the simulated GSMFs show reasonable agreement with observations. At higher redshifts~($z\sim6~\&~7$), the simulations start to underpredict the abundances of $\gtrsim10^{8}~M_{\odot}$ galaxies compared to the measurements. Additionally, we also know that at much higher redshifts~($z\gtrsim10$), TNG follow-up projects such as \texttt{THESAN}~\citep{2022MNRAS.511.4005K} and \texttt{Millennium-TNG}~\citep{2023MNRAS.524.2539P} also tend to underpredict the abundances of the galaxies observed by JWST~\citep{2023MNRAS.524.2594K}. Resolving these descrepancies would potentially require modifications to several aspects of our galaxy formation model such as star formation, metal enrichment and stellar feedback. This can also have substantial implications for BH seeding, which we shall explore in future work. For now, the fact that the \texttt{IllustrisTNG} galaxy formation model produces reasonable agreement between our simulations and JWST measurements at $z\sim4$ and 5, means that at least at these redshifts, any differences in the intrinsic $M_*$ vs. $M_{\rm bh}$ relations between the simulations and that inferred from observations, are likely to be much more readily attributable to the BH mass assembly rather than the galaxy stellar mass assembly.

\begin{figure}
\includegraphics[width=8 cm]{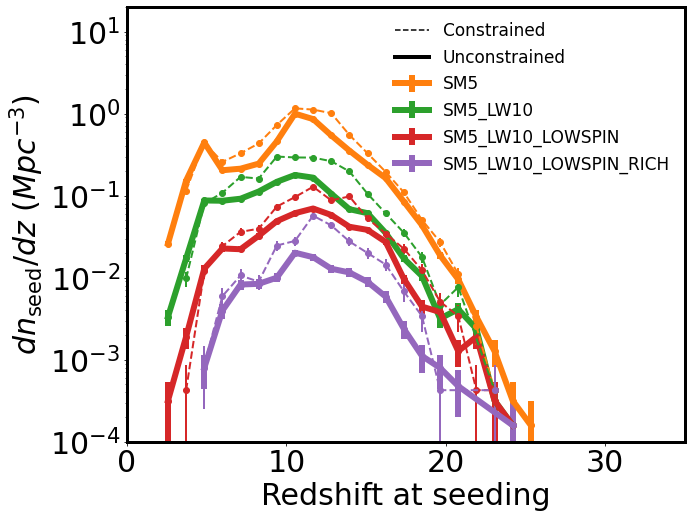}

\caption{Distribution of seed formation redshifts for the four different seed models. The solid and dashed lines show the unconstrained and constrained simulations respectively. As we stack up the different seeding criteria, the seed formation rates start to decrease. In the unconstrained simulations, at the peak of seed formation at $z\sim12$, the most lenient \texttt{SM5} seed model produces $\sim1$ seed per Mpc$^3$, whereas the strictest \texttt{SM5_LW10_LOWSPIN_RICH} seed model produces $\sim0.01$ seeds per Mpc$^3$.  }
\label{seed_formation_rates_fig}
\end{figure}

\begin{figure*}
\includegraphics[width=5.8 cm]{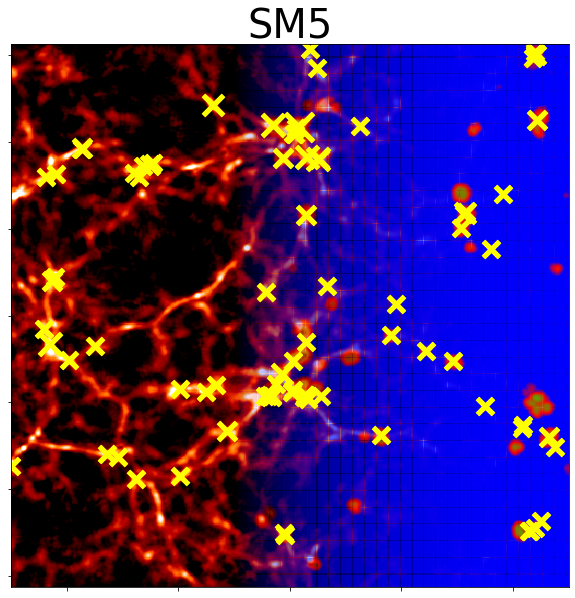}
\includegraphics[width=9 cm]{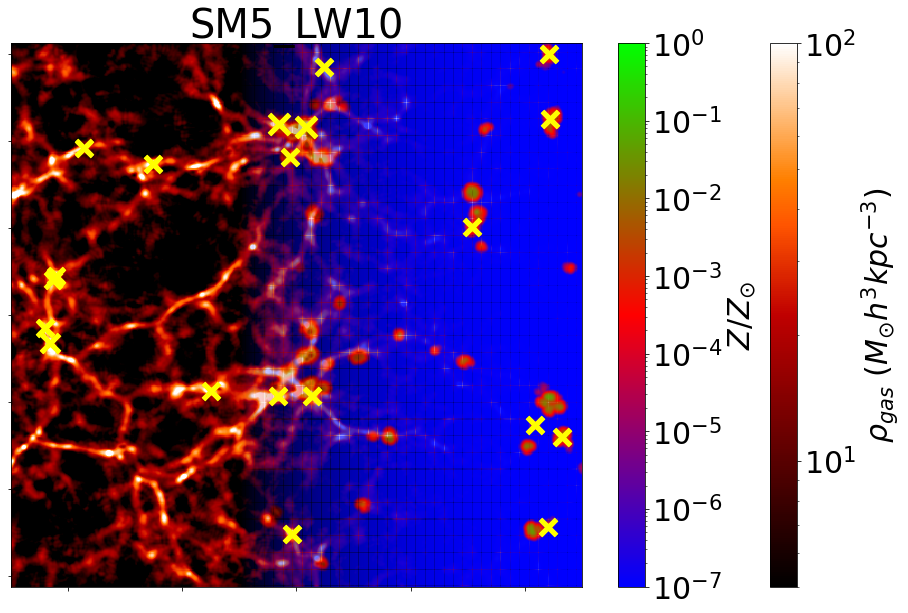} \\
\includegraphics[width=5.8 cm]{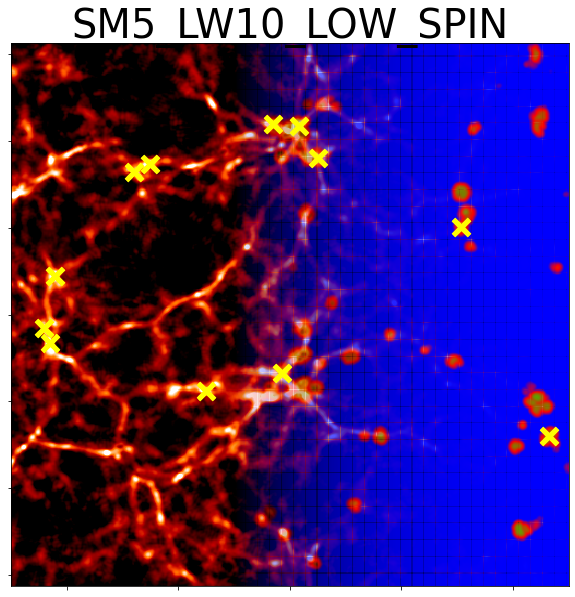}
\includegraphics[width=9 cm]{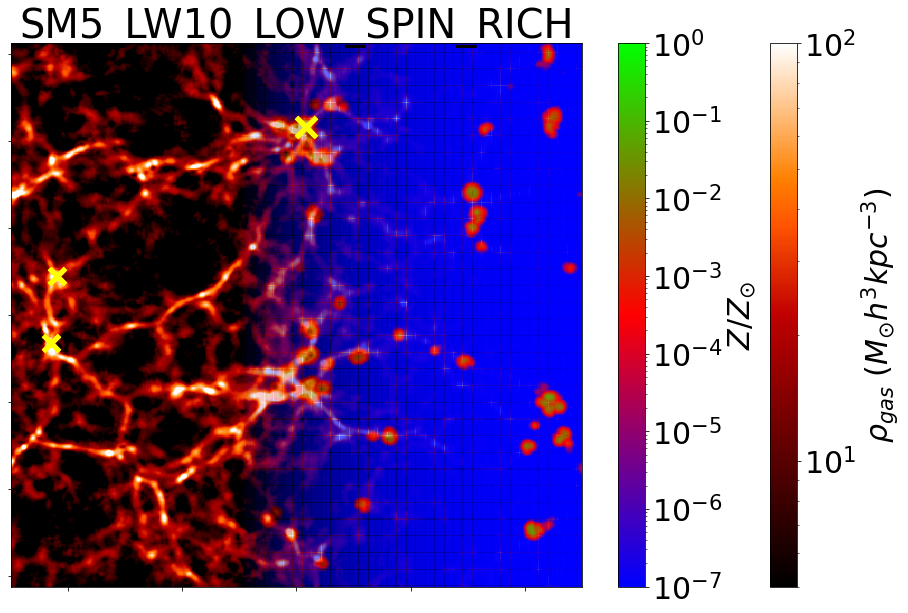}\\
\caption{Visualization of the 18 Mpc unconstrained simulated boxes on a $50$ kpc slice at $z=5$, for the four different seed models. The left side of the panels show the projected gas density profiles, which smoothly transitions in to the projected gas metallicity profiles on the right side of the panels. The yellow crosses show the positions of BHs. For the most lenient \texttt{SM5} seed model~(top left), BHs occupy a significant majority of the regions with dense gas. However, for the strictest seed model \texttt{SM5_LW10_LOWSPIN_RICH}, only a small fraction of them host BHs.}
\label{visualization_fig}

\includegraphics[width=4 cm]{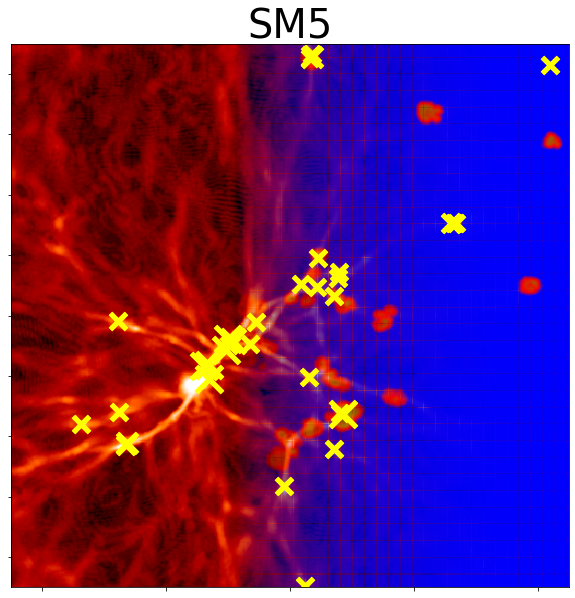}
\includegraphics[width=6 cm]{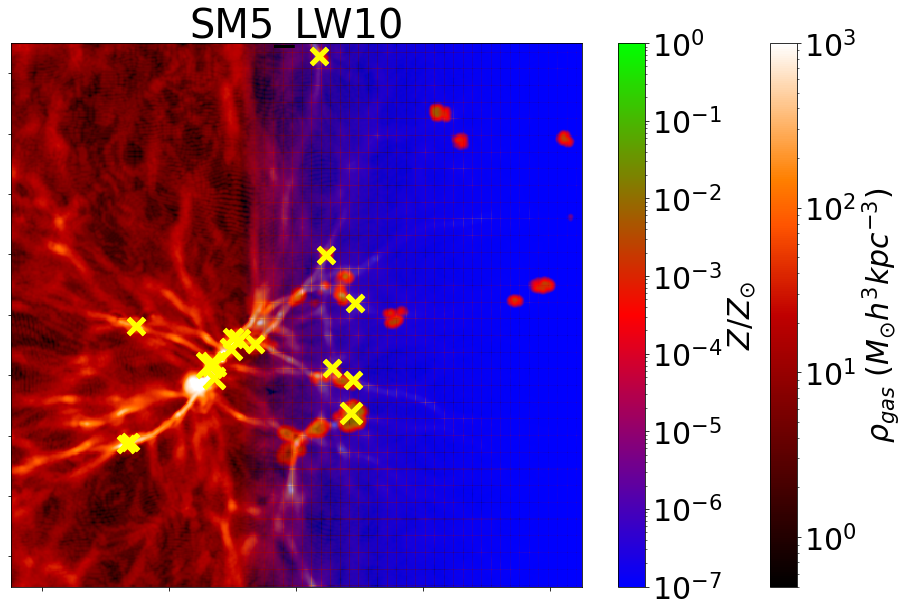} \\
\includegraphics[width=4 cm]{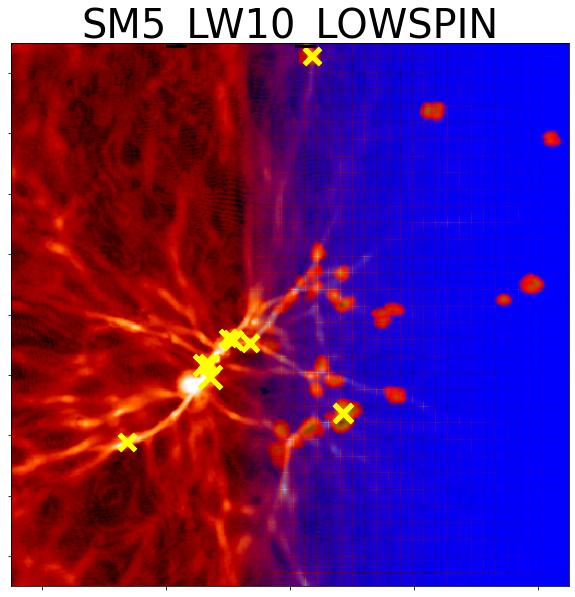}
\includegraphics[width=6 cm]{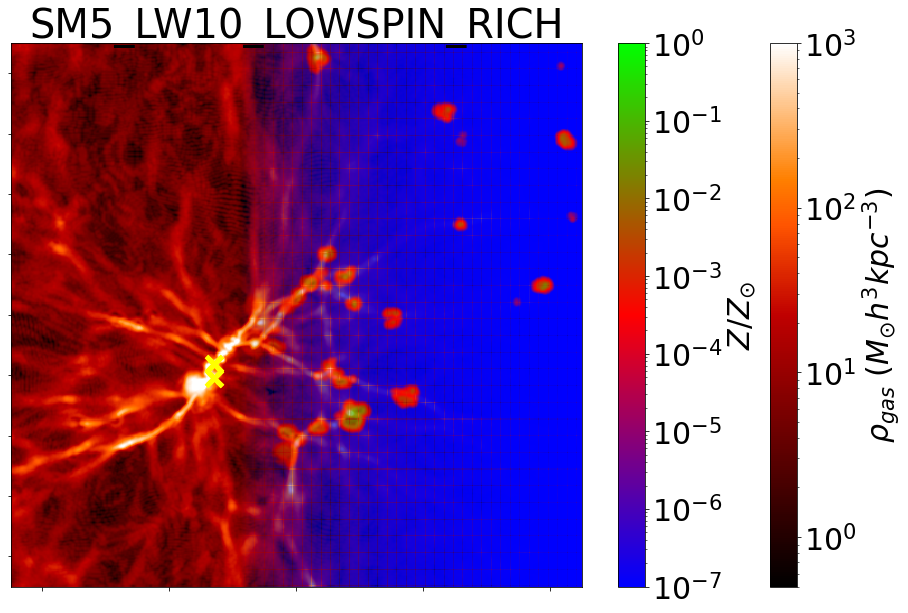}\\
\caption{Similar to the previous figure, but for the 13 Mpc constrained simulation boxes on a $50$ kpc slice at $z=5$, for the four different seed models. We can see that the BHs are much more strongly clustered compared to the unconstrained boxes, leading to enhancements in merger rates.}
\label{visualization_fig_2}
\end{figure*}

\subsection{Seed formation history}
Having established that the simulated galaxy populations in the \texttt{BRAHMA} boxes are broadly consistent with observations, we now focus on the BH assembly under different seeding assumptions. We begin by looking at the seed formation history of the four simulation boxes shown in Figure \ref{seed_formation_rates_fig}. For all the different seed models, the onset of seed formation occurs at $z\sim25$ in both the constrained and unconstrained boxes. This concides with the earliest  collapse of gas to densities greater than the star formation threshold~(hereafter referred to a ``dense gas") in a pristine universe. Continued onset of dense gas formation leads to a ramp-up of seed formation between $\sim25-12$. However, the formation of stars and the resulting stellar feedback drives metal enrichment of gas, which eventually slows down seed formation. The peak of seed formation occurs at $z\sim12$, after which their production is suppressed due to metal pollution. 

Let us first look at the seed formation rates for the different models in the unconstrained simulations~(solid lines in Figure \ref{seed_formation_rates_fig}). For the most lenient \texttt{SM5} model, the seed production peaks at $\sim1$ seed per Mpc$^{-3}$ per unit redshift. For the remaining \texttt{SM5_LW10}, \texttt{SM5_LW10_LOWSPIN} and \texttt{SM5_LW10_LOWSPIN_RICH} simulations, the peak seed production is reduced to $\sim0.1$, $\sim0.06$ and $\sim0.01$ Mpc$^{-3}$ per unit redshift, respectively. The differences in seed formation rates tend to be larger at lower redshifts. This is because the impact of the gas-spin and Lyman-Werner flux criteria becomes stronger with decreasing redshift, as demonstrated in \cite{2022MNRAS.510..177B}. For the Lyman-Werner flux criterion, the greater suppression of seed production at later times is because of the reduction of the fluxes due to Hubble expansion. For the gas-spin criterion, this may be due to the gradual build up of angular momentum of gas inside halos as time evolves. These seed model variations have strong implications for the $z\sim4-7$ BH populations, which can be readily seen in Figure \ref{visualization_fig}, which shows 2D projection plots of the gas density and gas metallicity fields at $z=5$. In the most lenient seed model \texttt{SM5}, BHs occupy a vast majority of the overdense regions. On the other hand, for the \texttt{SM5_LW10_LOWSPIN_RICH} model, only a tiny fraction of the overdense regions are occupied by BHs.  

For the constrained simulations~(dashed lines in Figure \ref{seed_formation_rates_fig}), the seed model variations are qualitatively similar to the unconstrained simulations. However, for a given seed model, the constrained simulations do form larger numbers of seeds overall. For the \texttt{SM5}, \texttt{SM5_LW10} and \texttt{SM5_LW10_LOWSPIN} simulations, the constrained boxes form $\sim2-3$ times higher number of seeds compared to their unconstrained counterparts. But notably, when the halo environment criterion is applied, the impact of the constrained ICs is slightly stronger, with the constrained simulations producing $\sim5$ times more seeds. This is not surprising given that the halo environment criterion favors seeding in rich environments, which are naturally more abundant within the constrained simulations. Figure \ref{visualization_fig_2} visualizes the constrained region, which clearly contains a strongly overdense peak close to the center of the box. This is in stark constrast to the unconstrained region~(top two rows) which contains relatively smaller overdensity peaks that are uniformly spread throughout the simulation volume. As a result, the BHs in the constrained simulations are much more strongly clustered than in the unconstrained simulations.

\subsection{BH growth: Mergers vs accretion}
\label{BH growth: Mergers vs accretion}
\begin{figure*}
\includegraphics[width=18 cm]{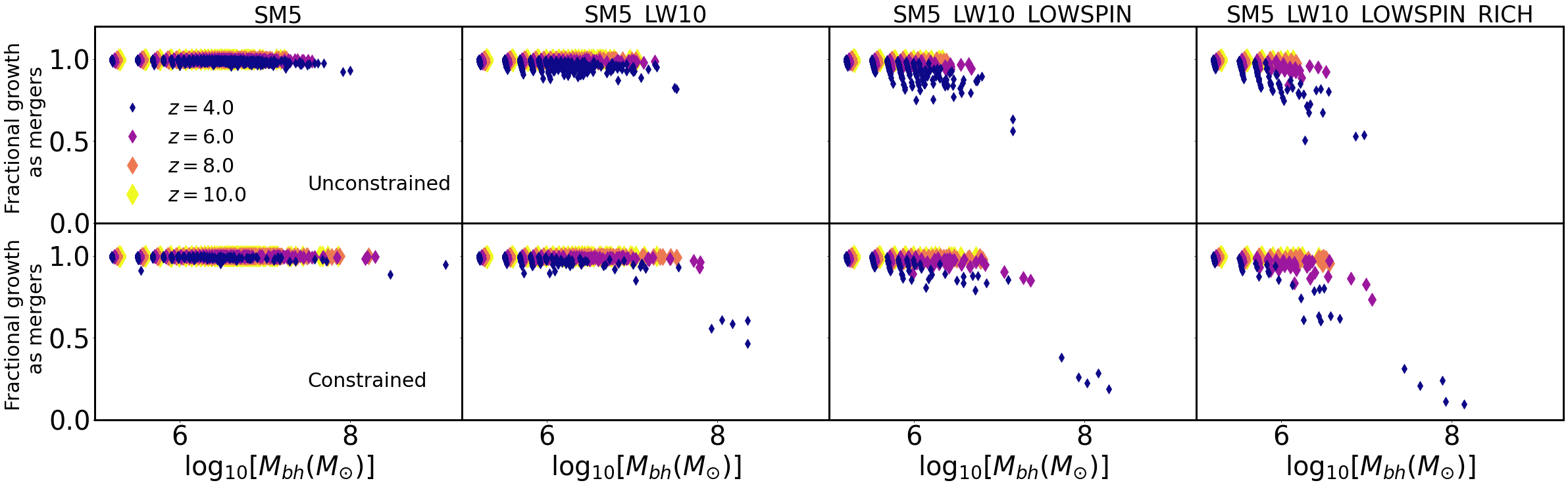}
\caption{The fraction of the overall BH mass that is contributed by merger driven BH growth. The top and bottom panels show the unconstrained and constrained simulations respectively. The different colors correspond to BHs at different redshift snapshots from $z=10$ to $z=4$. BH growth is generally dominated by mergers. As we make the seed model more restrictive, the relative contribution from gas accretion increases, particularly for more massive BHs at lower redshifts.}
\label{mergers_vs_accretion_fig}
\end{figure*}
Having discussed the formation history of seeds, we now consider their subsequent growth to form higher mass BHs. In particular, we quantify the relative contributions of BH mergers and gas accretion to the overall BH mass as shown in Figure \ref{mergers_vs_accretion_fig}. We show this for BH populations at snapshots $z=10,8,6~\&~4$. We can immediately see that in the unconstrained runs~(top panels), the most massive BH formed by $z=4$ with the most restrictive seed model~(\texttt{SM5_LW10_LOWSPIN_RICH}) is only $\sim10^7~M_{\odot}$, whereas the majority of the $z\sim4-7$ JWST BHs exceed that mass. This already hints that the combination of all the seeding criteria in Section \ref{Black hole seed models} makes DCBH formation too restrictive to assemble the JWST BHs. In fact, the JWST BHs have masses up to $\sim10^8~M_{\odot}$, which can only be achieved by the  least restrictive seed model~(\texttt{SM5}) in the unconstrained runs. The constrained runs~(bottom panels) do produce higher mass BHs. However, as we shall see in Section \ref{Stellar mass vs black hole mass relations: Comparison with JWST}, these BHs also live in higher mass galaxies. In other words, \textit{at fixed} galaxy stellar mass, the constrained runs do not produce higher BH masses compared to the unconstrained runs.

We find that regardless of the seed model, the BH mass accumulation is dominated by mergers down to $z=6$ for both constrained and unconstrained simulations. We also saw this in our earlier works based on these seed models~\citep{2021MNRAS.507.2012B,2023arXiv230915341B,2024arXiv240203626B}. In fact, for the most lenient seed model, the contribution from BH mergers continues to dominate~($\gtrsim95\%$) all the way down to $z=4$. It is only when we make the seed models more restrictive that there is a natural reduction in merger driven growth that increases the relative importance of gas accretion. Only for the most massive $z=4$ BHs formed by the most restrictive \texttt{SM5_LW10_LOWSPIN_RICH} seed model, the accretion driven growth contributes $\sim50\%$ and $\sim90\%$ of the BH mass within the unconstrained and constrained simulations, respectively. Nevertheless, for the vast majority of BHs at these redshifts, the mass growth is pre-dominantly driven by BH mergers. This is contributed by two things: First, the $M_{\rm bh}^2$ scaling of our Bondi accretion formulae naturally leads to very low accretion rates in lower mass BHs, which makes it difficult to grow them using gas accretion alone. Second, stellar feedback can significantly restrict the availability of enough gas to feed BHs, particularly in lower mass halos at high redshifts wherein the potential wells are relatively shallow~\citep{2017MNRAS.468.3935H}. In any case, one important implication of merger driven BH growth is that the build up of higher mass BHs relies on the availability of sufficient seeds to undergo mergers. As a result, the choice of our seed model has a substantial impact on the final BH masses at $z\sim4-7$ as mentioned in the previous paragraph. We shall discuss this further in Section \ref{Stellar mass vs black hole mass relations: Comparison with JWST}, wherein it will be evident that the merger-dominated BH growth is very consequential to the feasibility of different seed models in producing $M_*-M_{\rm bh}$ relations that are consistent with those inferred from the measured JWST BHs.

\subsection{AGN luminosity functions}
\label{AGN luminosity functions}
\begin{figure*}

\includegraphics[width=18 cm]{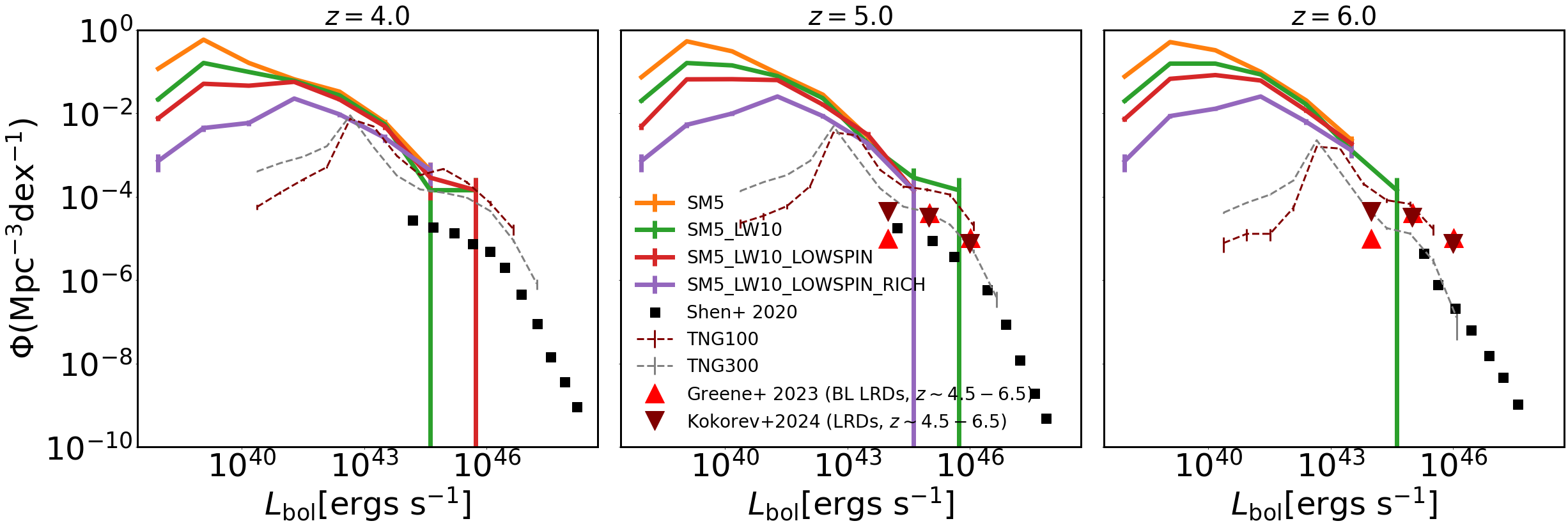}
\caption{AGN bolometric luminosity functions at $z\sim4,5~\&~6$ for the four different \texttt{BRAHMA} boxes compared against observational constraints from pre-JWST quasars~\citep{2020MNRAS.495.3252S} as well as from JWST AGN measured by \protect\cite{2024ApJ...968...38K} and \citealt{2023arXiv230905714G}~(lower limits). Here we show predictions from the unconstrained simulations. The maroon and grey dashed lines show the predictions from \texttt{TNG100} and \texttt{TNG300} respectively, that seed $10^6~M_{\odot}$ BHs based on a halo mass threshold of $7\times10^{10}~M_{\odot}$. The different seed models produce similar luminosity functions at $\gtrsim10^{43}~\mathrm{erg~s^{-1}}$. }
\label{luminosity_functions_fig}

\includegraphics[width=13 cm]{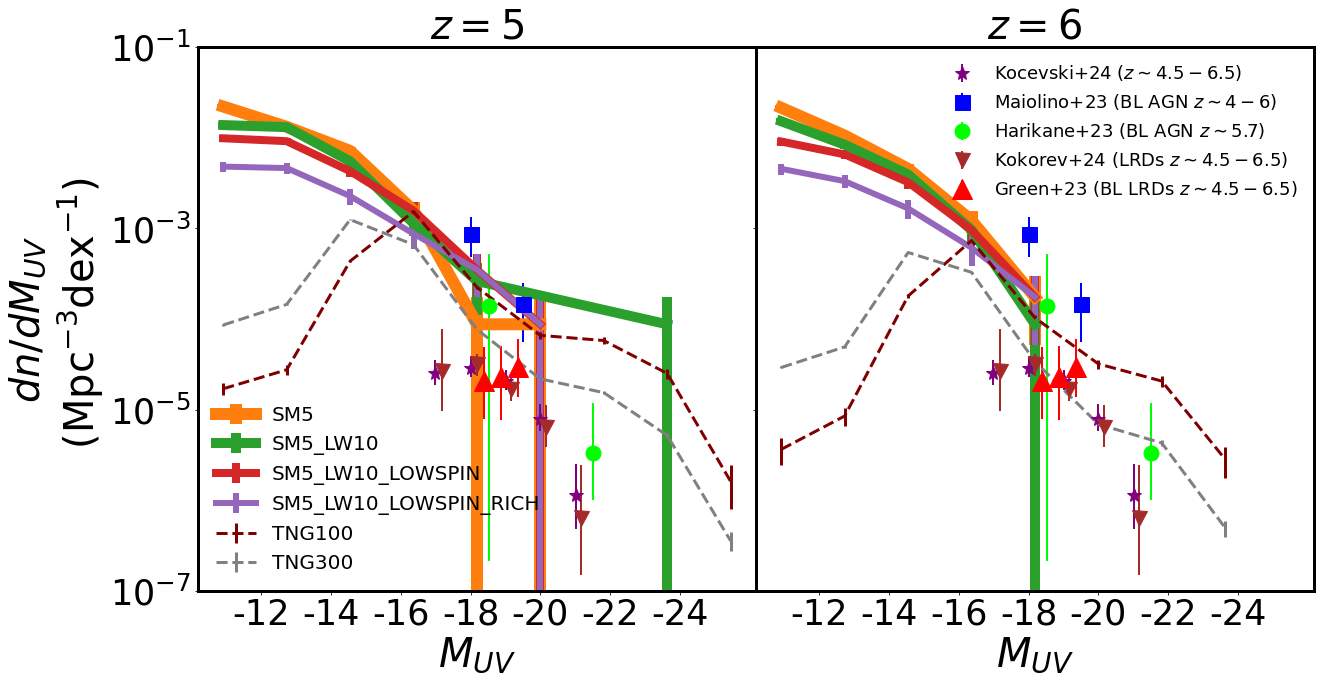}
\caption{AGN UV luminosity functions at $z\sim5~\&~6$ for the four different \texttt{BRAHMA} boxes compared against observational constraints from JWST AGNs. Here we show predictions from the unconstrained simulations. For the UV luminosities, the bolometric corrections were adopted from \protect\cite{2020MNRAS.495.3252S}. There is a significant spread in the observed measurements, but the simulation predictions are broadly consistent with them.}
\label{UV_luminosity_functions_fig}

\end{figure*}
As we find that gas accretion has a negligible contribution to the BH growth in our simulations at these high redshifts, it is instructive to look at the AGN luminosity functions~(LFs) and compare against available observational constraints. We start by looking at the bolometric LFs, since they can be most directly predicted by our simulations without any underlying assumptions about the AGN spectral energy distributions. In Figure \ref{luminosity_functions_fig}, we show the AGN bolometric LF predictions at $z=4,5~\&~6$ for our different seed models as predicted by our unconstrained simulations. The \texttt{BRAHMA} boxes probe bolometric luminosities up to $\sim10^{44}~\mathrm{erg~s^{-1}}$ before Poisson noise starts to dominate\footnote{While the constrained simulations do produce higher luminosities, they cannot be used to make volume independent AGN LF predictions as their IC realizations are not a representation of an average volume.}$^{,}$\footnote{The brightest AGN produced by the different boxes at a given snapshot can vary between $\sim10^{44}-10^{46}~\mathrm{erg~s^{-1}}$. However, these variations are simply due to the large time-variability of the AGN luminosities, and not necessarily due to the seed models.}. Before we focus on the comparison with observational constraints in the next paragraph, it is interesting to note here that the seed model variations are significant only for the faintest $L_{\mathrm{bol}}\lesssim10^{41}~\mathrm{erg~s^{-1}}$ AGN. At $\gtrsim10^{43}~\mathrm{erg~s^{-1}}$, the LFs are similar amongst the different \texttt{BRAHMA} seed models. The lack of seed model variations in the AGN LFs has also been shown in our previous papers for a wide range of seed models~\citep{2021MNRAS.507.2012B,2024arXiv240203626B}. To briefly summarize, we found that while the seed models produce differences in the overall number of BHs, the number of ``active" BHs remain similar as there are only a limited set of environments that provide enough gas to accrete and produce AGN at these high redshifts\footnote{Note that despite the merger dominated BH growth at $z\gtrsim3$, we also showed in \cite{2024arXiv240203626B} that BH mass assembly due to gas accretion starts to become comparable to mergers at $z\lesssim3$. Gas accretion eventually dominates the BH mass assembly at $z\sim0$, consistent with the Soltan argument.}. In fact, at $\gtrsim10^{43}~\mathrm{erg~s^{-1}}$, the \texttt{BRAHMA} seed model predictions are also similar to \texttt{TNG100} which simply seeds $10^6~M_{\odot}$ BHs when halos exceed $7\times10^{10}~M_{\odot}$\footnote{Despite having the same physics, \texttt{TNG300} and \texttt{TNG100} AGN LFs differ by factor of $3-5$, due to the different simulation resolutions. \texttt{TNG300} produces lower luminosities likely because it cannot resolve the high density peaks as effectively as \texttt{TNG100}.}. We also note that the \texttt{TNG} LFs sharply peak and fall off at luminosities below a few times $\sim10^{42}~\mathrm{erg~s^{-1}}$. This is due to the higher seed mass~($10^6~M_{\odot}$) and the halo mass seeding threshold~($7\times10^{10}~M_{\odot}$) in \texttt{TNG}. This sharp fall off does not occur in our \texttt{BRAHMA} boxes in which there is a large number of $\lesssim10^{42}~\mathrm{erg~s^{-1}}$ AGN that are fueled by $\sim10^5-10^6~M_{\odot}$ BHs residing in galaxies that are not massive enough to be seeded in \texttt{TNG}.

Due to the limited volume in our boxes, the overlap with observations is only at the brightest end of the simulation predictions~($\sim10^{43}-10^{44}~\mathrm{erg~s^{-1}}$ at $z\sim5$) wherein the seed model variations are very small. At this end, our predicted \texttt{BRAHMA} AGN LFs have a  higher normalization than the pre-JWST observational constraints~(black squares) from \cite{2020MNRAS.495.3252S} at $z\sim4~\&~5$. This is also the case for the \texttt{TNG100} and \texttt{TNG300} boxes which predict AGN LFs similar to our \texttt{BRAHMA} boxes despite having a very different seed model. As a result, the discrepancy with observations is unlikely to be originating from our seed models. In fact, a vast majority of simulations overpredict the AGN LFs compared to observational constraints at $z\sim0-4$~\citep{2022MNRAS.509.3015H}. In the future, we will investigate other aspects of the BH physics modeling~(such as BH accretion and dynamics) to explore the reasons for this discrepancy. At the same time, the discrepancy can also be due to uncertainties in the observational measurements within the modeling of AGN obscuration as well as bolometric corrections. This is hinted within the recent constraints from JWST AGN that are lower limits~(red triangles in Figure \ref{luminosity_functions_fig}) at $z\sim5$ derived by \cite{2023arXiv230905714G} as well as constraints from \cite{2024ApJ...968...38K}. We can see that these JWST based measurements are slightly higher than the pre-JWST measurements from \cite{2020MNRAS.495.3252S}, bringing them closer to the simulations. 

While very few measurements of the bolometric AGN LFs have been made due to the uncertainty in the bolometric corrections, there are several measurements in the rest frame UV band.  Given the greater availability of UV LF measurements, it is worthwhile to compare them against our simulations even though the conversion of the simulated bolometric luminosities to UV luminosities will carry similar bolometric correction uncertainties.  Therefore, in Figure \ref{UV_luminosity_functions_fig}, we convert the simulated bolometric LFs to rest frame UV LFs using the bolometric correction from \cite{2020MNRAS.495.3252S}. We compare the $z=5~\&~6$  snapshot predictions to the JWST measurements for $z\sim4.5-6.5$ AGN samples. Here again, the overlap between simulated and observed regimes is only over a very small range of $M_{UV}\sim-17$ to $-20$. There is also a significant spread amongst the observational constraints. However, it is still noteworthy that our simulations predict broadly consistent AGN abundances between $M_{UV}\sim-17$ to $-20$, that are well within the range of current observational measurements.

As future JWST observations lead to more precise constraints, it will shed further light on whether or not our simulations predict AGN LFs that are consistent with observations at these redshifts. Nevertheless, one clear outcome from this analysis is that despite our \texttt{BRAHMA} simulations exhibiting merger dominated BH growth at these redshifts, the predicted AGN activity is not substantially smaller than what is inferred from observations.    

\subsection{AGN-galaxy connection}
\label{AGN-galaxy connection}
Based on the galaxy stellar mass functions and AGN luminosity functions, it is clear that the number densities~($\sim10^{-2}~\mathrm{Mpc}^{-3}$) of typical galaxies with the measured stellar masses of JWST AGN hosts~($\gtrsim10^{7.5}~M_{\odot}$) are much larger than the inferred number densities of the AGNs themselves~(ranging from $\sim10^{-5}-10^{-3}~\mathrm{Mpc}^{-3}$). This could suggest that the JWST AGNs are observed at luminosities much higher than the typical population of BHs living in these galaxies. Given that our simulations are able to broadly reproduce the abundances of both galaxies and AGNs, it is instructive to also look at the simulations and the JWST observations on the stellar mass vs AGN luminosity~($M_*$ vs $L_{\mathrm{bol}}$) plane.

\begin{figure*}

\includegraphics[width=16 cm]{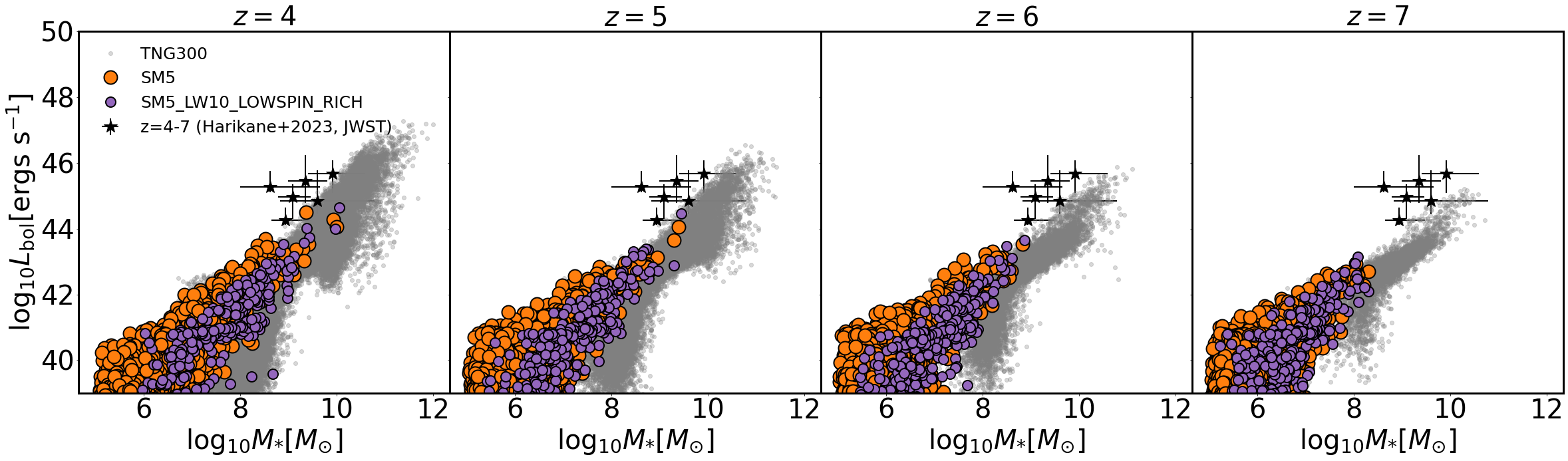}
\includegraphics[width=16 cm]{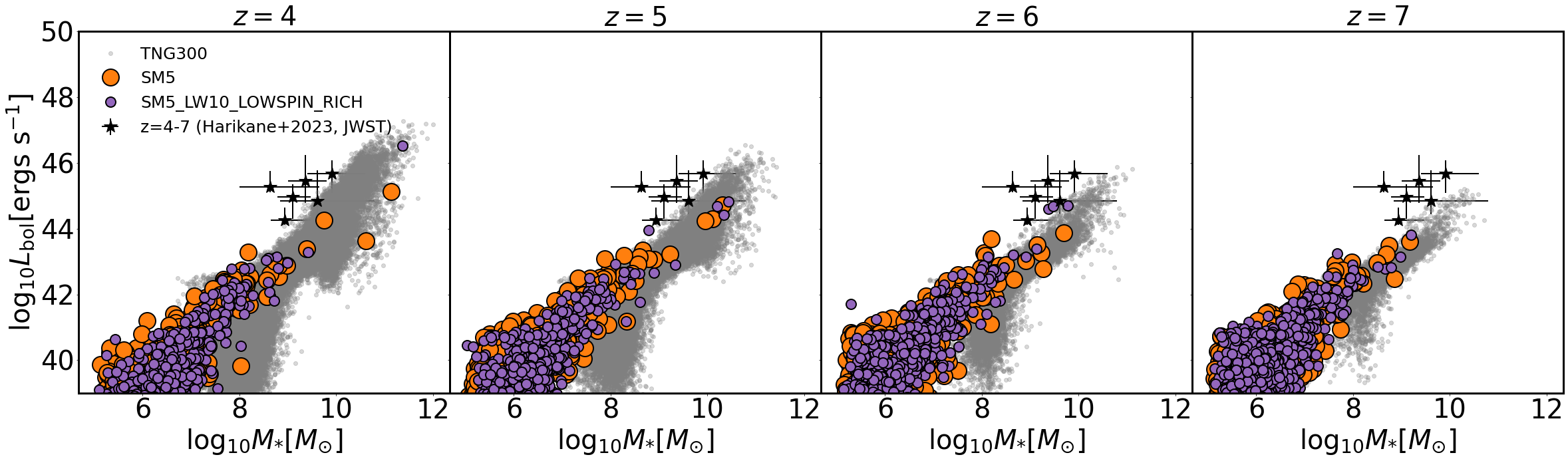}
\caption{Stellar mass vs AGN bolometric luminosity relations~($M_*$ vs $L_{\mathrm{bol}}$) at $z=4,5,6~\&~7$ produced by our simulations compared against the JWST observations at $z\sim4-7$. The top and bottom panels correspond to the constrained and unconstrained simulations respectively. At the observed luminosities, the simulated AGN host galaxies have higher stellar masses compared to observed measurements.}
\label{SM_Lum_fig}


\end{figure*}

Figure \ref{SM_Lum_fig} shows the $M_*$ vs $L_{\mathrm{bol}}$ relations for the simulations, plotted with observations from \cite{2023ApJ...959...39H}. The simulations show a clear positive correlation between $M_*$ and $L_{\mathrm{bol}}$. For the \texttt{BRAHMA} boxes, we only show results for the most lenient (\texttt{SM5}) and the most restrictive (\texttt{SM5_LW10_LOWSPIN_RICH}) seed models for clarity. Note that both these seed models produce broadly similar results that are also similar to the \texttt{TNG300} predictions~(except for the faintest $\lesssim10^{42}~\mathrm{erg~s^{-1}}$ AGNs living in the smallest galaxies). This shows that our seed models do not substantially impact the $M_*$ vs $L_{\mathrm{bol}}$ relations, which is expected given the earlier results from Figures \ref{luminosity_functions_fig} and \ref{UV_luminosity_functions_fig} that showed that AGN LFs are also not significantly sensitive to the seed model~(except the faintest end $\lesssim10^{41}~\mathrm{erg~s^{-1}}$).

Let us now focus on the comparison with observations. The \cite{2023ApJ...959...39H} sample contains several $\gtrsim10^{44}~\mathrm{erg~s^{-1}}$ AGN living in $M_*\gtrsim10^{8}~M_{\odot}$ galaxies. However, in our simulations, the typical AGN luminosities in $M_*\sim10^8~M_{\odot}$ galaxies are a few times $\sim10^{42}~\mathrm{erg~s^{-1}}$ at $z=4~\&~5$. In a similar vein, the simulated AGNs with $\gtrsim10^{44}~\mathrm{erg~s^{-1}}$ typically only live in $M_*\gtrsim10^9~M_{\odot}$ galaxies i.e. 
$\sim10$ times higher than the measurements for the JWST AGN hosts. This discrepancy in the simulated and observed $M_*$ vs $L_{\mathrm{bol}}$ relations is not surprising given that the AGN LFs are broadly consistent between them. This is because, given that the galaxy abundances are much higher than the AGNs, if the simulated $M_*$ vs $L_{\mathrm{bol}}$ relations were actually consistent with observations, the simulations would have dramatically overpredicted the AGN LFs. 

As mentioned in the introduction, it is possible that JWST is only observing the most luminous AGNs~(at fixed stellar mass) that are preferentially detected due to Lauer bias. While the scatter in the \texttt{BRAHMA} $M_*$ vs $L_{\mathrm{bol}}$ relations is not large enough to produce any up-scattered $\sim10^{44}~\mathrm{erg~s^{-1}}$ AGN in $\sim10^8~M_{\odot}$ galaxies, we acknowledge that this may be simply due to the limited volume of the \texttt{BRAHMA} boxes. However, we can also clearly see that even the TNG300 simulation~(see grey points in Figure \ref{SM_Lum_fig}) that has a volume much larger than the JWST fields, does not produce any upscattered AGNs that overlap with the observations. This firmly establishes that our galaxy formation model can only produce the observed AGN luminosities in galaxies with stellar masses significantly higher than the measurements for the JWST AGNs. 

The above analysis implies that if the discrepancy in the $M_*$ vs $L_{\mathrm{bol}}$ relations is solely due to Lauer bias, then the scatter in $M_*$ vs $L_{\mathrm{bol}}$ relations in the high-z Universe must be substantially higher than what our simulations predict. As more luminous AGNs tend to be powered by more massive BHs, the BH masses will also be impacted by Lauer bias~(which we address in the next section). However, in addition to Lauer bias, there could be contributions from other possible sources to this discrepancy that would impact the observed luminosities without necessarily impacting the  observed BH masses. First, there may be Eddington bias due to rapid variability in the AGN luminosities since the detection likelihood would be much higher during the peak luminosities. This would lead to observed luminosities being significantly higher than the actual time-averaged luminosities~(if they are below observational limits). Notably, this possibility has been explored in the galaxy sector i.e. to explain the excess of  highest-z~($z\gtrsim9$) JWST galaxies compared to theoretical predictions as a consequence of UV variability from bursty star formation ~\citep{2023MNRAS.525.3254S}. It is also well-known that AGN variability has an inverse scaling relation with BH mass~\citep{2012A&A...542A..83P,2013ApJ...779..187K}, making these JWST AGNs more susceptible to Eddington bias compared to the pre-JWST quasars. Second, the stellar masses of the AGN hosts may be underestimated as it is often difficult to separate the contribution from the AGN and the host galaxy within the observed light~\citep{2020MNRAS.499.4325R}. It is also important to note that while the current BH masses are also highly uncertain and are prone to systematic biases, the discrepancy in the simulated and observed $M_*$ vs $L_{\mathrm{bol}}$ relations is independent of the BH mass measurements. It will be interesting to revisit this discrepancy in future work as we anticipate the continued detection of new high-z AGNs and more precise stellar mass estimates. For now, it is nevertheless encouraging to see that the simulations are in broad agreement with both the galaxy stellar mass functions and AGN luminosity functions. For the remainder of the paper, we shall focus on the BH mass assembly, which is predominantly driven by mergers~(as shown in Section \ref{BH growth: Mergers vs accretion}).

\subsection{Black hole merger rates}
\begin{figure}
\includegraphics[width=8 cm]{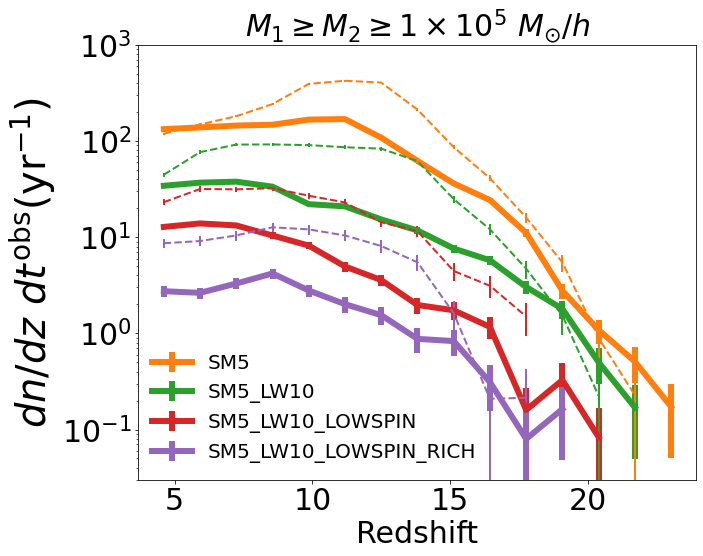}
\caption{Predicted merger rates of BH binaries in our different simulation boxes with different seed models. This is defined to be the number of mergers~($dn$) that occur within a comoving shell from $z$ to $z+dz$, the signal for which will reach an observer on Earth within a time interval $dt_{\mathrm{obs}}$. The solid lines correspond to the unconstrained simulations and the dashed lines show the constrained simulations. The merger rates are strongly sensitive to the seed model. The unconstrained simulations produce peak merger rates of $\sim200$ per year at $z\sim12$ for our most optimistic seed model, and $\sim4$ per year at $z\sim9$ for the strictest seed model.}
\label{merger_rates_fig}
\end{figure}
In Figure \ref{merger_rates_fig}, we show the rates at which BHs merge within our four different simulations. These mergers are expected to produce gravitational waves detectable with LISA. However, due to our BH repositioning scheme, we are implicitly assuming the most optimistic scenario wherein for every galaxy merger, the BHs merge instantaneously. In reality, we expect these mergers to occur after a finite binary inspiral time, which could be a significant fraction of the Hubble time in some cases; this would depend on the eccentricities of the orbiting binaries as well as the effectiveness of processes that contribute to the hardening of the binaries at sub-kpc scales such as stellar loss-cone scattering, and drag due to circumbinary discs~\citep{2017MNRAS.464.3131K,2021MNRAS.501.2531S,2023MNRAS.522.2707S,2024arXiv240308871S}. Merger remnants can also get kicked out of the galaxies due to gravitational recoil~\citep{2008ApJ...687L..57V,2008MNRAS.390.1311B,2008ApJ...686..829H,2016MNRAS.456..961B,2016PhRvL.117a1101G,2020ApJ...896...72D}, which can impact future mergers. Due to all these reasons, the results from  Figure \ref{merger_rates_fig} should only be interpreted as upper limits. Not surprisingly, we can see that the BH merger rates are strongly dependent on the seed model. For the unconstrained simulations, the most optimistic \texttt{SM5} model predicts peak merger rates of $\sim200$ per year. The remaining \texttt{SM5_LW10}, \texttt{SM5_LW10_LOWSPIN} and \texttt{SM5_LW10_LOWSPIN_RICH} predict peak merger rates of 30, 10 and 4 per year, respectively. The constrained simulations generally predict $\sim5$ times higher merger rates compared to their unconstrained counterparts for all seed models~(solid vs. dashed lines in Figure \ref{merger_rates_fig}). Notably, the constrained ICs enhance the merger rates more strongly than they enhance the seeding rates~(revisit solid vs dashed lines in Figure \ref{seed_formation_rates_fig}). This is because the BHs in the constrained ICs are much more clustered, allowing them to merge more efficiently compared to the unconstrained ICs. 

Finally, as the BH growth is dominated by mergers at $z\gtrsim4$, the differences in the merger rates for the different seed models have strong implications for the final BH masses accumulated at different redshifts, which we study in detail in the next section.

\subsection{Stellar mass vs black hole mass relations: Comparison with JWST observations}
\label{Stellar mass vs black hole mass relations: Comparison with JWST}
\begin{figure*}
\includegraphics[width=18 cm]{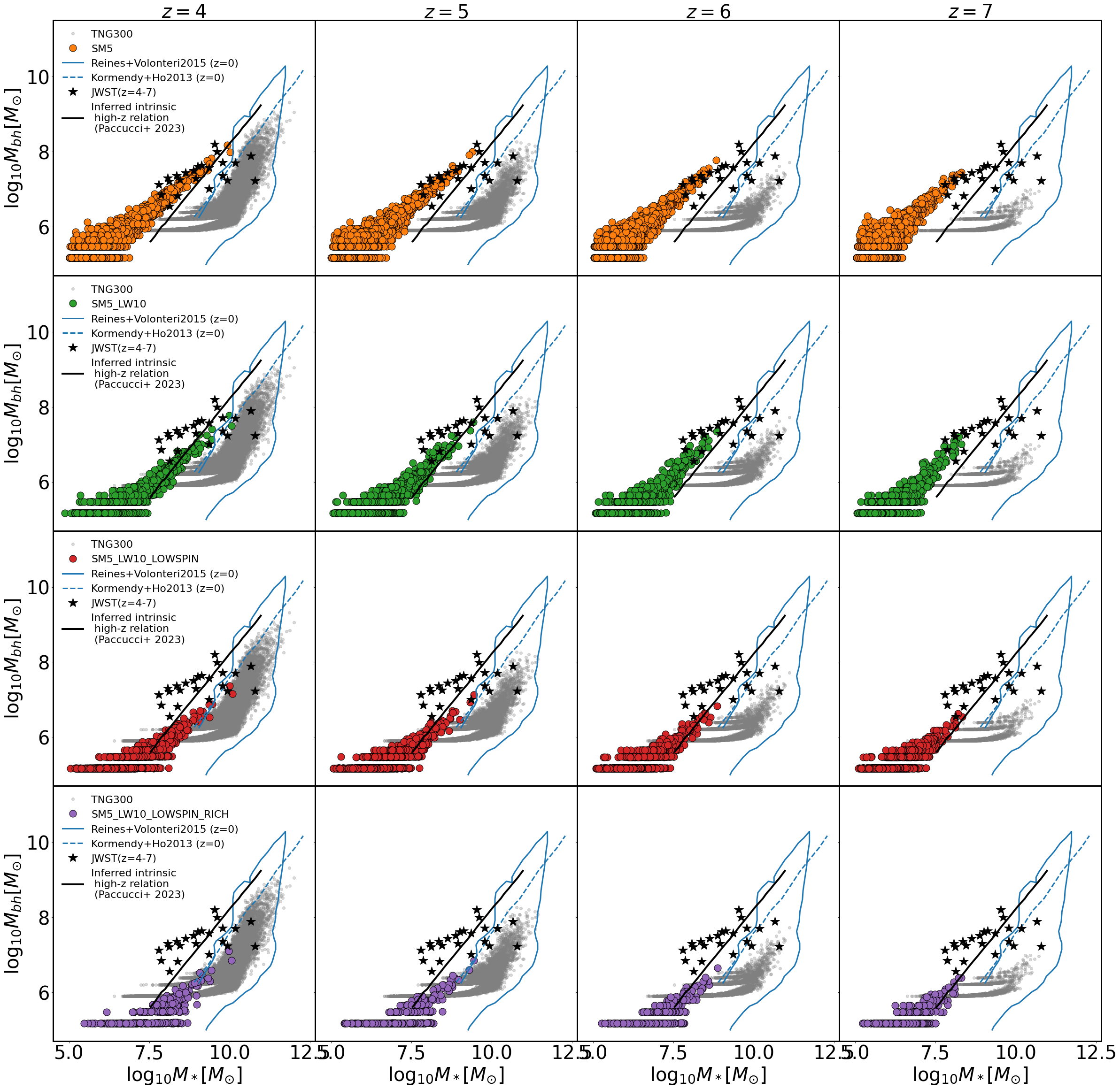}
\caption{Stellar mass vs. BH mass~($M_*-M_{\rm bh}$) relation predictions from our unconstrained simulations with different seed models~(colored circles). The solid black line shows the inferred intrinsic high-z relations inferred by \protect\cite{2023ApJ...957L...3P} based on JWST measurements from several works prior to it~(black stars). These measurements have uncertainties~(not shown for clarity) around $\sim1~\mathrm{dex}$ for both stellar and BH masses. The grey solid line approximately represents the scatter in local $M_*-M_{\rm bh}$ relation from \protect\cite{2015ApJ...813...82R}. The grey dashed and dotted lines are the local measurments from \protect\cite{2013ARA&A..51..511K} and \protect\cite{2016ApJ...830L..12T} respectively. At $z\sim4-7$, the most lenient seed model~(\texttt{SM5}) has a substantial overlap with the JWST BHs. However, the most restrictive seed model that applies all the seeding criteria required for DCBHs significantly underpredicts the BH masses compared to the JWST BHs.}
\label{SM_BHM_fig}
\end{figure*}

\begin{figure*}
\includegraphics[width=18 cm]{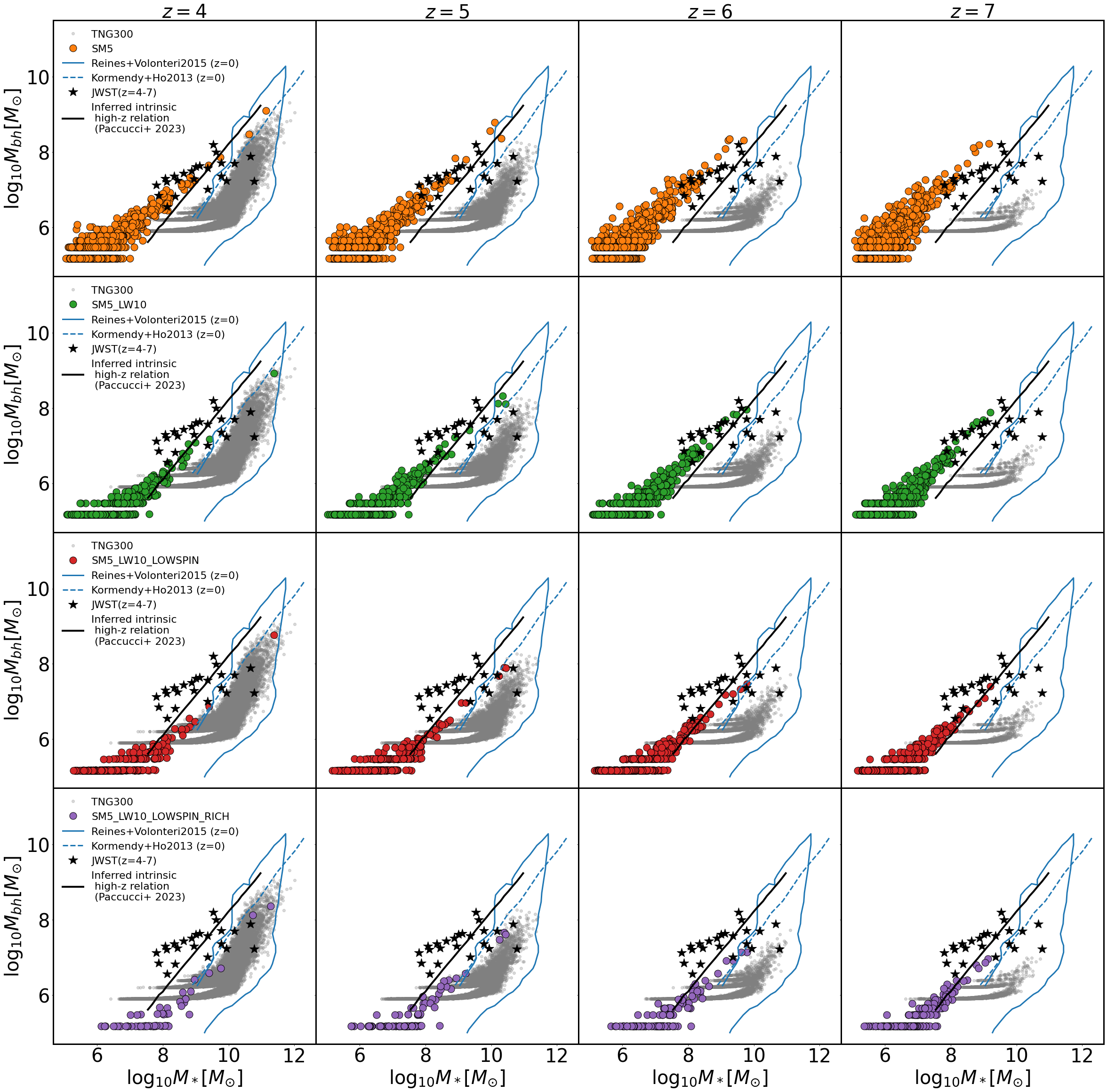}
\caption{Similar to Figure \ref{SM_BHM_fig}, but for the constrained simulations. For a given seed model, the $M_*$-$M_{\rm bh}$ relations are similar between the  constrained and unconstrained runs. But the constrained runs are able to produce higher mass galaxies, fully covering the range of JWST host galaxy stellar masses at $z=4~\&~5$.}
\label{SM_BHM_constrained_fig}
\end{figure*}

We finally look at the $M_*$ vs. $M_{\rm bh}$ relation plotted in Figures \ref{SM_BHM_fig}~(unconstrained) and \ref{SM_BHM_constrained_fig}~(constrained) and compare them against the estimates based on JWST observations. However, we must be wary of the fact that 1) the existing BH mass measurements are highly uncertain~(in addition to the stellar mass measurements as discussed earlier), and 2) the observed BH masses are expected to be higher than the \textit{intrinsic} $M_*$ vs. $M_{\rm bh}$ relations due to Lauer bias. If the bias is substantial, a direct comparison between the JWST measurements~(shown as black points) and our complete sample of simulated BH populations would not be even-handed. As mentioned in the introduction, \cite{2023ApJ...957L...3P} used an MCMC approach to estimate the intrinsic $M_*$ vs. $M_{\rm bh}$ relations~(parameterized by a power law) using combined data from \cite{2023ApJ...959...39H}, \cite{2023arXiv230801230M}, \cite{2023ApJ...946L..13F} and \cite{2023A&A...677A.145U}. Notably, the data only included those BHs that are spectroscopically confirmed with NIRSpec,
and their black hole masses are estimated with the H$\alpha$ line~\citep{2005ApJ...630..122G}. Based on the measurements and   H$\alpha$ FWHM detection limits, \cite{2023ApJ...957L...3P} accounted for the Lauer bias and inferred a high-z intrinsic $M_*$ vs. $M_{\rm bh}$ relation~(shown as solid black lines in Figures \ref{SM_BHM_fig} and \ref{SM_BHM_constrained_fig}) that is still higher than the local relations by $>3\sigma$. This relation~(hereafter P23 relation) may be directly compared to our simulation predictions. Recall also that \cite{2024arXiv240300074L} used a similar approach~(but using the flux limits instead of the H$\alpha$ FWHM limits) to infer a high-z $M_*$ vs. $M_{\rm bh}$ relation that is consistent with the local BHs. While it is not firmly established whether high-z BHs indeed are systematically overmassive, it is a crucial time to use our simulations to explore implications of possible ``overmassive-ness" on the feasibility of different seed models.

We start by noting that even though the constrained simulations probe a much rarer overdense region and produce higher numbers of galaxies and BHs, their $M_*$ vs $M_{\rm bh}$ relations are very similar to the unconstrained simulations. This implies that the volume limitations in our simulations do not significantly impact the $M_*$ vs. $M_{\rm bh}$ predictions. This enables us to robustly probe the impact of seed models on the $M_*$ vs $M_{\rm bh}$ relations. As clearly seen in Figures \ref{SM_BHM_fig} and \ref{SM_BHM_constrained_fig}, the merger dominated BH growth in our simulations leads to the final BH masses~(at fixed stellar mass) being substantially impacted by the seed model. More specifically, we find that as the seed models become more restrictive, our simulations produce smaller BH masses at fixed stellar mass. We also note that for a given seed model, as we go from $z=7$ to $z=4$, the $M_*$ vs $M_{\rm bh}$ relations shift rightward as galaxy growth is faster than BH growth at these redshifts. As these observed samples continue to grow in the future, we shall hopefully be able to infer their redshift evolution and compare with our predictions. But for now, since most of the AGNs comprising the $z\sim4-7$ composite P23 sample are actually at $z\sim4-5$, we shall mostly focus on comparing these results with our $z=4~\&~5$ snapshot predictions.

For the most optimistic seed model~(\texttt{SM5}: 1st row of Figures \ref{SM_BHM_fig} and \ref{SM_BHM_constrained_fig}) that produces the highest number of seeds, the BH masses are about $\sim100$ times larger than implied by the local scaling relations~(black dashed line). In $\sim10^8-10^9~M_{\odot}$ galaxies, this seed model produces BH populations that readily overlap with the JWST BHs on the $M_*-M_{\rm bh}$ plane at $z=4~\&~5$. When we include the LW flux criterion in \texttt{SM5_LW10} model, the resulting $M_*$ vs $M_{\rm bh}$ relations shift downward, but continue to have some overlap with the JWST BHs. When we further add the gas spin and the halo environment criteria~(\texttt{SM5_LW10_LOWSPIN} and \texttt{SM5_LW10_LOWSPIN_RICH}), the resulting BH populations are significantly below the JWST BHs particularly within $M_*\lesssim10^9~M_{\odot}$ galaxies. However, if these JWST measurements are subject to Lauer bias and are indeed up-scattered, comparing them to the full BH populations in our simulations would not be fair.
Additionally, even if the underlying scatter in the $M_*$ vs $M_{\rm bh}$ relation is significant, our small simulation boxes will not capture BHs that are substantially up-scattered. Here again~(as we did earlier for the $M_*$ vs $L_{\mathrm{bol}}$ relations), we can use the much larger  \texttt{TNG300} simulations to evaluate how much scatter we can expect if our seed models were applied to much larger volumes\footnote{we caution however that the scatter in the $M_*$ vs $M_{\rm bh}$ relations produced in our seed models may be very different from the TNG seed model}. For the majority of the range of galaxy masses captured by \texttt{TNG300}, the scatter is roughly $\sim1.5~\mathrm{dex}$\footnote{the $M_*$ vs $M_{\rm bh}$ relations in \texttt{TNG300}  flattens at $M_*\lesssim10^{9}~M_{\odot}$ as a consequence of its underlying seed model~($\sim10^{6}~M_{\odot}$ BHs seeded in halos above $\gtrsim10^{10}~M_{\odot}$ halos)}. Note that our most restrictive \texttt{SM5_LW10_LOWSPIN_RICH} seed model predicts a mean relation $\sim1~\mathrm{dex}$ below the JWST observations. For this model, if we assume a scatter of $\sim1.5~\mathrm{dex}$, it would be difficult to produce BHs up-scattered enough to overlap with the JWST BHs even in a \texttt{TNG300}-like volume. We should also note that the JWST fields are substantially smaller than \texttt{TNG300}. All this suggests that assuming a $\sim1.5~\mathrm{dex}$ scatter, the mean $M_*$ vs $M_{\rm bh}$ relation predicted by \texttt{SM5_LW10_LOWSPIN_RICH} would imply a very small likelihood of JWST surveys containing these observed overmassive BHs. Therefore, even if we account for Lauer bias, \texttt{SM5_LW10_LOWSPIN_RICH} likely underpredicts the BH growth compared to what is required to produce the JWST measurements. At the other end, the most optimistic \texttt{SM5} seed model already produces a mean $M_*$ vs $M_{\rm bh}$ relation that is overlapping with the presumably up-scattered JWST observations; this implies that in the event of significant Lauer bias, this model overpredicts the BH growth compared to what is inferred from JWST measurements. 

We must also bear in mind that if the scatter in the $M_*$ vs $M_{\rm bh}$~(and the $M_*$ vs $L_{\rm bol}$) relations were large enough, one could have a significant likelihood of detecting these JWST AGNs regardless of the location of the \textit{mean} relation. Therefore, the possible confirmation of these JWST measurements could also imply that the scatter in these relations at high-z is substantially larger than our simulation predictions. Recall that we reached similar conclusions for the scatter in the $M_*$ vs $L_{\rm bol}$ relations in Section \ref{AGN-galaxy connection}.

To summarize the above arguments, assuming that the high-z scatter in the  $M_*$ vs $M_{\rm bh}$ relations is not much larger than the TNG300 prediction of $\sim1.5~\mathrm{dex}$, our most restrictive \texttt{SM5_LW10_LOWSPIN_RICH} seed model likely under-predicts the BH growth whereas our most optimistic \texttt{SM5} seed model likely over-predicts the BH growth. We shall now see how the foregoing is consistent with an ``apples-to-apples" comparison of our simulation predictions with the intrinsic P23 relation. The \texttt{SM5} model predictions are higher than the P23 relations at $z=4~\&~5$. 
 This indeed implies that the model produces too many seeds which merge with one another to produce $z\sim4-5$ BH populations that are too massive. This is not unexpected as \texttt{SM5} assumes that a heavy $\sim10^5~M_{\odot}$ DCBH seed is produced in any region with sufficient dense and metal poor gas; this is unlikely to happen in most environments where the cooling due to molecular hydrogen will fragment the gas and prevent DCBH formation. When we include the LW flux criterion in the \texttt{SM5_LW10} model, the resulting relations are very close to the P23 relation at $z=4~\&~5$~(particularly at $z=4$). However, when we also include the gas spin and halo environment criteria~(\texttt{SM5_LW10_LOWSPIN} and \texttt{SM5_LW10_LOWSPIN_RICH}), the $M_*$ vs $M_{\rm bh}$ relations shift further downward. The most restrictive \texttt{SM5_LW10_LOWSPIN_RICH} model substantially underpredicts the BH masses compared to the P23 relation by factors of $\sim10$. Finally, the \texttt{TNG300} simulation shows the maximum disagreement with the P23 relation as the simulated BHs are already at the local scaling relations by $z\sim7$. 

Overall, we find that when all the potential preconditions for DCBH formation are included in our seed model, the simulated $M_*$ vs $M_{\rm bh}$ relations are no longer consistent with what is inferred from the JWST observations. Recall that with our BH repositioning scheme, we are already assuming the most optimistic scenario for the merging efficiency of BH binaries. Therefore, our results suggest that in order to produce sufficiently overmassive BHs~(as suggested by P23) using merger driven BH growth, we need additional heavy seeding channels compared to the standard DCBH scenario. These additional channels need to either form heavy seeds in higher numbers, or seeds that are much more massive than our assumption of $\sim10^5~M_{\odot}$. The presence of light seeds can also boost the merger driven BH growth by merging with heavy seeds. Alternatively, one can also boost the BH growth due to gas accretion by allowing for accretion rates higher those inferred from the standard Bondi-Hoyle accretion formula, as recently explored in \cite{2024arXiv240218773J}. Although not shown in any figures, we re-ran our most restrictive seed model with a Bondi boost factor of 100, as well as a reduced radiative efficiency of 0.1, but they only make a small difference to the final BH masses~(up to factors of $\sim2$) that are inconsequential to our main conclusions. This may be contributed by the difficulty in growing low mass BHs due to the $M_{bh}^2$ scaling of the Bondi accretion rate. In the future we plan to explore other accretion models beyond the Bondi-Hoyle model~(that scale differently with BH mass), and also explore other aspects of our galaxy formation model that can influence the BH accretion rates~(for example, stellar feedback).   

\subsection{Impact of delayed BH mergers}
\begin{figure}
\includegraphics[width=8 cm]{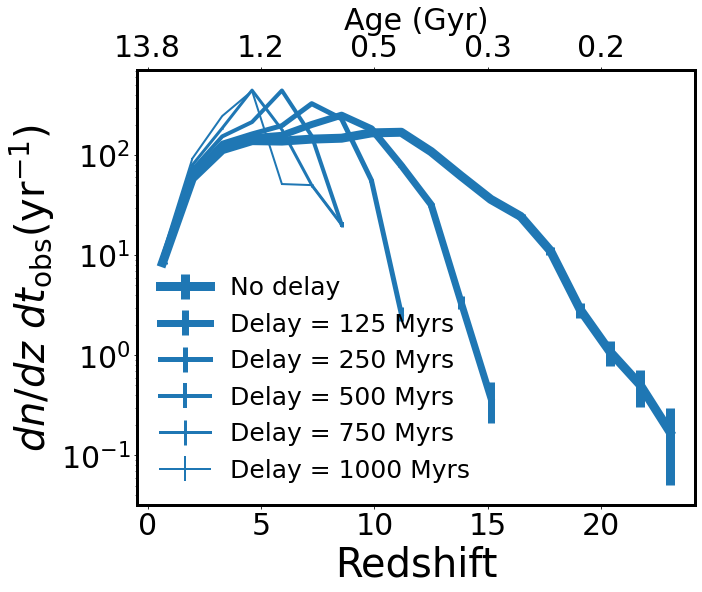}
\caption{Predicted merger rates for the \texttt{SM5} model under different assumptions for the delay times for the \textit{true merger} (compared to the simulated merger). These are shown for the unconstrained simulations. As we increase the delay times, the merger rates are strongly suppressed at the highest redshifts, and the peak of the distribution occurs at lower redshifts.}
\label{merger_rates_delay}
\end{figure}
\begin{figure*}
\includegraphics[width=16 cm]{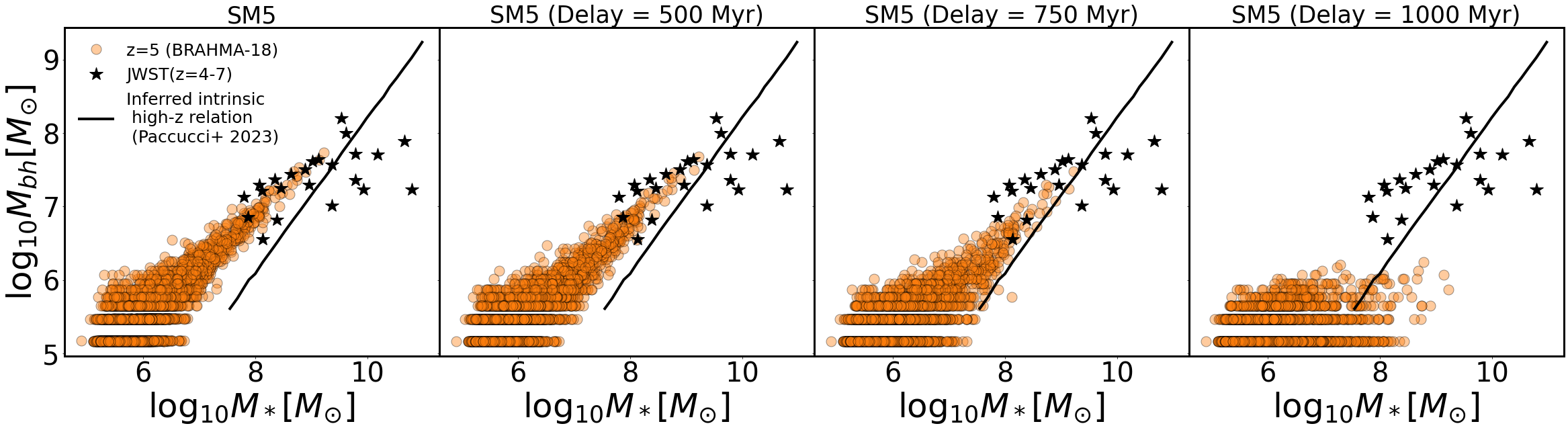}
\includegraphics[width=16 cm]{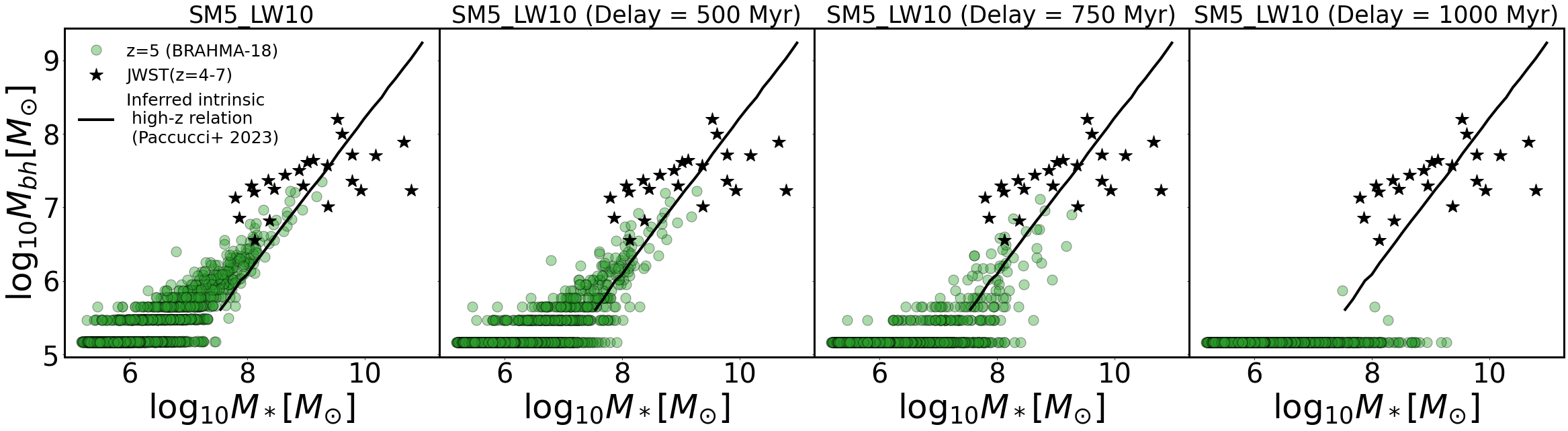}
\caption{Revised $z=5$ $M_*-M_{\rm bh}$ relations for simulations with the two most lenient seed models \texttt{SM5} and \texttt{SM5_LW10}, under different assumptions of delay times. The leftmost panel assumes no delay time, the remaining panels assume non-zero delay time. These are shown for the unconstrained simulations. Since accretion driven BH growth is negligible, $M_{\rm bh}$ is calculated by simply multiplying the seed mass by the number of merger progenitors before $z=5$. To produce BHs with masses similar to the JWST AGNs, the delay times need to be $\lesssim750~\mathrm{Myr}$.}

\label{SMBHM_delay}
\end{figure*}
\label{Impact of delayed BH mergers}
In the previous subsection, we found that only our most optimistic seed models~(\texttt{SM5} and \texttt{SM5_LW10}) produce BH populations that are broadly comparable to the JWST measurements as they currently stand. Additionally, the \texttt{SM5} model overpredicts the P23 intrinsic high-z $M_*$ vs $M_{\rm bh}$ relation, while the \texttt{SM5_LW10} is comparable to the P23 relation at $z=4~\&~5$. However, even for these models, the merger driven growth in BH mass is likely overestimated as our repositioning scheme promptly merges the BHs soon after their host galaxies merge. Additionally, prompt mergers could also overestimate merger rate predictions for LISA. In this section, we study the implications of a possible time delay between the BH mergers and galaxy mergers. We consider a simple model that assumes a uniform time delay~($\tau$), and reconstruct the merger histories of all the BHs at a given snapshot. In Figure \ref{merger_rates_delay}, we consider delay times of $\tau=125,250,500,750~\&~1000~\mathrm{Myr}$, and show the resulting merger rates for the \texttt{SM5} model in the unconstrained simulations. Of course in reality, the delay times are not expected to be fixed and will likely scale with the dynamical time of the host halo. But due to the limited snapshot resolution, we are unable to track the host halos of all the BH mergers at the exact time they merge. Nevertheless, for our goal of simply estimating the typical merging times required by these seed models to reproduce the observations, a simple model with fixed delay time suffices. As we can see, the time-delay causes the merger rates at the highest redshifts~($z\gtrsim10$) to be strongly suppressed. Because all these earliest mergers are pushed to later times, the merger rates are enhanced at lower redshifts. 

Since the BH growth is merger-dominated, the corrected final mass of a BH at a given redshift snapshot $z$ is then given by 
\begin{equation}
M_{\rm bh} (z) = N_{\mathrm{prog}} (z,\tau) \times M_{\mathrm{seed}}
\end{equation}
where $N_{\mathrm{prog}} (z,\tau)$ is the number of progenitors contributed by all merger events before redshift $z$ for a given delay time $\tau$. In Figure \ref{SMBHM_delay}, we show the impact of different delay times on the $M_*$ vs. $M_{\rm bh}$ relations at $z=5$ for the \texttt{SM5} and \texttt{SM5_LW10} seed models in the unconstrained simulations. Not surprisingly, the time-delay decreases the BH masses accumulated at $z=5$, as the mergers that are delayed to $z < 5$ no longer contribute to the $z=5$ BH mass. The \texttt{SM5_LW10} seed model continues to be consistent with the P23 relation up to a delay time of $750~\mathrm{Myrs}$. But for a delay time of  $1000~\mathrm{Myrs}$, the predicted BH masses are strongly suppressed to values significantly below the JWST measurements. This is because at delay times of $\gtrsim1000~\mathrm{Myrs}$ or more, the majority of mergers occur after $z=5$~(revisit Figure \ref{merger_rates_delay}). 

Overall, we find that even for our most optimistic seed models, we would be able to reproduce the current JWST measurements only if the delay times between BH mergers and galaxy mergers are $\lesssim750~\mathrm{Myr}$. However, the delay times at these high redshifts are highly uncertain, and several recent works are finding that it is difficult to sink BHs to the halo centers within low mass halos at high redshifts~\citep{2018ApJ...857L..22T,2021MNRAS.503.6098R,2021MNRAS.508.1973M,2021MNRAS.505.5129B,2023arXiv231008079P}. Even at low redshifts, it is not clear whether SMBH pairs are able to effectively harden once they are at separations below a few parsecs; this is commonly known as the `final parsec problem'~\citep{1980Natur.287..307B, 2003ApJ...596..860M}. Nevertheless, mechanisms such as drag to circumbinary gas disks, stellar loss cone scattering and BH triple interactions could potentially solve the final parsec problem. To that end, recent detection of the stochastic gravitational wave background by the various Pulsar Timing Array~(PTA) collaborations such as North American Nanohertz Observatory for Gravitational Waves~(NANOGrav, \citealt{2023ApJ...951L...8A}), European and Indian PTA~(EPTA + InPTA,~\citealt{2023arXiv230616227A}), Chinese PTA~(CPTA,~\citealt{Xu_2023}), and the Parkes PTA~(PPTA,~\citealt{2023ApJ...951L...6R}), serve as the first possible hint that SMBHs do merge. In the future, we will trace the sub-resolution dynamics of our inspiraling BHs and investigate the role of these processes at high redshifts. We will do this using post-processing models similar to the ones developed in \cite{2017MNRAS.464.3131K}, and thereby estimate these merging delay times. 

\subsection{Predictions at cosmic noon: Implications of recently observed $z\sim1-3$ overmassive BHs}
\begin{figure}
\includegraphics[width=8 cm]{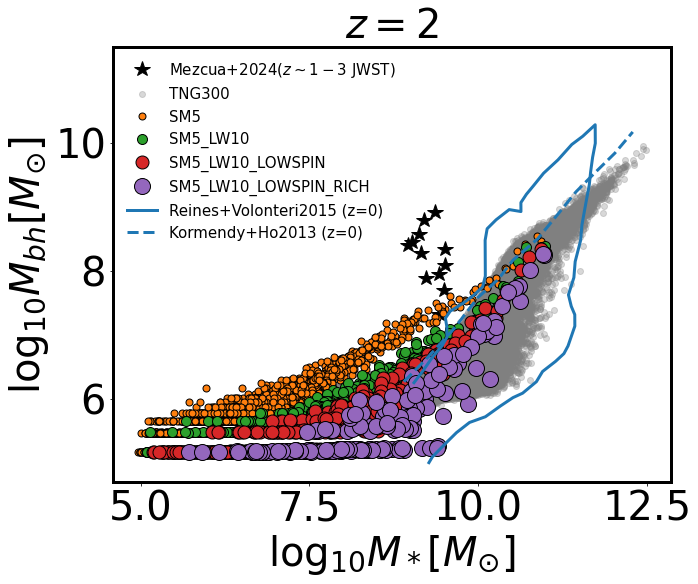}
\caption{$M_*-M_{\rm bh}$ relations at $z=2$ produced by our unconstrained simulations compared against the recent observations of overmassive BHs at cosmic noon~\protect\citep{2024arXiv240405793M} with the assumption of zero delay time. Our simulated BHs are close to the local scaling relations by $z\sim2$ for all the seed models. Therefore, the simulated BHs are significantly smaller than the JWST observations for all the seed models.}

\label{SMBHM_z2}
\end{figure}

While this paper largely focuses on the $z\sim4-7$ BH populations, very recently, \cite{2024arXiv240405793M} reported JWST observations of 12 SMBHs at $z\sim1-3$ that are also overmassive compared to the local scaling relations. In Figure \ref{SMBHM_z2}, we compare these observations to our unconstrained  simulation predictions at the $z=2$ snapshot for all the seed models. We find that even for the most optimistic seed model~(\texttt{SM5}) under zero delay time for the BH mergers~($\tau=0$), the predicted $M_*-M_{\rm bh}$ relations are already very close to the local measurements at $z=2$. Therefore, while JWST observes $\sim10^7-10^9~M_{\odot}$ BHs within galaxies with stellar masses $\sim10^9~M_{\odot}$, the \texttt{SM5} model predicts $\sim10^6-10^7~M_{\odot}$ BHs within similarly massive simulated galaxies. The strictest \texttt{SM5_LW10_LOWSPIN_RICH} seed model predicts BHs that are even smaller than \texttt{SM5} by a factor of $\sim10$. Overall, because galaxy growth is significantly faster than BH growth between $z\sim4$ and $z\sim2$ in our simulations, none of our seed models~(including the most optimistic one) produce BHs that overlap the overmassive BHs at cosmic noon reported by \cite{2024arXiv240405793M}. 

As with the $z\sim4-7$ AGNs, the BH mass measurements of the cosmic noon AGN populations may also be impacted by Lauer bias. Here again, our simulations are not large enough to produce any substantially up-scattered BHs that are significantly above the mean relations. Given the $1.5-2~\mathrm{dex}$ scatter in the \texttt{TNG300} simulation~(grey points in Figure \ref{SMBHM_z2}), we could expect the \texttt{SM5} seed model to produce a significant number of up-scattered BHs consistent with the  \cite{2024arXiv240405793M} observations if it was run over a larger volume. However, for the most restrictive \texttt{SM5_LW10_LOWSPIN_RICH} seed model, wherein the mean relation is $\sim2$ dex below the observations, it would very difficult to produce BHs up-scattered enough to overlap with the observations even in a \texttt{TNG300}-like volume. Here again, recall that the JWST surveys are substantially smaller than \texttt{TNG300}. Overall, the detection of these overmassive $z\sim1-3$ BHs further adds to our longstanding puzzle of SMBH origins particularly at high-z, as they are even more difficult to produce with our seed models than the $z\sim4-7$ JWST AGN and the $z\gtrsim6$ pre-JWST quasars~\citep[see][]{2022MNRAS.516..138B}. This further echoes the need to explore alternative BH seeding and growth scenarios described at the end of Section \ref{Stellar mass vs black hole mass relations: Comparison with JWST}.

 \section{Summary and Conclusions}
 \label{conclusions}
Recent JWST measurements hint at the possible existence of overmassive BHs that are $\sim10-100$ times higher on the $M_*$ vs $M_{\rm bh}$ plane compared to the local scaling relations. To understand the possible implications of these developments on BH seeding, we have studied the growth of SMBHs at high redshift~($z\sim4-7$) under systematic variations of heavy seeding scenarios by running a set of cosmological hydrodynamic simulations as part of the \texttt{BRAHMA} suite. 

We used the following seeding criteria to identify halos and place $1.5\times10^5~M_{\odot}$ seeds: The \textit{dense and metal poor gas mass criterion} ensures that halos have a minimum amount of dense and metal-poor gas mass of $5$ times the seed mass. The \textit{LW flux criterion} ensures that the dense and metal-poor gas mass is also illuminated by LW radiation above a critical flux of $10~J_{21}$. The \textit{gas spin criterion} ensures that the net spin of the gas within the seed forming halos is less than the Toomre instability threshold. Finally, the \textit{environmental richness criterion} requires the seed forming halos to be in rich environments with at least one neighboring halo of comparable or higher mass. This criterion emulates the expectation that DCBH formation can be aided by dynamical heating of gas during major galaxy mergers that are more frequent within rich environments. We ran simulation boxes that sequentially add the above seeding criteria: The first box~(\texttt{SM5}) assumes the most optimistic seeding efficiency by only applying the \textit{dense and metal poor gas mass criterion}. The second~(\texttt{SM5_LW10}) and third~(\texttt{SM5_LW10_LOWSPIN}) boxes include the \textit{LW-flux criterion} and \textit{gas-spin criterion}. Only the fourth and final box~(\texttt{SM5_LW10_LOWSPIN_RICH}) includes all the seeding criteria considered in this work. Our main conclusions are summarized as follows:

\subsection*{Seed models have no significant impact on high-z galaxy populations and AGN luminosity functions. These predictions are also broadly consistent with observations.}
Notably, our seed models have a negligible consequence on the galaxy populations. This is likely because the BH accretion rates are too small for AGN feedback to impact the galaxies. Additionally, the seed models also do not significantly impact the relatively bright end~($L_{\mathrm{bol}}\gtrsim10^{43}~\mathrm{erg~s^{-1}}$) of our simulated AGN populations. As also noted in our previous papers, this is because regardless of how many seeds are produced, there are only a limited set of environments that support enough gas accretion to produce luminous AGN. Concurrently, we also find that the simulations are in broad agreement with JWST galaxy and AGN populations for the galaxy luminosity functions, galaxy stellar mass functions, as well as the AGN luminosity functions. This is particularly encouraging as it not only serves as a validation, but also as a benchmark to understand the implications of measurements that are much more challenging to observationally probe for the JWST AGNs and their host galaxies, for example the $M_{\rm bh}$ vs $M_*$ relations.   

\subsection*{Simulations do not overlap with the JWST observations on the galaxy stellar mass vs AGN luminosity plane.}

Despite the simulations being in simultaneous agreement with the galaxy stellar mass functions as well as the AGN luminosity functions, the simulated AGNs do not overlap with the JWST AGNs on the $M_*$ vs $L_{\mathrm{bol}}$ plane. More specifically, the simulated AGNs with $L_{\mathrm{bol}}\gtrsim10^{44}~\mathrm{erg~s^{-1}}$ live in $M_*\gtrsim10^{9}~M_{\odot}$ galaxies, which are $\sim10$ times higher than the measured stellar masses of the JWST AGN hosts. Importantly, this discrepancy is independent of the potential uncertainties in the observed BH masses.  Possible reasons for this discrepancy include: 1) AGN variability at short time scales could lead to Eddington bias i.e.  their preferential detection at luminosities significantly higher than their \textit{time averaged values}. 2) Potential underestimation of the observed stellar masses due to the difficulty in subtracting the AGN contribution from the observed light. Lastly, 3) due to Lauer bias, we may preferentially observe only the significantly up-scattered AGN having luminosities significantly higher than the mean $M_*$ vs $L_{\mathrm{bol}}$ relation; however, this would imply that the scatter in the $M_*$ vs $L_{\mathrm{bol}}$ relations is significantly higher than the predictions from the simulations.

\subsection*{BH growth at high-z is dominated by mergers. Therefore, BH masses are substantially impacted by the seed model.}

In our simulations, the BH growth is predominantly contributed by mergers at these high redshifts, with accretion driven BH growth being relatively small. Due to this, our seed models have a substantial impact on the BH masses even though the AGN luminosities are minimally impacted. To that end, by repositioning the BHs to the local potential minima, we assume the most optimistic scenario for the merging efficiencies wherein there is zero time delay between BH mergers and galaxy mergers. As we explain next, despite this assumption, our restrictive seed models significantly underpredict the $M_*$ vs $M_{\rm bh}$ relations compared to JWST.
\subsection*{Comparing the simulation predictions against the JWST measurements on the $M_*$ vs $M_{\rm bh}$ plane:}
Amongst our seed models, only the two most optimistic ones~(\texttt{SM5} and \texttt{SM5_LW10}) produce $M_*$ vs $M_{\rm bh}$ relations that have some overlap with JWST AGNs. On the other hand, for the most restrictive restrictive seed model~(\texttt{SM5_LW10_LOWSPIN_RICH}) where all the seeding criteria are included, the predictions are substantially below the JWST AGNs. However, if the Lauer bias is substantial in the observed BH populations, the resulting BH masses may also be significantly up-scattered on the $M_*$ vs $M_{\rm bh}$ plane, as more massive BHs typically power more luminous AGN. This makes it difficult to directly compare these results with the \texttt{BRAHMA} boxes since they are not large enough to produce BHs that are substantially upscattered. 

\subsection*{Comparing the simulation predictions against the intrinsic high-z $M_*$ vs $M_{\rm bh}$ relation derived by P23:}

P23 found that even after accounting for possible systematic biases and measurement uncertainties in the BH mass and host stellar masses, the JWST AGNs~(as they currently stand) implied an intrinsic $M_*$ vs $M_{\rm bh}$ relation that lies above the local scaling relations at a $>3\sigma$ confidence level. Making an ``apples-to-apples" comparison of this high-z relation against the simulations, provides strong implications on our seed models. For our most restrictive seed model (\texttt{SM5_LW10_LOWSPIN_RICH}), the BH masses are lower than the P23 relation by a factor of $\sim10$ at $z=4~\&~5$. This is because not enough seeds are produced to fuel the merger driven BH growth. In addition, any time-delay between BH mergers and galaxy mergers would further compromise the growth of these BHs. Therefore, if the inferred high-z $M_*$ vs. $M_{\rm bh}$ relation from P23 proves to be robust in the future~(when we have larger samples of high-z AGN), this could potentially rule out heavy seeds formed via standard direct collapse scenarios~(as considered in this paper) as their sole seeding origins.      

Only the simulation~(\texttt{SM5_LW10}) that excludes the gas spin and environmental richness criteria, predicts a $M_*$ vs. $M_{\rm bh}$ relation consistent with the P23 relation. If we also exclude the LW flux criterion~(\texttt{SM5}), the simulation overpredicts the BH masses compared to the P23 relation. However, these simulations are rather optimistic: For example, not all regions with sufficient dense and metal poor gas are expected to form DCBHs~(as assumed by the \texttt{SM5} simulation). This is because the molecular hydrogen will cool and fragment the gas to form Pop III stars instead. Moreover, even when we include the LW flux criterion to restrict seeding to those halos wherein the radiation can suppress the molecular Hydrogen cooling, our choice of $J_{\mathrm{crit}}=10~J_{21}$ is much lower than the predictions from small scale hydrodynamic simulations~($\gtrsim1000~J_{21}$). While such low $J_{\mathrm{crit}}$ values may be feasible if the gas is subjected to dynamical heating during halo mergers, restricting the seed formation to rich environments~(where these major mergers are expected to occur) leads to BHs significantly less massive than the P23 relation.

We further determined that even with our most optimistic simulations~(\texttt{SM5} and \texttt{SM5_LW10}), one could potentially produce the JWST AGNs and the P23 relation only if the typical delay times between the BH mergers and galaxy mergers are $\lesssim750~\mathrm{Myr}$. However, several recent works are finding that sinking BHs to halo centers is challenging within low mass halos at high redshifts~\citep{2021MNRAS.508.1973M,2023arXiv231008079P}. Additionally, at low redshifts, the estimated delay times for the merger events can be up to several Gyr~\citep{2017MNRAS.464.3131K}. Therefore, the feasibility of delay times being $\lesssim750~\mathrm{Myr}$ at higher redshifts will require further investigation in the future. 

\subsection*{Possibility of Lauer bias in the BH mass measurements would imply that the simulations underpredict the scatter in the BH-galaxy scaling relations.}

Our work also reveals that the BH mass and host stellar mass measurements of these high-z JWST AGNs may have significant implications not just on the mean trends, but also the underlying scatter within the high-z $M_*$ vs $M_{\rm bh}$~(and $M_*$ vs $L_{\rm bol}$)  relations. In general, having a larger scatter in the intrinsic scaling relations enhances the likelihood of detecting up-scattered objects  in smaller volume surveys. Therefore, if the JWST BH mass measurements are indeed up-scattered due to Lauer bias, it would imply that the high-z scatter is significantly larger than predicted by our simulations. Due to the limited volume, the \texttt{BRAHMA} boxes do not effectively probe the full scatter of the $M_*$ vs $M_{\rm bh}$ relations, as they cannot capture the formation of rare up-scattered BHs. In the future, we plan to explore the scatter in more detail by running some of our seed models within much larger simulation volumes.

\subsection*{Final remarks:}

 Overall, our work hints that if it is confirmed in the future that high-z BH populations are systematically overmassive, we would need to consider alternative paths to their assembly using heavy seeds. First, there may be alternative heavy seeding channels that may be more efficient than the ones considered in this work. For instance, in this work, the seeding efficiency is limited by the mass resolution of the seed forming halos and the eEOS description of the ISM; this prevents us from exploring seed formation in halos that are significantly below our halo mass resolution limit of $\sim5 \times 10^7~M_{\odot}$. However, higher resolution simulations with an explicitly resolved ISM will allow us to explore heavy seed formation in lower mass halos. Additionally, light seeds are expected to form in much higher numbers compared to heavy seeds. These light seeds could also potentially boost the merger driven BH growth of heavy seeds. Another possibility is that the initial seed masses could be much higher than $\sim10^5~M_{\odot}$. For example, \cite{2024ApJ...961...76M} demonstrated that dynamical heating caused by major mergers could lead to direct formation of $\sim10^8~M_{\odot}$ BHs even in metal enriched regions that are concurrently undergoing starbursts. Finally, the assembly of overmassive high-z BHs could be driven by a select few seeds that grow much more rapidly via more efficient gas accretion channels compared to the Bondi-accretion model considered in this work~\citep{2024arXiv240218773J}. In the near future, we plan to explore all above avenues in detail and investigate the joint implications of possible high-z overmassive BHs on seeding, growth and dynamics.




\section*{Acknowledgements}
LB acknowledges support from NSF awards AST-1909933 \& AST-2307171 and Cottrell Scholar Award \#27553 from the Research Corporation for Science Advancement.
PT acknowledges support from NSF-AST 2008490.
RW acknowledges funding of a Leibniz Junior Research Group (project number J131/2022).  LH acknowledges support by the Simons Collaboration on ``Learning the Universe''. 
\section*{Data availablity}
The underlying data used in this work shall be made available upon reasonable request to the corresponding author.

\bibliography{references}

\begin{thebibliography}{}
\makeatletter
\relax
\def\mn@urlcharsother{\let\do\@makeother \do\$\do\&\do\#\do\^\do\_\do\%\do\~}
\def\mn@doi{\begingroup\mn@urlcharsother \@ifnextchar [ {\mn@doi@} {\mn@doi@[]}}
\def\mn@doi@[#1]#2{\def\@tempa{#1}\ifx\@tempa\@empty \href {http://dx.doi.org/#2} {doi:#2}\else \href {http://dx.doi.org/#2} {#1}\fi \endgroup}
\def\mn@eprint#1#2{\mn@eprint@#1:#2::\@nil}
\def\mn@eprint@arXiv#1{\href {http://arxiv.org/abs/#1} {{\tt arXiv:#1}}}
\def\mn@eprint@dblp#1{\href {http://dblp.uni-trier.de/rec/bibtex/#1.xml} {dblp:#1}}
\def\mn@eprint@#1:#2:#3:#4\@nil{\def\@tempa {#1}\def\@tempb {#2}\def\@tempc {#3}\ifx \@tempc \@empty \let \@tempc \@tempb \let \@tempb \@tempa \fi \ifx \@tempb \@empty \def\@tempb {arXiv}\fi \@ifundefined {mn@eprint@\@tempb}{\@tempb:\@tempc}{\expandafter \expandafter \csname mn@eprint@\@tempb\endcsname \expandafter{\@tempc}}}

\bibitem[\protect\citeauthoryear{{Agarwal}, {Davis}, {Khochfar}, {Natarajan}  \& {Dunlop}}{{Agarwal} et~al.}{2013}]{2013MNRAS.432.3438A}
{Agarwal} B.,  {Davis} A.~J.,  {Khochfar} S.,  {Natarajan} P.,   {Dunlop} J.~S.,  2013, \mn@doi [\mnras] {10.1093/mnras/stt696}, \href {https://ui.adsabs.harvard.edu/abs/2013MNRAS.432.3438A} {432, 3438}

\bibitem[\protect\citeauthoryear{{Agazie} et~al.,}{{Agazie} et~al.}{2023}]{2023ApJ...951L...8A}
{Agazie} G.,  et~al., 2023, \mn@doi [\apjl] {10.3847/2041-8213/acdac6}, \href {https://ui.adsabs.harvard.edu/abs/2023ApJ...951L...8A} {951, L8}

\bibitem[\protect\citeauthoryear{{Amaro-Seoane} et~al.,}{{Amaro-Seoane} et~al.}{2017}]{2017arXiv170200786A}
{Amaro-Seoane} P.,  et~al., 2017, \mn@doi [arXiv e-prints] {10.48550/arXiv.1702.00786}, \href {https://ui.adsabs.harvard.edu/abs/2017arXiv170200786A} {p. arXiv:1702.00786}

\bibitem[\protect\citeauthoryear{{Andika} et~al.,}{{Andika} et~al.}{2024}]{2024A&A...685A..25A}
{Andika} I.~T.,  et~al., 2024, \mn@doi [\aap] {10.1051/0004-6361/202349025}, \href {https://ui.adsabs.harvard.edu/abs/2024A&A...685A..25A} {685, A25}

\bibitem[\protect\citeauthoryear{{Antoniadis} et~al.,}{{Antoniadis} et~al.}{2023}]{2023arXiv230616227A}
{Antoniadis} J.,  et~al., 2023, \mn@doi [arXiv e-prints] {10.48550/arXiv.2306.16227}, \href {https://ui.adsabs.harvard.edu/abs/2023arXiv230616227A} {p. arXiv:2306.16227}

\bibitem[\protect\citeauthoryear{{Ba{\~n}ados} et~al.,}{{Ba{\~n}ados} et~al.}{2018}]{2018Natur.553..473B}
{Ba{\~n}ados} E.,  et~al., 2018, \mn@doi [\nat] {10.1038/nature25180}, \href {https://ui.adsabs.harvard.edu/abs/2018Natur.553..473B} {553, 473}

\bibitem[\protect\citeauthoryear{{Barnes} \& {Hut}}{{Barnes} \& {Hut}}{1986}]{1986Natur.324..446B}
{Barnes} J.,  {Hut} P.,  1986, \mn@doi [\nat] {10.1038/324446a0}, \href {https://ui.adsabs.harvard.edu/abs/1986Natur.324..446B} {324, 446}

\bibitem[\protect\citeauthoryear{Bañados et~al.,}{Bañados et~al.}{2016}]{2016Banados}
Bañados E.,  et~al., 2016, \mn@doi [The Astrophysical Journal Supplement Series] {10.3847/0067-0049/227/1/11}, 227, 11

\bibitem[\protect\citeauthoryear{{Begelman}}{{Begelman}}{2010}]{2010MNRAS.402..673B}
{Begelman} M.~C.,  2010, \mn@doi [\mnras] {10.1111/j.1365-2966.2009.15916.x}, \href {https://ui.adsabs.harvard.edu/abs/2010MNRAS.402..673B} {402, 673}

\bibitem[\protect\citeauthoryear{{Begelman} \& {Silk}}{{Begelman} \& {Silk}}{2023}]{2023MNRAS.526L..94B}
{Begelman} M.~C.,  {Silk} J.,  2023, \mn@doi [\mnras] {10.1093/mnrasl/slad124}, \href {https://ui.adsabs.harvard.edu/abs/2023MNRAS.526L..94B} {526, L94}

\bibitem[\protect\citeauthoryear{{Begelman}, {Blandford}  \& {Rees}}{{Begelman} et~al.}{1980}]{1980Natur.287..307B}
{Begelman} M.~C.,  {Blandford} R.~D.,   {Rees} M.~J.,  1980, \mn@doi [\nat] {10.1038/287307a0}, \href {https://ui.adsabs.harvard.edu/abs/1980Natur.287..307B} {287, 307}

\bibitem[\protect\citeauthoryear{{Begelman}, {Volonteri}  \& {Rees}}{{Begelman} et~al.}{2006}]{2006MNRAS.370..289B}
{Begelman} M.~C.,  {Volonteri} M.,   {Rees} M.~J.,  2006, \mn@doi [\mnras] {10.1111/j.1365-2966.2006.10467.x}, \href {https://ui.adsabs.harvard.edu/abs/2006MNRAS.370..289B} {370, 289}

\bibitem[\protect\citeauthoryear{{Bellovary}, {Cleary}, {Munshi}, {Tremmel}, {Christensen}, {Brooks}  \& {Quinn}}{{Bellovary} et~al.}{2019}]{2019MNRAS.482.2913B}
{Bellovary} J.~M.,  {Cleary} C.~E.,  {Munshi} F.,  {Tremmel} M.,  {Christensen} C.~R.,  {Brooks} A.,   {Quinn} T.~R.,  2019, \mn@doi [\mnras] {10.1093/mnras/sty2842}, \href {https://ui.adsabs.harvard.edu/abs/2019MNRAS.482.2913B} {482, 2913}

\bibitem[\protect\citeauthoryear{{Bellovary} et~al.,}{{Bellovary} et~al.}{2021}]{2021MNRAS.505.5129B}
{Bellovary} J.~M.,  et~al., 2021, \mn@doi [\mnras] {10.1093/mnras/stab1665}, \href {https://ui.adsabs.harvard.edu/abs/2021MNRAS.505.5129B} {505, 5129}

\bibitem[\protect\citeauthoryear{{Bhowmick} et~al.,}{{Bhowmick} et~al.}{2021}]{2021MNRAS.507.2012B}
{Bhowmick} A.~K.,  et~al., 2021, \mn@doi [\mnras] {10.1093/mnras/stab2204}, \href {https://ui.adsabs.harvard.edu/abs/2021MNRAS.507.2012B} {507, 2012}

\bibitem[\protect\citeauthoryear{{Bhowmick}, {Blecha}, {Torrey}, {Kelley}, {Vogelsberger}, {Nelson}, {Weinberger}  \& {Hernquist}}{{Bhowmick} et~al.}{2022a}]{2022MNRAS.510..177B}
{Bhowmick} A.~K.,  {Blecha} L.,  {Torrey} P.,  {Kelley} L.~Z.,  {Vogelsberger} M.,  {Nelson} D.,  {Weinberger} R.,   {Hernquist} L.,  2022a, \mn@doi [\mnras] {10.1093/mnras/stab3439}, \href {https://ui.adsabs.harvard.edu/abs/2022MNRAS.510..177B} {510, 177}

\bibitem[\protect\citeauthoryear{{Bhowmick} et~al.,}{{Bhowmick} et~al.}{2022b}]{2022MNRAS.516..138B}
{Bhowmick} A.~K.,  et~al., 2022b, \mn@doi [\mnras] {10.1093/mnras/stac2238}, \href {https://ui.adsabs.harvard.edu/abs/2022MNRAS.516..138B} {516, 138}

\bibitem[\protect\citeauthoryear{{Bhowmick}, {Blecha}, {Torrey}, {Weinberger}, {Kelley}, {Vogelsberger}, {Hernquist}  \& {Somerville}}{{Bhowmick} et~al.}{2023}]{2023arXiv230915341B}
{Bhowmick} A.~K.,  {Blecha} L.,  {Torrey} P.,  {Weinberger} R.,  {Kelley} L.~Z.,  {Vogelsberger} M.,  {Hernquist} L.,   {Somerville} R.~S.,  2023, \mn@doi [arXiv e-prints] {10.48550/arXiv.2309.15341}, \href {https://ui.adsabs.harvard.edu/abs/2023arXiv230915341B} {p. arXiv:2309.15341}

\bibitem[\protect\citeauthoryear{{Bhowmick} et~al.,}{{Bhowmick} et~al.}{2024}]{2024arXiv240203626B}
{Bhowmick} A.~K.,  et~al., 2024, \mn@doi [arXiv e-prints] {10.48550/arXiv.2402.03626}, \href {https://ui.adsabs.harvard.edu/abs/2024arXiv240203626B} {p. arXiv:2402.03626}

\bibitem[\protect\citeauthoryear{{Blecha} \& {Loeb}}{{Blecha} \& {Loeb}}{2008}]{2008MNRAS.390.1311B}
{Blecha} L.,  {Loeb} A.,  2008, \mn@doi [\mnras] {10.1111/j.1365-2966.2008.13790.x}, \href {https://ui.adsabs.harvard.edu/abs/2008MNRAS.390.1311B} {390, 1311}

\bibitem[\protect\citeauthoryear{{Blecha} et~al.,}{{Blecha} et~al.}{2016}]{2016MNRAS.456..961B}
{Blecha} L.,  et~al., 2016, \mn@doi [\mnras] {10.1093/mnras/stv2646}, \href {https://ui.adsabs.harvard.edu/abs/2016MNRAS.456..961B} {456, 961}

\bibitem[\protect\citeauthoryear{{Bogd{\'a}n} et~al.,}{{Bogd{\'a}n} et~al.}{2024}]{2024NatAs...8..126B}
{Bogd{\'a}n} {\'A}.,  et~al., 2024, \mn@doi [Nature Astronomy] {10.1038/s41550-023-02111-9}, \href {https://ui.adsabs.harvard.edu/abs/2024NatAs...8..126B} {8, 126}

\bibitem[\protect\citeauthoryear{{Bouwens}, {Illingworth}, {van Dokkum}, {Oesch}, {Stefanon}  \& {Ribeiro}}{{Bouwens} et~al.}{2022a}]{2022ApJ...927...81B}
{Bouwens} R.~J.,  {Illingworth} G.~D.,  {van Dokkum} P.~G.,  {Oesch} P.~A.,  {Stefanon} M.,   {Ribeiro} B.,  2022a, \mn@doi [\apj] {10.3847/1538-4357/ac4791}, \href {https://ui.adsabs.harvard.edu/abs/2022ApJ...927...81B} {927, 81}

\bibitem[\protect\citeauthoryear{{Bouwens}, {Illingworth}, {Ellis}, {Oesch}  \& {Stefanon}}{{Bouwens} et~al.}{2022b}]{2022ApJ...940...55B}
{Bouwens} R.~J.,  {Illingworth} G.,  {Ellis} R.~S.,  {Oesch} P.,   {Stefanon} M.,  2022b, \mn@doi [\apj] {10.3847/1538-4357/ac86d1}, \href {https://ui.adsabs.harvard.edu/abs/2022ApJ...940...55B} {940, 55}

\bibitem[\protect\citeauthoryear{{Bromm} \& {Loeb}}{{Bromm} \& {Loeb}}{2003}]{2003ApJ...596...34B}
{Bromm} V.,  {Loeb} A.,  2003, \mn@doi [\apj] {10.1086/377529}, \href {https://ui.adsabs.harvard.edu/abs/2003ApJ...596...34B} {596, 34}

\bibitem[\protect\citeauthoryear{{Chabrier}}{{Chabrier}}{2003}]{2003PASP..115..763C}
{Chabrier} G.,  2003, \mn@doi [\pasp] {10.1086/376392}, \href {https://ui.adsabs.harvard.edu/abs/2003PASP..115..763C} {115, 763}

\bibitem[\protect\citeauthoryear{{Das}, {Schleicher}, {Basu}  \& {Boekholt}}{{Das} et~al.}{2021a}]{2021MNRAS.tmp.1381D}
{Das} A.,  {Schleicher} D. R.~G.,  {Basu} S.,   {Boekholt} T. C.~N.,  2021a, \mn@doi [\mnras] {10.1093/mnras/stab1428}, \href {https://ui.adsabs.harvard.edu/abs/2021MNRAS.tmp.1381D} {}

\bibitem[\protect\citeauthoryear{{Das}, {Schleicher}, {Leigh}  \& {Boekholt}}{{Das} et~al.}{2021b}]{2021MNRAS.503.1051D}
{Das} A.,  {Schleicher} D. R.~G.,  {Leigh} N. W.~C.,   {Boekholt} T. C.~N.,  2021b, \mn@doi [\mnras] {10.1093/mnras/stab402}, \href {https://ui.adsabs.harvard.edu/abs/2021MNRAS.503.1051D} {503, 1051}

\bibitem[\protect\citeauthoryear{{Davies}, {Miller}  \& {Bellovary}}{{Davies} et~al.}{2011}]{2011ApJ...740L..42D}
{Davies} M.~B.,  {Miller} M.~C.,   {Bellovary} J.~M.,  2011, \mn@doi [\apjl] {10.1088/2041-8205/740/2/L42}, \href {https://ui.adsabs.harvard.edu/abs/2011ApJ...740L..42D} {740, L42}

\bibitem[\protect\citeauthoryear{{Davis}, {Efstathiou}, {Frenk}  \& {White}}{{Davis} et~al.}{1985}]{1985ApJ...292..371D}
{Davis} M.,  {Efstathiou} G.,  {Frenk} C.~S.,   {White} S.~D.~M.,  1985, \mn@doi [\apj] {10.1086/163168}, \href {https://ui.adsabs.harvard.edu/abs/1985ApJ...292..371D} {292, 371}

\bibitem[\protect\citeauthoryear{{DeGraf} \& {Sijacki}}{{DeGraf} \& {Sijacki}}{2020}]{2020MNRAS.491.4973D}
{DeGraf} C.,  {Sijacki} D.,  2020, \mn@doi [\mnras] {10.1093/mnras/stz3309}, \href {https://ui.adsabs.harvard.edu/abs/2020MNRAS.491.4973D} {491, 4973}

\bibitem[\protect\citeauthoryear{{Di Matteo}, {Khandai}, {DeGraf}, {Feng}, {Croft}, {Lopez}  \& {Springel}}{{Di Matteo} et~al.}{2012}]{2012ApJ...745L..29D}
{Di Matteo} T.,  {Khandai} N.,  {DeGraf} C.,  {Feng} Y.,  {Croft} R.~A.~C.,  {Lopez} J.,   {Springel} V.,  2012, \mn@doi [\apjl] {10.1088/2041-8205/745/2/L29}, \href {https://ui.adsabs.harvard.edu/abs/2012ApJ...745L..29D} {745, L29}

\bibitem[\protect\citeauthoryear{{Dijkstra}, {Haiman}, {Mesinger}  \& {Wyithe}}{{Dijkstra} et~al.}{2008}]{2008MNRAS.391.1961D}
{Dijkstra} M.,  {Haiman} Z.,  {Mesinger} A.,   {Wyithe} J. S.~B.,  2008, \mn@doi [\mnras] {10.1111/j.1365-2966.2008.14031.x}, \href {https://ui.adsabs.harvard.edu/abs/2008MNRAS.391.1961D} {391, 1961}

\bibitem[\protect\citeauthoryear{{Ding} et~al.,}{{Ding} et~al.}{2023}]{2023Natur.621...51D}
{Ding} X.,  et~al., 2023, \mn@doi [\nat] {10.1038/s41586-023-06345-5}, \href {https://ui.adsabs.harvard.edu/abs/2023Natur.621...51D} {621, 51}

\bibitem[\protect\citeauthoryear{{Donnari} et~al.,}{{Donnari} et~al.}{2019}]{2019MNRAS.485.4817D}
{Donnari} M.,  et~al., 2019, \mn@doi [\mnras] {10.1093/mnras/stz712}, \href {https://ui.adsabs.harvard.edu/abs/2019MNRAS.485.4817D} {485, 4817}

\bibitem[\protect\citeauthoryear{{Dunn}, {Holley-Bockelmann}  \& {Bellovary}}{{Dunn} et~al.}{2020}]{2020ApJ...896...72D}
{Dunn} G.,  {Holley-Bockelmann} K.,   {Bellovary} J.,  2020, \mn@doi [\apj] {10.3847/1538-4357/ab7cd2}, \href {https://ui.adsabs.harvard.edu/abs/2020ApJ...896...72D} {896, 72}

\bibitem[\protect\citeauthoryear{{Durodola}, {Pacucci}  \& {Hickox}}{{Durodola} et~al.}{2024}]{2024arXiv240610329D}
{Durodola} E.,  {Pacucci} F.,   {Hickox} R.~C.,  2024, \mn@doi [arXiv e-prints] {10.48550/arXiv.2406.10329}, \href {https://ui.adsabs.harvard.edu/abs/2024arXiv240610329D} {p. arXiv:2406.10329}

\bibitem[\protect\citeauthoryear{{Fan} et~al.,}{{Fan} et~al.}{2001}]{2001AJ....122.2833F}
{Fan} X.,  et~al., 2001, \mn@doi [\aj] {10.1086/324111}, \href {https://ui.adsabs.harvard.edu/abs/2001AJ....122.2833F} {122, 2833}

\bibitem[\protect\citeauthoryear{{Feng}, {Di-Matteo}, {Croft}, {Bird}, {Battaglia}  \& {Wilkins}}{{Feng} et~al.}{2016}]{2016MNRAS.455.2778F}
{Feng} Y.,  {Di-Matteo} T.,  {Croft} R.~A.,  {Bird} S.,  {Battaglia} N.,   {Wilkins} S.,  2016, \mn@doi [\mnras] {10.1093/mnras/stv2484}, \href {https://ui.adsabs.harvard.edu/abs/2016MNRAS.455.2778F} {455, 2778}

\bibitem[\protect\citeauthoryear{{Finkelstein} et~al.,}{{Finkelstein} et~al.}{2023}]{2023ApJ...946L..13F}
{Finkelstein} S.~L.,  et~al., 2023, \mn@doi [\apjl] {10.3847/2041-8213/acade4}, \href {https://ui.adsabs.harvard.edu/abs/2023ApJ...946L..13F} {946, L13}

\bibitem[\protect\citeauthoryear{{Fryer}, {Woosley}  \& {Heger}}{{Fryer} et~al.}{2001}]{2001ApJ...550..372F}
{Fryer} C.~L.,  {Woosley} S.~E.,   {Heger} A.,  2001, \mn@doi [\apj] {10.1086/319719}, \href {https://ui.adsabs.harvard.edu/abs/2001ApJ...550..372F} {550, 372}

\bibitem[\protect\citeauthoryear{{Genel} et~al.,}{{Genel} et~al.}{2018}]{2018MNRAS.474.3976G}
{Genel} S.,  et~al., 2018, \mn@doi [\mnras] {10.1093/mnras/stx3078}, \href {https://ui.adsabs.harvard.edu/abs/2018MNRAS.474.3976G} {474, 3976}

\bibitem[\protect\citeauthoryear{{Gerosa} \& {Moore}}{{Gerosa} \& {Moore}}{2016}]{2016PhRvL.117a1101G}
{Gerosa} D.,  {Moore} C.~J.,  2016, \mn@doi [\prl] {10.1103/PhysRevLett.117.011101}, \href {https://ui.adsabs.harvard.edu/abs/2016PhRvL.117a1101G} {117, 011101}

\bibitem[\protect\citeauthoryear{{Greene} \& {Ho}}{{Greene} \& {Ho}}{2005}]{2005ApJ...630..122G}
{Greene} J.~E.,  {Ho} L.~C.,  2005, \mn@doi [\apj] {10.1086/431897}, \href {https://ui.adsabs.harvard.edu/abs/2005ApJ...630..122G} {630, 122}

\bibitem[\protect\citeauthoryear{{Greene} et~al.,}{{Greene} et~al.}{2023}]{2023arXiv230905714G}
{Greene} J.~E.,  et~al., 2023, \mn@doi [arXiv e-prints] {10.48550/arXiv.2309.05714}, \href {https://ui.adsabs.harvard.edu/abs/2023arXiv230905714G} {p. arXiv:2309.05714}

\bibitem[\protect\citeauthoryear{{Habouzit}, {Volonteri}, {Latif}, {Dubois}  \& {Peirani}}{{Habouzit} et~al.}{2016}]{2016MNRAS.463..529H}
{Habouzit} M.,  {Volonteri} M.,  {Latif} M.,  {Dubois} Y.,   {Peirani} S.,  2016, \mn@doi [\mnras] {10.1093/mnras/stw1924}, \href {https://ui.adsabs.harvard.edu/abs/2016MNRAS.463..529H} {463, 529}

\bibitem[\protect\citeauthoryear{{Habouzit}, {Volonteri}  \& {Dubois}}{{Habouzit} et~al.}{2017}]{2017MNRAS.468.3935H}
{Habouzit} M.,  {Volonteri} M.,   {Dubois} Y.,  2017, \mn@doi [\mnras] {10.1093/mnras/stx666}, \href {https://ui.adsabs.harvard.edu/abs/2017MNRAS.468.3935H} {468, 3935}

\bibitem[\protect\citeauthoryear{{Habouzit} et~al.,}{{Habouzit} et~al.}{2019}]{2019MNRAS.484.4413H}
{Habouzit} M.,  et~al., 2019, \mn@doi [\mnras] {10.1093/mnras/stz102}, \href {https://ui.adsabs.harvard.edu/abs/2019MNRAS.484.4413H} {484, 4413}

\bibitem[\protect\citeauthoryear{{Habouzit} et~al.,}{{Habouzit} et~al.}{2021}]{2021MNRAS.503.1940H}
{Habouzit} M.,  et~al., 2021, \mn@doi [\mnras] {10.1093/mnras/stab496}, \href {https://ui.adsabs.harvard.edu/abs/2021MNRAS.503.1940H} {503, 1940}

\bibitem[\protect\citeauthoryear{{Habouzit} et~al.,}{{Habouzit} et~al.}{2022}]{2022MNRAS.509.3015H}
{Habouzit} M.,  et~al., 2022, \mn@doi [\mnras] {10.1093/mnras/stab3147}, \href {https://ui.adsabs.harvard.edu/abs/2022MNRAS.509.3015H} {509, 3015}

\bibitem[\protect\citeauthoryear{{Haemmerl{\'e}}, {Klessen}, {Mayer}  \& {Zwick}}{{Haemmerl{\'e}} et~al.}{2021}]{2021A&A...652L...7H}
{Haemmerl{\'e}} L.,  {Klessen} R.~S.,  {Mayer} L.,   {Zwick} L.,  2021, \mn@doi [\aap] {10.1051/0004-6361/202141376}, \href {https://ui.adsabs.harvard.edu/abs/2021A&A...652L...7H} {652, L7}

\bibitem[\protect\citeauthoryear{{Hahn} \& {Abel}}{{Hahn} \& {Abel}}{2011}]{2011MNRAS.415.2101H}
{Hahn} O.,  {Abel} T.,  2011, \mn@doi [\mnras] {10.1111/j.1365-2966.2011.18820.x}, \href {https://ui.adsabs.harvard.edu/abs/2011MNRAS.415.2101H} {415, 2101}

\bibitem[\protect\citeauthoryear{{Harikane} et~al.,}{{Harikane} et~al.}{2022}]{2022ApJS..259...20H}
{Harikane} Y.,  et~al., 2022, \mn@doi [\apjs] {10.3847/1538-4365/ac3dfc}, \href {https://ui.adsabs.harvard.edu/abs/2022ApJS..259...20H} {259, 20}

\bibitem[\protect\citeauthoryear{{Harikane} et~al.,}{{Harikane} et~al.}{2023}]{2023ApJ...959...39H}
{Harikane} Y.,  et~al., 2023, \mn@doi [\apj] {10.3847/1538-4357/ad029e}, \href {https://ui.adsabs.harvard.edu/abs/2023ApJ...959...39H} {959, 39}

\bibitem[\protect\citeauthoryear{{Holley-Bockelmann}, {G{\"u}ltekin}, {Shoemaker}  \& {Yunes}}{{Holley-Bockelmann} et~al.}{2008}]{2008ApJ...686..829H}
{Holley-Bockelmann} K.,  {G{\"u}ltekin} K.,  {Shoemaker} D.,   {Yunes} N.,  2008, \mn@doi [\apj] {10.1086/591218}, \href {https://ui.adsabs.harvard.edu/abs/2008ApJ...686..829H} {686, 829}

\bibitem[\protect\citeauthoryear{{Hosokawa}, {Omukai}  \& {Yorke}}{{Hosokawa} et~al.}{2012}]{2012ApJ...756...93H}
{Hosokawa} T.,  {Omukai} K.,   {Yorke} H.~W.,  2012, \mn@doi [\apj] {10.1088/0004-637X/756/1/93}, \href {https://ui.adsabs.harvard.edu/abs/2012ApJ...756...93H} {756, 93}

\bibitem[\protect\citeauthoryear{{Hosokawa}, {Yorke}, {Inayoshi}, {Omukai}  \& {Yoshida}}{{Hosokawa} et~al.}{2013}]{2013ApJ...778..178H}
{Hosokawa} T.,  {Yorke} H.~W.,  {Inayoshi} K.,  {Omukai} K.,   {Yoshida} N.,  2013, \mn@doi [\apj] {10.1088/0004-637X/778/2/178}, \href {https://ui.adsabs.harvard.edu/abs/2013ApJ...778..178H} {778, 178}

\bibitem[\protect\citeauthoryear{{Huang}, {Di Matteo}, {Bhowmick}, {Feng}  \& {Ma}}{{Huang} et~al.}{2018}]{2018MNRAS.478.5063H}
{Huang} K.-W.,  {Di Matteo} T.,  {Bhowmick} A.~K.,  {Feng} Y.,   {Ma} C.-P.,  2018, \mn@doi [\mnras] {10.1093/mnras/sty1329}, \href {https://ui.adsabs.harvard.edu/abs/2018MNRAS.478.5063H} {478, 5063}

\bibitem[\protect\citeauthoryear{{Jeon}, {Bromm}, {Liu}  \& {Finkelstein}}{{Jeon} et~al.}{2024}]{2024arXiv240218773J}
{Jeon} J.,  {Bromm} V.,  {Liu} B.,   {Finkelstein} S.~L.,  2024, \mn@doi [arXiv e-prints] {10.48550/arXiv.2402.18773}, \href {https://ui.adsabs.harvard.edu/abs/2024arXiv240218773J} {p. arXiv:2402.18773}

\bibitem[\protect\citeauthoryear{{Jiang} et~al.,}{{Jiang} et~al.}{2016}]{2016ApJ...833..222J}
{Jiang} L.,  et~al., 2016, \mn@doi [\apj] {10.3847/1538-4357/833/2/222}, \href {https://ui.adsabs.harvard.edu/abs/2016ApJ...833..222J} {833, 222}

\bibitem[\protect\citeauthoryear{{Kannan}, {Vogelsberger}, {Pfrommer}, {Weinberger}, {Springel}, {Hernquist}, {Puchwein}  \& {Pakmor}}{{Kannan} et~al.}{2017}]{2017ApJ...837L..18K}
{Kannan} R.,  {Vogelsberger} M.,  {Pfrommer} C.,  {Weinberger} R.,  {Springel} V.,  {Hernquist} L.,  {Puchwein} E.,   {Pakmor} R.,  2017, \mn@doi [\apjl] {10.3847/2041-8213/aa624b}, \href {https://ui.adsabs.harvard.edu/abs/2017ApJ...837L..18K} {837, L18}

\bibitem[\protect\citeauthoryear{{Kannan}, {Garaldi}, {Smith}, {Pakmor}, {Springel}, {Vogelsberger}  \& {Hernquist}}{{Kannan} et~al.}{2022}]{2022MNRAS.511.4005K}
{Kannan} R.,  {Garaldi} E.,  {Smith} A.,  {Pakmor} R.,  {Springel} V.,  {Vogelsberger} M.,   {Hernquist} L.,  2022, \mn@doi [\mnras] {10.1093/mnras/stab3710}, \href {https://ui.adsabs.harvard.edu/abs/2022MNRAS.511.4005K} {511, 4005}

\bibitem[\protect\citeauthoryear{{Kannan} et~al.,}{{Kannan} et~al.}{2023}]{2023MNRAS.524.2594K}
{Kannan} R.,  et~al., 2023, \mn@doi [\mnras] {10.1093/mnras/stac3743}, \href {https://ui.adsabs.harvard.edu/abs/2023MNRAS.524.2594K} {524, 2594}

\bibitem[\protect\citeauthoryear{{Katz}, {Weinberg}  \& {Hernquist}}{{Katz} et~al.}{1996}]{1996ApJS..105...19K}
{Katz} N.,  {Weinberg} D.~H.,   {Hernquist} L.,  1996, \mn@doi [\apjs] {10.1086/192305}, \href {https://ui.adsabs.harvard.edu/abs/1996ApJS..105...19K} {105, 19}

\bibitem[\protect\citeauthoryear{{Kaviraj} et~al.,}{{Kaviraj} et~al.}{2017}]{2017MNRAS.467.4739K}
{Kaviraj} S.,  et~al., 2017, \mn@doi [\mnras] {10.1093/mnras/stx126}, \href {https://ui.adsabs.harvard.edu/abs/2017MNRAS.467.4739K} {467, 4739}

\bibitem[\protect\citeauthoryear{{Kelley}, {Blecha}  \& {Hernquist}}{{Kelley} et~al.}{2017}]{2017MNRAS.464.3131K}
{Kelley} L.~Z.,  {Blecha} L.,   {Hernquist} L.,  2017, \mn@doi [\mnras] {10.1093/mnras/stw2452}, \href {https://ui.adsabs.harvard.edu/abs/2017MNRAS.464.3131K} {464, 3131}

\bibitem[\protect\citeauthoryear{{Kelly}, {Treu}, {Malkan}, {Pancoast}  \& {Woo}}{{Kelly} et~al.}{2013}]{2013ApJ...779..187K}
{Kelly} B.~C.,  {Treu} T.,  {Malkan} M.,  {Pancoast} A.,   {Woo} J.-H.,  2013, \mn@doi [\apj] {10.1088/0004-637X/779/2/187}, \href {https://ui.adsabs.harvard.edu/abs/2013ApJ...779..187K} {779, 187}

\bibitem[\protect\citeauthoryear{{Khandai}, {Di Matteo}, {Croft}, {Wilkins}, {Feng}, {Tucker}, {DeGraf}  \& {Liu}}{{Khandai} et~al.}{2015}]{2015MNRAS.450.1349K}
{Khandai} N.,  {Di Matteo} T.,  {Croft} R.,  {Wilkins} S.,  {Feng} Y.,  {Tucker} E.,  {DeGraf} C.,   {Liu} M.-S.,  2015, \mn@doi [\mnras] {10.1093/mnras/stv627}, \href {https://ui.adsabs.harvard.edu/abs/2015MNRAS.450.1349K} {450, 1349}

\bibitem[\protect\citeauthoryear{{Killi} et~al.,}{{Killi} et~al.}{2023}]{2023arXiv231203065K}
{Killi} M.,  et~al., 2023, \mn@doi [arXiv e-prints] {10.48550/arXiv.2312.03065}, \href {https://ui.adsabs.harvard.edu/abs/2023arXiv231203065K} {p. arXiv:2312.03065}

\bibitem[\protect\citeauthoryear{{Kocevski} et~al.,}{{Kocevski} et~al.}{2023}]{2023ApJ...954L...4K}
{Kocevski} D.~D.,  et~al., 2023, \mn@doi [\apjl] {10.3847/2041-8213/ace5a0}, \href {https://ui.adsabs.harvard.edu/abs/2023ApJ...954L...4K} {954, L4}

\bibitem[\protect\citeauthoryear{{Kocevski} et~al.,}{{Kocevski} et~al.}{2024}]{2024arXiv240403576K}
{Kocevski} D.~D.,  et~al., 2024, \mn@doi [arXiv e-prints] {10.48550/arXiv.2404.03576}, \href {https://ui.adsabs.harvard.edu/abs/2024arXiv240403576K} {p. arXiv:2404.03576}

\bibitem[\protect\citeauthoryear{{Kokorev} et~al.,}{{Kokorev} et~al.}{2023}]{2023ApJ...957L...7K}
{Kokorev} V.,  et~al., 2023, \mn@doi [\apjl] {10.3847/2041-8213/ad037a}, \href {https://ui.adsabs.harvard.edu/abs/2023ApJ...957L...7K} {957, L7}

\bibitem[\protect\citeauthoryear{{Kokorev} et~al.,}{{Kokorev} et~al.}{2024}]{2024ApJ...968...38K}
{Kokorev} V.,  et~al., 2024, \mn@doi [\apj] {10.3847/1538-4357/ad4265}, \href {https://ui.adsabs.harvard.edu/abs/2024ApJ...968...38K} {968, 38}

\bibitem[\protect\citeauthoryear{{Kormendy} \& {Ho}}{{Kormendy} \& {Ho}}{2013}]{2013ARA&A..51..511K}
{Kormendy} J.,  {Ho} L.~C.,  2013, \mn@doi [\araa] {10.1146/annurev-astro-082708-101811}, \href {https://ui.adsabs.harvard.edu/abs/2013ARA&A..51..511K} {51, 511}

\bibitem[\protect\citeauthoryear{{Kroupa}, {Subr}, {Jerabkova}  \& {Wang}}{{Kroupa} et~al.}{2020}]{2020MNRAS.498.5652K}
{Kroupa} P.,  {Subr} L.,  {Jerabkova} T.,   {Wang} L.,  2020, \mn@doi [\mnras] {10.1093/mnras/staa2276}, \href {https://ui.adsabs.harvard.edu/abs/2020MNRAS.498.5652K} {498, 5652}

\bibitem[\protect\citeauthoryear{{Langeroodi} \& {Hjorth}}{{Langeroodi} \& {Hjorth}}{2023}]{2023ApJ...957L..27L}
{Langeroodi} D.,  {Hjorth} J.,  2023, \mn@doi [\apjl] {10.3847/2041-8213/acfeec}, \href {https://ui.adsabs.harvard.edu/abs/2023ApJ...957L..27L} {957, L27}

\bibitem[\protect\citeauthoryear{{Larson} et~al.,}{{Larson} et~al.}{2023}]{2023ApJ...953L..29L}
{Larson} R.~L.,  et~al., 2023, \mn@doi [\apjl] {10.3847/2041-8213/ace619}, \href {https://ui.adsabs.harvard.edu/abs/2023ApJ...953L..29L} {953, L29}

\bibitem[\protect\citeauthoryear{{Latif}, {Schleicher}  \& {Hartwig}}{{Latif} et~al.}{2016}]{2016MNRAS.458..233L}
{Latif} M.~A.,  {Schleicher} D.~R.~G.,   {Hartwig} T.,  2016, \mn@doi [\mnras] {10.1093/mnras/stw297}, \href {https://ui.adsabs.harvard.edu/abs/2016MNRAS.458..233L} {458, 233}

\bibitem[\protect\citeauthoryear{{Lauer}, {Tremaine}, {Richstone}  \& {Faber}}{{Lauer} et~al.}{2007}]{2007ApJ...670..249L}
{Lauer} T.~R.,  {Tremaine} S.,  {Richstone} D.,   {Faber} S.~M.,  2007, \mn@doi [\apj] {10.1086/522083}, \href {https://ui.adsabs.harvard.edu/abs/2007ApJ...670..249L} {670, 249}

\bibitem[\protect\citeauthoryear{{Li} et~al.,}{{Li} et~al.}{2020}]{2020ApJ...895..102L}
{Li} Y.,  et~al., 2020, \mn@doi [\apj] {10.3847/1538-4357/ab8f8d}, \href {https://ui.adsabs.harvard.edu/abs/2020ApJ...895..102L} {895, 102}

\bibitem[\protect\citeauthoryear{{Li} et~al.,}{{Li} et~al.}{2024}]{2024arXiv240300074L}
{Li} J.,  et~al., 2024, \mn@doi [arXiv e-prints] {10.48550/arXiv.2403.00074}, \href {https://ui.adsabs.harvard.edu/abs/2024arXiv240300074L} {p. arXiv:2403.00074}

\bibitem[\protect\citeauthoryear{{Lodato} \& {Natarajan}}{{Lodato} \& {Natarajan}}{2006}]{2006MNRAS.371.1813L}
{Lodato} G.,  {Natarajan} P.,  2006, \mn@doi [\mnras] {10.1111/j.1365-2966.2006.10801.x}, \href {https://ui.adsabs.harvard.edu/abs/2006MNRAS.371.1813L} {371, 1813}

\bibitem[\protect\citeauthoryear{{Luo}, {Ardaneh}, {Shlosman}, {Nagamine}, {Wise}  \& {Begelman}}{{Luo} et~al.}{2018}]{2018MNRAS.476.3523L}
{Luo} Y.,  {Ardaneh} K.,  {Shlosman} I.,  {Nagamine} K.,  {Wise} J.~H.,   {Begelman} M.~C.,  2018, \mn@doi [\mnras] {10.1093/mnras/sty362}, \href {https://ui.adsabs.harvard.edu/abs/2018MNRAS.476.3523L} {476, 3523}

\bibitem[\protect\citeauthoryear{{Luo}, {Shlosman}, {Nagamine}  \& {Fang}}{{Luo} et~al.}{2020}]{2020MNRAS.492.4917L}
{Luo} Y.,  {Shlosman} I.,  {Nagamine} K.,   {Fang} T.,  2020, \mn@doi [\mnras] {10.1093/mnras/staa153}, \href {https://ui.adsabs.harvard.edu/abs/2020MNRAS.492.4917L} {492, 4917}

\bibitem[\protect\citeauthoryear{{Lupi}, {Colpi}, {Devecchi}, {Galanti}  \& {Volonteri}}{{Lupi} et~al.}{2014}]{2014MNRAS.442.3616L}
{Lupi} A.,  {Colpi} M.,  {Devecchi} B.,  {Galanti} G.,   {Volonteri} M.,  2014, \mn@doi [\mnras] {10.1093/mnras/stu1120}, \href {https://ui.adsabs.harvard.edu/abs/2014MNRAS.442.3616L} {442, 3616}

\bibitem[\protect\citeauthoryear{{Ma}, {Hopkins}, {Ma}, {Angl{\'e}s-Alc{\'a}zar}, {Faucher-Gigu{\`e}re}  \& {Kelley}}{{Ma} et~al.}{2021}]{2021MNRAS.508.1973M}
{Ma} L.,  {Hopkins} P.~F.,  {Ma} X.,  {Angl{\'e}s-Alc{\'a}zar} D.,  {Faucher-Gigu{\`e}re} C.-A.,   {Kelley} L.~Z.,  2021, \mn@doi [\mnras] {10.1093/mnras/stab2713}, \href {https://ui.adsabs.harvard.edu/abs/2021MNRAS.508.1973M} {508, 1973}

\bibitem[\protect\citeauthoryear{{Madau} \& {Dickinson}}{{Madau} \& {Dickinson}}{2014}]{MD:2014}
{Madau} P.,  {Dickinson} M.,  2014, \mn@doi [\araa] {10.1146/annurev-astro-081811-125615}, \href {https://ui.adsabs.harvard.edu/abs/2014ARA&A..52..415M} {52, 415}

\bibitem[\protect\citeauthoryear{{Madau} \& {Rees}}{{Madau} \& {Rees}}{2001}]{2001ApJ...551L..27M}
{Madau} P.,  {Rees} M.~J.,  2001, \mn@doi [\apjl] {10.1086/319848}, \href {https://ui.adsabs.harvard.edu/abs/2001ApJ...551L..27M} {551, L27}

\bibitem[\protect\citeauthoryear{{Maiolino} et~al.,}{{Maiolino} et~al.}{2023}]{2023arXiv230801230M}
{Maiolino} R.,  et~al., 2023, \mn@doi [arXiv e-prints] {10.48550/arXiv.2308.01230}, \href {https://ui.adsabs.harvard.edu/abs/2023arXiv230801230M} {p. arXiv:2308.01230}

\bibitem[\protect\citeauthoryear{{Marinacci} et~al.,}{{Marinacci} et~al.}{2018}]{2018MNRAS.480.5113M}
{Marinacci} F.,  et~al., 2018, \mn@doi [\mnras] {10.1093/mnras/sty2206}, \href {https://ui.adsabs.harvard.edu/abs/2018MNRAS.480.5113M} {480, 5113}

\bibitem[\protect\citeauthoryear{{Matsuoka} et~al.,}{{Matsuoka} et~al.}{2018}]{2018ApJS..237....5M}
{Matsuoka} Y.,  et~al., 2018, \mn@doi [\apjs] {10.3847/1538-4365/aac724}, \href {https://ui.adsabs.harvard.edu/abs/2018ApJS..237....5M} {237, 5}

\bibitem[\protect\citeauthoryear{{Matsuoka} et~al.,}{{Matsuoka} et~al.}{2019}]{2019ApJ...872L...2M}
{Matsuoka} Y.,  et~al., 2019, \mn@doi [\apjl] {10.3847/2041-8213/ab0216}, \href {https://ui.adsabs.harvard.edu/abs/2019ApJ...872L...2M} {872, L2}

\bibitem[\protect\citeauthoryear{{Matthee} et~al.,}{{Matthee} et~al.}{2024}]{2024ApJ...963..129M}
{Matthee} J.,  et~al., 2024, \mn@doi [\apj] {10.3847/1538-4357/ad2345}, \href {https://ui.adsabs.harvard.edu/abs/2024ApJ...963..129M} {963, 129}

\bibitem[\protect\citeauthoryear{{Mayer}, {Capelo}, {Zwick}  \& {Di Matteo}}{{Mayer} et~al.}{2023}]{2023arXiv230402066M}
{Mayer} L.,  {Capelo} P.~R.,  {Zwick} L.,   {Di Matteo} T.,  2023, \mn@doi [arXiv e-prints] {10.48550/arXiv.2304.02066}, \href {https://ui.adsabs.harvard.edu/abs/2023arXiv230402066M} {p. arXiv:2304.02066}

\bibitem[\protect\citeauthoryear{{Mayer}, {Capelo}, {Zwick}  \& {Di Matteo}}{{Mayer} et~al.}{2024}]{2024ApJ...961...76M}
{Mayer} L.,  {Capelo} P.~R.,  {Zwick} L.,   {Di Matteo} T.,  2024, \mn@doi [\apj] {10.3847/1538-4357/ad11cf}, \href {https://ui.adsabs.harvard.edu/abs/2024ApJ...961...76M} {961, 76}

\bibitem[\protect\citeauthoryear{{Mezcua}, {Pacucci}, {Suh}, {Siudek}  \& {Natarajan}}{{Mezcua} et~al.}{2024}]{2024arXiv240405793M}
{Mezcua} M.,  {Pacucci} F.,  {Suh} H.,  {Siudek} M.,   {Natarajan} P.,  2024, \mn@doi [arXiv e-prints] {10.48550/arXiv.2404.05793}, \href {https://ui.adsabs.harvard.edu/abs/2024arXiv240405793M} {p. arXiv:2404.05793}

\bibitem[\protect\citeauthoryear{{Milosavljevi{\'c}} \& {Merritt}}{{Milosavljevi{\'c}} \& {Merritt}}{2003}]{2003ApJ...596..860M}
{Milosavljevi{\'c}} M.,  {Merritt} D.,  2003, \mn@doi [\apj] {10.1086/378086}, \href {https://ui.adsabs.harvard.edu/abs/2003ApJ...596..860M} {596, 860}

\bibitem[\protect\citeauthoryear{{Mortlock} et~al.,}{{Mortlock} et~al.}{2011}]{2011Natur.474..616M}
{Mortlock} D.~J.,  et~al., 2011, \mn@doi [\nat] {10.1038/nature10159}, \href {https://ui.adsabs.harvard.edu/abs/2011Natur.474..616M} {474, 616}

\bibitem[\protect\citeauthoryear{{Naiman} et~al.,}{{Naiman} et~al.}{2018}]{2018MNRAS.477.1206N}
{Naiman} J.~P.,  et~al., 2018, \mn@doi [\mnras] {10.1093/mnras/sty618}, \href {https://ui.adsabs.harvard.edu/abs/2018MNRAS.477.1206N} {477, 1206}

\bibitem[\protect\citeauthoryear{{Natarajan} \& {Volonteri}}{{Natarajan} \& {Volonteri}}{2012}]{2012MNRAS.422.2051N}
{Natarajan} P.,  {Volonteri} M.,  2012, \mn@doi [\mnras] {10.1111/j.1365-2966.2012.20708.x}, \href {https://ui.adsabs.harvard.edu/abs/2012MNRAS.422.2051N} {422, 2051}

\bibitem[\protect\citeauthoryear{{Natarajan}, {Pacucci}, {Ferrara}, {Agarwal}, {Ricarte}, {Zackrisson}  \& {Cappelluti}}{{Natarajan} et~al.}{2017}]{2017ApJ...838..117N}
{Natarajan} P.,  {Pacucci} F.,  {Ferrara} A.,  {Agarwal} B.,  {Ricarte} A.,  {Zackrisson} E.,   {Cappelluti} N.,  2017, \mn@doi [\apj] {10.3847/1538-4357/aa6330}, \href {https://ui.adsabs.harvard.edu/abs/2017ApJ...838..117N} {838, 117}

\bibitem[\protect\citeauthoryear{{Natarajan}, {Pacucci}, {Ricarte}, {Bogd{\'a}n}, {Goulding}  \& {Cappelluti}}{{Natarajan} et~al.}{2024}]{2024ApJ...960L...1N}
{Natarajan} P.,  {Pacucci} F.,  {Ricarte} A.,  {Bogd{\'a}n} {\'A}.,  {Goulding} A.~D.,   {Cappelluti} N.,  2024, \mn@doi [\apjl] {10.3847/2041-8213/ad0e76}, \href {https://ui.adsabs.harvard.edu/abs/2024ApJ...960L...1N} {960, L1}

\bibitem[\protect\citeauthoryear{{Navarro-Carrera}, {Rinaldi}, {Caputi}, {Iani}, {Kokorev}  \& {van Mierlo}}{{Navarro-Carrera} et~al.}{2024}]{2024ApJ...961..207N}
{Navarro-Carrera} R.,  {Rinaldi} P.,  {Caputi} K.~I.,  {Iani} E.,  {Kokorev} V.,   {van Mierlo} S.~E.,  2024, \mn@doi [\apj] {10.3847/1538-4357/ad0df6}, \href {https://ui.adsabs.harvard.edu/abs/2024ApJ...961..207N} {961, 207}

\bibitem[\protect\citeauthoryear{{Nelson} et~al.,}{{Nelson} et~al.}{2015}]{2015A&C....13...12N}
{Nelson} D.,  et~al., 2015, \mn@doi [Astronomy and Computing] {10.1016/j.ascom.2015.09.003}, \href {https://ui.adsabs.harvard.edu/abs/2015A&C....13...12N} {13, 12}

\bibitem[\protect\citeauthoryear{{Nelson} et~al.,}{{Nelson} et~al.}{2018}]{2018MNRAS.475..624N}
{Nelson} D.,  et~al., 2018, \mn@doi [\mnras] {10.1093/mnras/stx3040}, \href {https://ui.adsabs.harvard.edu/abs/2018MNRAS.475..624N} {475, 624}

\bibitem[\protect\citeauthoryear{{Nelson} et~al.,}{{Nelson} et~al.}{2019a}]{2019ComAC...6....2N}
{Nelson} D.,  et~al., 2019a, \mn@doi [Computational Astrophysics and Cosmology] {10.1186/s40668-019-0028-x}, \href {https://ui.adsabs.harvard.edu/abs/2019ComAC...6....2N} {6, 2}

\bibitem[\protect\citeauthoryear{{Nelson} et~al.,}{{Nelson} et~al.}{2019b}]{2019MNRAS.490.3234N}
{Nelson} D.,  et~al., 2019b, \mn@doi [\mnras] {10.1093/mnras/stz2306}, \href {https://ui.adsabs.harvard.edu/abs/2019MNRAS.490.3234N} {490, 3234}

\bibitem[\protect\citeauthoryear{{Ni}, {Di Matteo}  \& {Feng}}{{Ni} et~al.}{2022}]{2022MNRAS.509.3043N}
{Ni} Y.,  {Di Matteo} T.,   {Feng} Y.,  2022, \mn@doi [\mnras] {10.1093/mnras/stab3162}, \href {https://ui.adsabs.harvard.edu/abs/2022MNRAS.509.3043N} {509, 3043}

\bibitem[\protect\citeauthoryear{{Onoue} et~al.,}{{Onoue} et~al.}{2023}]{2023ApJ...942L..17O}
{Onoue} M.,  et~al., 2023, \mn@doi [\apjl] {10.3847/2041-8213/aca9d3}, \href {https://ui.adsabs.harvard.edu/abs/2023ApJ...942L..17O} {942, L17}

\bibitem[\protect\citeauthoryear{{Pacucci}, {Nguyen}, {Carniani}, {Maiolino}  \& {Fan}}{{Pacucci} et~al.}{2023}]{2023ApJ...957L...3P}
{Pacucci} F.,  {Nguyen} B.,  {Carniani} S.,  {Maiolino} R.,   {Fan} X.,  2023, \mn@doi [\apjl] {10.3847/2041-8213/ad0158}, \href {https://ui.adsabs.harvard.edu/abs/2023ApJ...957L...3P} {957, L3}

\bibitem[\protect\citeauthoryear{{Pakmor}, {Bauer}  \& {Springel}}{{Pakmor} et~al.}{2011}]{2011MNRAS.418.1392P}
{Pakmor} R.,  {Bauer} A.,   {Springel} V.,  2011, \mn@doi [\mnras] {10.1111/j.1365-2966.2011.19591.x}, \href {https://ui.adsabs.harvard.edu/abs/2011MNRAS.418.1392P} {418, 1392}

\bibitem[\protect\citeauthoryear{{Pakmor}, {Pfrommer}, {Simpson}, {Kannan}  \& {Springel}}{{Pakmor} et~al.}{2016}]{2016MNRAS.462.2603P}
{Pakmor} R.,  {Pfrommer} C.,  {Simpson} C.~M.,  {Kannan} R.,   {Springel} V.,  2016, \mn@doi [\mnras] {10.1093/mnras/stw1761}, \href {https://ui.adsabs.harvard.edu/abs/2016MNRAS.462.2603P} {462, 2603}

\bibitem[\protect\citeauthoryear{{Pakmor} et~al.,}{{Pakmor} et~al.}{2023}]{2023MNRAS.524.2539P}
{Pakmor} R.,  et~al., 2023, \mn@doi [\mnras] {10.1093/mnras/stac3620}, \href {https://ui.adsabs.harvard.edu/abs/2023MNRAS.524.2539P} {524, 2539}

\bibitem[\protect\citeauthoryear{{Partmann}, {Naab}, {Rantala}, {Genina}, {Mannerkoski}  \& {Johansson}}{{Partmann} et~al.}{2023}]{2023arXiv231008079P}
{Partmann} C.,  {Naab} T.,  {Rantala} A.,  {Genina} A.,  {Mannerkoski} M.,   {Johansson} P.~H.,  2023, \mn@doi [arXiv e-prints] {10.48550/arXiv.2310.08079}, \href {https://ui.adsabs.harvard.edu/abs/2023arXiv231008079P} {p. arXiv:2310.08079}

\bibitem[\protect\citeauthoryear{{P{\'e}rez-Gonz{\'a}lez} et~al.,}{{P{\'e}rez-Gonz{\'a}lez} et~al.}{2024}]{2024arXiv240108782P}
{P{\'e}rez-Gonz{\'a}lez} P.~G.,  et~al., 2024, \mn@doi [arXiv e-prints] {10.48550/arXiv.2401.08782}, \href {https://ui.adsabs.harvard.edu/abs/2024arXiv240108782P} {p. arXiv:2401.08782}

\bibitem[\protect\citeauthoryear{{Pillepich} et~al.,}{{Pillepich} et~al.}{2018a}]{2018MNRAS.473.4077P}
{Pillepich} A.,  et~al., 2018a, \mn@doi [\mnras] {10.1093/mnras/stx2656}, \href {https://ui.adsabs.harvard.edu/abs/2018MNRAS.473.4077P} {473, 4077}

\bibitem[\protect\citeauthoryear{{Pillepich} et~al.,}{{Pillepich} et~al.}{2018b}]{2018MNRAS.475..648P}
{Pillepich} A.,  et~al., 2018b, \mn@doi [\mnras] {10.1093/mnras/stx3112}, \href {https://ui.adsabs.harvard.edu/abs/2018MNRAS.475..648P} {475, 648}

\bibitem[\protect\citeauthoryear{{Pillepich} et~al.,}{{Pillepich} et~al.}{2019}]{2019MNRAS.490.3196P}
{Pillepich} A.,  et~al., 2019, \mn@doi [\mnras] {10.1093/mnras/stz2338}, \href {https://ui.adsabs.harvard.edu/abs/2019MNRAS.490.3196P} {490, 3196}

\bibitem[\protect\citeauthoryear{{Planck Collaboration} et~al.,}{{Planck Collaboration} et~al.}{2016}]{2016A&A...594A..13P}
{Planck Collaboration} et~al., 2016, \mn@doi [\aap] {10.1051/0004-6361/201525830}, \href {https://ui.adsabs.harvard.edu/abs/2016A&A...594A..13P} {594, A13}

\bibitem[\protect\citeauthoryear{{Ponti}, {Papadakis}, {Bianchi}, {Guainazzi}, {Matt}, {Uttley}  \& {Bonilla}}{{Ponti} et~al.}{2012}]{2012A&A...542A..83P}
{Ponti} G.,  {Papadakis} I.,  {Bianchi} S.,  {Guainazzi} M.,  {Matt} G.,  {Uttley} P.,   {Bonilla} N.~F.,  2012, \mn@doi [\aap] {10.1051/0004-6361/201118326}, \href {https://ui.adsabs.harvard.edu/abs/2012A&A...542A..83P} {542, A83}

\bibitem[\protect\citeauthoryear{{Ramos Padilla}, {Ashby}, {Smith}, {Mart{\'\i}nez-Galarza}, {Beverage}, {Dietrich}, {Higuera-G.}  \& {Weiner}}{{Ramos Padilla} et~al.}{2020}]{2020MNRAS.499.4325R}
{Ramos Padilla} A.~F.,  {Ashby} M.~L.~N.,  {Smith} H.~A.,  {Mart{\'\i}nez-Galarza} J.~R.,  {Beverage} A.~G.,  {Dietrich} J.,  {Higuera-G.} M.-A.,   {Weiner} A.~S.,  2020, \mn@doi [\mnras] {10.1093/mnras/staa2813}, \href {https://ui.adsabs.harvard.edu/abs/2020MNRAS.499.4325R} {499, 4325}

\bibitem[\protect\citeauthoryear{{Reardon} et~al.,}{{Reardon} et~al.}{2023}]{2023ApJ...951L...6R}
{Reardon} D.~J.,  et~al., 2023, \mn@doi [\apjl] {10.3847/2041-8213/acdd02}, \href {https://ui.adsabs.harvard.edu/abs/2023ApJ...951L...6R} {951, L6}

\bibitem[\protect\citeauthoryear{{Reed} et~al.,}{{Reed} et~al.}{2017}]{2017MNRAS.468.4702R}
{Reed} S.~L.,  et~al., 2017, \mn@doi [\mnras] {10.1093/mnras/stx728}, \href {https://ui.adsabs.harvard.edu/abs/2017MNRAS.468.4702R} {468, 4702}

\bibitem[\protect\citeauthoryear{{Regan}, {Johansson}  \& {Wise}}{{Regan} et~al.}{2014}]{2014ApJ...795..137R}
{Regan} J.~A.,  {Johansson} P.~H.,   {Wise} J.~H.,  2014, \mn@doi [\apj] {10.1088/0004-637X/795/2/137}, \href {https://ui.adsabs.harvard.edu/abs/2014ApJ...795..137R} {795, 137}

\bibitem[\protect\citeauthoryear{{Regan}, {Haiman}, {Wise}, {O'Shea}  \& {Norman}}{{Regan} et~al.}{2020a}]{2020OJAp....3E...9R}
{Regan} J.~A.,  {Haiman} Z.,  {Wise} J.~H.,  {O'Shea} B.~W.,   {Norman} M.~L.,  2020a, \mn@doi [The Open Journal of Astrophysics] {10.21105/astro.2006.14625}, \href {https://ui.adsabs.harvard.edu/abs/2020OJAp....3E...9R} {3, E9}

\bibitem[\protect\citeauthoryear{{Regan}, {Wise}, {Woods}, {Downes}, {O'Shea}  \& {Norman}}{{Regan} et~al.}{2020b}]{2020OJAp....3E..15R}
{Regan} J.~A.,  {Wise} J.~H.,  {Woods} T.~E.,  {Downes} T.~P.,  {O'Shea} B.~W.,   {Norman} M.~L.,  2020b, \mn@doi [The Open Journal of Astrophysics] {10.21105/astro.2008.08090}, \href {https://ui.adsabs.harvard.edu/abs/2020OJAp....3E..15R} {3, 15}

\bibitem[\protect\citeauthoryear{{Regan}, {Wise}, {O'Shea}  \& {Norman}}{{Regan} et~al.}{2020c}]{2020MNRAS.492.3021R}
{Regan} J.~A.,  {Wise} J.~H.,  {O'Shea} B.~W.,   {Norman} M.~L.,  2020c, \mn@doi [\mnras] {10.1093/mnras/staa035}, \href {https://ui.adsabs.harvard.edu/abs/2020MNRAS.492.3021R} {492, 3021}

\bibitem[\protect\citeauthoryear{{Reines} \& {Volonteri}}{{Reines} \& {Volonteri}}{2015}]{2015ApJ...813...82R}
{Reines} A.~E.,  {Volonteri} M.,  2015, \mn@doi [\apj] {10.1088/0004-637X/813/2/82}, \href {https://ui.adsabs.harvard.edu/abs/2015ApJ...813...82R} {813, 82}

\bibitem[\protect\citeauthoryear{{Ricarte}, {Tremmel}, {Natarajan}, {Zimmer}  \& {Quinn}}{{Ricarte} et~al.}{2021}]{2021MNRAS.503.6098R}
{Ricarte} A.,  {Tremmel} M.,  {Natarajan} P.,  {Zimmer} C.,   {Quinn} T.,  2021, \mn@doi [\mnras] {10.1093/mnras/stab866}, \href {https://ui.adsabs.harvard.edu/abs/2021MNRAS.503.6098R} {503, 6098}

\bibitem[\protect\citeauthoryear{{Rodriguez-Gomez} et~al.,}{{Rodriguez-Gomez} et~al.}{2019}]{2019MNRAS.483.4140R}
{Rodriguez-Gomez} V.,  et~al., 2019, \mn@doi [\mnras] {10.1093/mnras/sty3345}, \href {https://ui.adsabs.harvard.edu/abs/2019MNRAS.483.4140R} {483, 4140}

\bibitem[\protect\citeauthoryear{{Sayeb}, {Blecha}, {Kelley}, {Gerosa}, {Kesden}  \& {Thomas}}{{Sayeb} et~al.}{2021}]{2021MNRAS.501.2531S}
{Sayeb} M.,  {Blecha} L.,  {Kelley} L.~Z.,  {Gerosa} D.,  {Kesden} M.,   {Thomas} J.,  2021, \mn@doi [\mnras] {10.1093/mnras/staa3826}, \href {https://ui.adsabs.harvard.edu/abs/2021MNRAS.501.2531S} {501, 2531}

\bibitem[\protect\citeauthoryear{{Schleicher}, {Palla}, {Ferrara}, {Galli}  \& {Latif}}{{Schleicher} et~al.}{2013}]{2013A&A...558A..59S}
{Schleicher} D. R.~G.,  {Palla} F.,  {Ferrara} A.,  {Galli} D.,   {Latif} M.,  2013, \mn@doi [\aap] {10.1051/0004-6361/201321949}, \href {https://ui.adsabs.harvard.edu/abs/2013A&A...558A..59S} {558, A59}

\bibitem[\protect\citeauthoryear{{Shang}, {Bryan}  \& {Haiman}}{{Shang} et~al.}{2010}]{2010MNRAS.402.1249S}
{Shang} C.,  {Bryan} G.~L.,   {Haiman} Z.,  2010, \mn@doi [\mnras] {10.1111/j.1365-2966.2009.15960.x}, \href {https://ui.adsabs.harvard.edu/abs/2010MNRAS.402.1249S} {402, 1249}

\bibitem[\protect\citeauthoryear{{Shen}, {Hopkins}, {Faucher-Gigu{\`e}re}, {Alexander}, {Richards}, {Ross}  \& {Hickox}}{{Shen} et~al.}{2020}]{2020MNRAS.495.3252S}
{Shen} X.,  {Hopkins} P.~F.,  {Faucher-Gigu{\`e}re} C.-A.,  {Alexander} D.~M.,  {Richards} G.~T.,  {Ross} N.~P.,   {Hickox} R.~C.,  2020, \mn@doi [\mnras] {10.1093/mnras/staa1381}, \href {https://ui.adsabs.harvard.edu/abs/2020MNRAS.495.3252S} {495, 3252}

\bibitem[\protect\citeauthoryear{{Shen}, {Vogelsberger}, {Boylan-Kolchin}, {Tacchella}  \& {Kannan}}{{Shen} et~al.}{2023}]{2023MNRAS.525.3254S}
{Shen} X.,  {Vogelsberger} M.,  {Boylan-Kolchin} M.,  {Tacchella} S.,   {Kannan} R.,  2023, \mn@doi [\mnras] {10.1093/mnras/stad2508}, \href {https://ui.adsabs.harvard.edu/abs/2023MNRAS.525.3254S} {525, 3254}

\bibitem[\protect\citeauthoryear{{Sijacki}, {Vogelsberger}, {Genel}, {Springel}, {Torrey}, {Snyder}, {Nelson}  \& {Hernquist}}{{Sijacki} et~al.}{2015}]{2015MNRAS.452..575S}
{Sijacki} D.,  {Vogelsberger} M.,  {Genel} S.,  {Springel} V.,  {Torrey} P.,  {Snyder} G.~F.,  {Nelson} D.,   {Hernquist} L.,  2015, \mn@doi [\mnras] {10.1093/mnras/stv1340}, \href {https://ui.adsabs.harvard.edu/abs/2015MNRAS.452..575S} {452, 575}

\bibitem[\protect\citeauthoryear{{Siwek}, {Weinberger}  \& {Hernquist}}{{Siwek} et~al.}{2023}]{2023MNRAS.522.2707S}
{Siwek} M.,  {Weinberger} R.,   {Hernquist} L.,  2023, \mn@doi [\mnras] {10.1093/mnras/stad1131}, \href {https://ui.adsabs.harvard.edu/abs/2023MNRAS.522.2707S} {522, 2707}

\bibitem[\protect\citeauthoryear{{Siwek}, {Kelley}  \& {Hernquist}}{{Siwek} et~al.}{2024}]{2024arXiv240308871S}
{Siwek} M.,  {Kelley} L.~Z.,   {Hernquist} L.,  2024, \mn@doi [arXiv e-prints] {10.48550/arXiv.2403.08871}, \href {https://ui.adsabs.harvard.edu/abs/2024arXiv240308871S} {p. arXiv:2403.08871}

\bibitem[\protect\citeauthoryear{{Smith}, {Sigurdsson}  \& {Abel}}{{Smith} et~al.}{2008}]{2008MNRAS.385.1443S}
{Smith} B.,  {Sigurdsson} S.,   {Abel} T.,  2008, \mn@doi [\mnras] {10.1111/j.1365-2966.2008.12922.x}, \href {https://ui.adsabs.harvard.edu/abs/2008MNRAS.385.1443S} {385, 1443}

\bibitem[\protect\citeauthoryear{{Smith}, {Regan}, {Downes}, {Norman}, {O'Shea}  \& {Wise}}{{Smith} et~al.}{2018}]{2018MNRAS.480.3762S}
{Smith} B.~D.,  {Regan} J.~A.,  {Downes} T.~P.,  {Norman} M.~L.,  {O'Shea} B.~W.,   {Wise} J.~H.,  2018, \mn@doi [\mnras] {10.1093/mnras/sty2103}, \href {https://ui.adsabs.harvard.edu/abs/2018MNRAS.480.3762S} {480, 3762}

\bibitem[\protect\citeauthoryear{{Springel}}{{Springel}}{2010}]{2010MNRAS.401..791S}
{Springel} V.,  2010, \mn@doi [\mnras] {10.1111/j.1365-2966.2009.15715.x}, \href {https://ui.adsabs.harvard.edu/abs/2010MNRAS.401..791S} {401, 791}

\bibitem[\protect\citeauthoryear{{Springel} \& {Hernquist}}{{Springel} \& {Hernquist}}{2003}]{2003MNRAS.339..289S}
{Springel} V.,  {Hernquist} L.,  2003, \mn@doi [\mnras] {10.1046/j.1365-8711.2003.06206.x}, \href {https://ui.adsabs.harvard.edu/abs/2003MNRAS.339..289S} {339, 289}

\bibitem[\protect\citeauthoryear{{Springel}, {White}, {Tormen}  \& {Kauffmann}}{{Springel} et~al.}{2001}]{2001MNRAS.328..726S}
{Springel} V.,  {White} S. D.~M.,  {Tormen} G.,   {Kauffmann} G.,  2001, \mn@doi [\mnras] {10.1046/j.1365-8711.2001.04912.x}, \href {https://ui.adsabs.harvard.edu/abs/2001MNRAS.328..726S} {328, 726}

\bibitem[\protect\citeauthoryear{{Springel} et~al.,}{{Springel} et~al.}{2018}]{2018MNRAS.475..676S}
{Springel} V.,  et~al., 2018, \mn@doi [\mnras] {10.1093/mnras/stx3304}, \href {https://ui.adsabs.harvard.edu/abs/2018MNRAS.475..676S} {475, 676}

\bibitem[\protect\citeauthoryear{{Stone}, {Lyu}, {Rieke}, {Alberts}  \& {Hainline}}{{Stone} et~al.}{2023}]{2023arXiv231018395S}
{Stone} M.~A.,  {Lyu} J.,  {Rieke} G.~H.,  {Alberts} S.,   {Hainline} K.~N.,  2023, \mn@doi [arXiv e-prints] {10.48550/arXiv.2310.18395}, \href {https://ui.adsabs.harvard.edu/abs/2023arXiv231018395S} {p. arXiv:2310.18395}

\bibitem[\protect\citeauthoryear{{Sugimura}, {Omukai}  \& {Inoue}}{{Sugimura} et~al.}{2014}]{2014MNRAS.445..544S}
{Sugimura} K.,  {Omukai} K.,   {Inoue} A.~K.,  2014, \mn@doi [\mnras] {10.1093/mnras/stu1778}, \href {https://ui.adsabs.harvard.edu/abs/2014MNRAS.445..544S} {445, 544}

\bibitem[\protect\citeauthoryear{{Taylor} \& {Kobayashi}}{{Taylor} \& {Kobayashi}}{2015}]{2015MNRAS.448.1835T}
{Taylor} P.,  {Kobayashi} C.,  2015, \mn@doi [\mnras] {10.1093/mnras/stv139}, \href {https://ui.adsabs.harvard.edu/abs/2015MNRAS.448.1835T} {448, 1835}

\bibitem[\protect\citeauthoryear{{Terrazas}, {Bell}, {Henriques}, {White}, {Cattaneo}  \& {Woo}}{{Terrazas} et~al.}{2016}]{2016ApJ...830L..12T}
{Terrazas} B.~A.,  {Bell} E.~F.,  {Henriques} B. M.~B.,  {White} S. D.~M.,  {Cattaneo} A.,   {Woo} J.,  2016, \mn@doi [\apjl] {10.3847/2041-8205/830/1/L12}, \href {https://ui.adsabs.harvard.edu/abs/2016ApJ...830L..12T} {830, L12}

\bibitem[\protect\citeauthoryear{{Torrey} et~al.,}{{Torrey} et~al.}{2019}]{2019MNRAS.484.5587T}
{Torrey} P.,  et~al., 2019, \mn@doi [\mnras] {10.1093/mnras/stz243}, \href {https://ui.adsabs.harvard.edu/abs/2019MNRAS.484.5587T} {484, 5587}

\bibitem[\protect\citeauthoryear{{Tremmel}, {Karcher}, {Governato}, {Volonteri}, {Quinn}, {Pontzen}, {Anderson}  \& {Bellovary}}{{Tremmel} et~al.}{2017}]{2017MNRAS.470.1121T}
{Tremmel} M.,  {Karcher} M.,  {Governato} F.,  {Volonteri} M.,  {Quinn} T.~R.,  {Pontzen} A.,  {Anderson} L.,   {Bellovary} J.,  2017, \mn@doi [\mnras] {10.1093/mnras/stx1160}, \href {https://ui.adsabs.harvard.edu/abs/2017MNRAS.470.1121T} {470, 1121}

\bibitem[\protect\citeauthoryear{{Tremmel}, {Governato}, {Volonteri}, {Pontzen}  \& {Quinn}}{{Tremmel} et~al.}{2018}]{2018ApJ...857L..22T}
{Tremmel} M.,  {Governato} F.,  {Volonteri} M.,  {Pontzen} A.,   {Quinn} T.~R.,  2018, \mn@doi [\apjl] {10.3847/2041-8213/aabc0a}, \href {https://ui.adsabs.harvard.edu/abs/2018ApJ...857L..22T} {857, L22}

\bibitem[\protect\citeauthoryear{{{\"U}bler} et~al.,}{{{\"U}bler} et~al.}{2021}]{2021MNRAS.500.4597U}
{{\"U}bler} H.,  et~al., 2021, \mn@doi [\mnras] {10.1093/mnras/staa3464}, \href {https://ui.adsabs.harvard.edu/abs/2021MNRAS.500.4597U} {500, 4597}

\bibitem[\protect\citeauthoryear{{{\"U}bler} et~al.,}{{{\"U}bler} et~al.}{2023}]{2023A&A...677A.145U}
{{\"U}bler} H.,  et~al., 2023, \mn@doi [\aap] {10.1051/0004-6361/202346137}, \href {https://ui.adsabs.harvard.edu/abs/2023A&A...677A.145U} {677, A145}

\bibitem[\protect\citeauthoryear{{Venemans} et~al.,}{{Venemans} et~al.}{2015}]{2015MNRAS.453.2259V}
{Venemans} B.~P.,  et~al., 2015, \mn@doi [\mnras] {10.1093/mnras/stv1774}, \href {https://ui.adsabs.harvard.edu/abs/2015MNRAS.453.2259V} {453, 2259}

\bibitem[\protect\citeauthoryear{{Vogelsberger}, {Genel}, {Sijacki}, {Torrey}, {Springel}  \& {Hernquist}}{{Vogelsberger} et~al.}{2013}]{2013MNRAS.436.3031V}
{Vogelsberger} M.,  {Genel} S.,  {Sijacki} D.,  {Torrey} P.,  {Springel} V.,   {Hernquist} L.,  2013, \mn@doi [\mnras] {10.1093/mnras/stt1789}, \href {https://ui.adsabs.harvard.edu/abs/2013MNRAS.436.3031V} {436, 3031}

\bibitem[\protect\citeauthoryear{{Vogelsberger} et~al.,}{{Vogelsberger} et~al.}{2014a}]{2014MNRAS.444.1518V}
{Vogelsberger} M.,  et~al., 2014a, \mn@doi [\mnras] {10.1093/mnras/stu1536}, \href {https://ui.adsabs.harvard.edu/abs/2014MNRAS.444.1518V} {444, 1518}

\bibitem[\protect\citeauthoryear{{Vogelsberger} et~al.,}{{Vogelsberger} et~al.}{2014b}]{2014Natur.509..177V}
{Vogelsberger} M.,  et~al., 2014b, \mn@doi [\nat] {10.1038/nature13316}, \href {https://ui.adsabs.harvard.edu/abs/2014Natur.509..177V} {509, 177}

\bibitem[\protect\citeauthoryear{{Vogelsberger}, {Marinacci}, {Torrey}  \& {Puchwein}}{{Vogelsberger} et~al.}{2020a}]{2020NatRP...2...42V}
{Vogelsberger} M.,  {Marinacci} F.,  {Torrey} P.,   {Puchwein} E.,  2020a, \mn@doi [Nature Reviews Physics] {10.1038/s42254-019-0127-2}, \href {https://ui.adsabs.harvard.edu/abs/2020NatRP...2...42V} {2, 42}

\bibitem[\protect\citeauthoryear{{Vogelsberger} et~al.,}{{Vogelsberger} et~al.}{2020b}]{2020MNRAS.492.5167V}
{Vogelsberger} M.,  et~al., 2020b, \mn@doi [\mnras] {10.1093/mnras/staa137}, \href {https://ui.adsabs.harvard.edu/abs/2020MNRAS.492.5167V} {492, 5167}

\bibitem[\protect\citeauthoryear{{Volonteri} \& {Madau}}{{Volonteri} \& {Madau}}{2008}]{2008ApJ...687L..57V}
{Volonteri} M.,  {Madau} P.,  2008, \mn@doi [\apjl] {10.1086/593353}, \href {https://ui.adsabs.harvard.edu/abs/2008ApJ...687L..57V} {687, L57}

\bibitem[\protect\citeauthoryear{{Wang} et~al.,}{{Wang} et~al.}{2018}]{2018ApJ...869L...9W}
{Wang} F.,  et~al., 2018, \mn@doi [\apjl] {10.3847/2041-8213/aaf1d2}, \href {https://ui.adsabs.harvard.edu/abs/2018ApJ...869L...9W} {869, L9}

\bibitem[\protect\citeauthoryear{{Wang} et~al.,}{{Wang} et~al.}{2021}]{2021ApJ...907L...1W}
{Wang} F.,  et~al., 2021, \mn@doi [\apjl] {10.3847/2041-8213/abd8c6}, \href {https://ui.adsabs.harvard.edu/abs/2021ApJ...907L...1W} {907, L1}

\bibitem[\protect\citeauthoryear{{Weibel} et~al.,}{{Weibel} et~al.}{2024}]{2024arXiv240308872W}
{Weibel} A.,  et~al., 2024, \mn@doi [arXiv e-prints] {10.48550/arXiv.2403.08872}, \href {https://ui.adsabs.harvard.edu/abs/2024arXiv240308872W} {p. arXiv:2403.08872}

\bibitem[\protect\citeauthoryear{{Weinberger} et~al.,}{{Weinberger} et~al.}{2017}]{2017MNRAS.465.3291W}
{Weinberger} R.,  et~al., 2017, \mn@doi [\mnras] {10.1093/mnras/stw2944}, \href {https://ui.adsabs.harvard.edu/abs/2017MNRAS.465.3291W} {465, 3291}

\bibitem[\protect\citeauthoryear{{Weinberger} et~al.,}{{Weinberger} et~al.}{2018}]{2018MNRAS.479.4056W}
{Weinberger} R.,  et~al., 2018, \mn@doi [\mnras] {10.1093/mnras/sty1733}, \href {https://ui.adsabs.harvard.edu/abs/2018MNRAS.479.4056W} {479, 4056}

\bibitem[\protect\citeauthoryear{{Weinberger}, {Springel}  \& {Pakmor}}{{Weinberger} et~al.}{2020}]{2020ApJS..248...32W}
{Weinberger} R.,  {Springel} V.,   {Pakmor} R.,  2020, \mn@doi [\apjs] {10.3847/1538-4365/ab908c}, \href {https://ui.adsabs.harvard.edu/abs/2020ApJS..248...32W} {248, 32}

\bibitem[\protect\citeauthoryear{{Willott} et~al.,}{{Willott} et~al.}{2010}]{2010AJ....139..906W}
{Willott} C.~J.,  et~al., 2010, \mn@doi [\aj] {10.1088/0004-6256/139/3/906}, \href {https://ui.adsabs.harvard.edu/abs/2010AJ....139..906W} {139, 906}

\bibitem[\protect\citeauthoryear{{Wise}, {Regan}, {O'Shea}, {Norman}, {Downes}  \& {Xu}}{{Wise} et~al.}{2019}]{2019Natur.566...85W}
{Wise} J.~H.,  {Regan} J.~A.,  {O'Shea} B.~W.,  {Norman} M.~L.,  {Downes} T.~P.,   {Xu} H.,  2019, \mn@doi [\nat] {10.1038/s41586-019-0873-4}, \href {https://ui.adsabs.harvard.edu/abs/2019Natur.566...85W} {566, 85}

\bibitem[\protect\citeauthoryear{{Wolcott-Green}, {Haiman}  \& {Bryan}}{{Wolcott-Green} et~al.}{2017}]{2017MNRAS.469.3329W}
{Wolcott-Green} J.,  {Haiman} Z.,   {Bryan} G.~L.,  2017, \mn@doi [\mnras] {10.1093/mnras/stx167}, \href {https://ui.adsabs.harvard.edu/abs/2017MNRAS.469.3329W} {469, 3329}

\bibitem[\protect\citeauthoryear{{Xu}, {Wise}  \& {Norman}}{{Xu} et~al.}{2013}]{2013ApJ...773...83X}
{Xu} H.,  {Wise} J.~H.,   {Norman} M.~L.,  2013, \mn@doi [\apj] {10.1088/0004-637X/773/2/83}, \href {https://ui.adsabs.harvard.edu/abs/2013ApJ...773...83X} {773, 83}

\bibitem[\protect\citeauthoryear{Xu et~al.,}{Xu et~al.}{2023}]{Xu_2023}
Xu H.,  et~al., 2023, \mn@doi [Research in Astronomy and Astrophysics] {10.1088/1674-4527/acdfa5}, 23, 075024

\bibitem[\protect\citeauthoryear{{Yang} et~al.,}{{Yang} et~al.}{2019}]{2019AJ....157..236Y}
{Yang} J.,  et~al., 2019, \mn@doi [\aj] {10.3847/1538-3881/ab1be1}, \href {https://ui.adsabs.harvard.edu/abs/2019AJ....157..236Y} {157, 236}

\bibitem[\protect\citeauthoryear{{Yue} et~al.,}{{Yue} et~al.}{2023}]{2023arXiv230904614Y}
{Yue} M.,  et~al., 2023, \mn@doi [arXiv e-prints] {10.48550/arXiv.2309.04614}, \href {https://ui.adsabs.harvard.edu/abs/2023arXiv230904614Y} {p. arXiv:2309.04614}

\bibitem[\protect\citeauthoryear{{Zinger} et~al.,}{{Zinger} et~al.}{2020}]{2020MNRAS.499..768Z}
{Zinger} E.,  et~al., 2020, \mn@doi [\mnras] {10.1093/mnras/staa2607}, \href {https://ui.adsabs.harvard.edu/abs/2020MNRAS.499..768Z} {499, 768}

\makeatother
\end{thebibliography}
\end{document}